\documentclass[manuscript]{aastex631}
\usepackage[titletoc,title]{appendix}
\usepackage{CJKutf8}
\usepackage{color}
\usepackage{colortbl}
\usepackage{comment}
\usepackage{gensymb}
\usepackage{latexsym, bm}
\usepackage{makecell}
\usepackage{multirow}
\usepackage{tikz}
\usetikzlibrary{shapes.geometric, arrows}
\usepackage{txfonts}
\usepackage{url}
\usepackage{xparse}

\ExplSyntaxOn
\NewDocumentCommand{\mylist}{m}
 {
  \seq_set_from_clist:Nn \l_tmpa_seq { #1 }
  \int_set:Nn \l_tmpa_int {\seq_count:N \l_tmpa_seq}
  \int_case:nnF {\l_tmpa_int}
   {
    {1} { \seq_item:Nn \l_tmpa_seq {1} }
    {2} { \seq_item:Nn \l_tmpa_seq {1} \ and\ \seq_item:Nn \l_tmpa_seq {2} }
   }
  {
    \seq_pop_right:NN \l_tmpa_seq \l_tmpb_tl
    \seq_map_function:NN \l_tmpa_seq \__mylist_format:n
    \int_compare:nT {\l_tmpa_int > 1} {\ and\ }
    \l_tmpb_tl
  }
 }
\cs_new:Npn \__mylist_format:n #1 {#1,\ }
\ExplSyntaxOff

\graphicspath{{./}{figures/}}

\usepackage[titletoc,title]{appendix}

\shorttitle{Power laws}
\shortauthors{WANG AND QIN}

\begin{document}
\begin{CJK*}{UTF8}{gbsn}

\newcommand{\qinemail}{qingang@hit.edu.cn}
\newcommand{\gqin}{G. Qin (秦刚)}
\newcommand{\gqincode}{\author[0000-0002-3437-3716]{\gqin}}

\newcommand{\wangyangemail}{ywangsz@hit.edu.cn}
\newcommand{\ywang}{Y. Wang (汪洋)}
\newcommand{\ywangcode}{\author[0000-0002-4581-9242]{\ywang}}

\newcommand{\jfwang}{J.-F. Wang (王俊芳)}
\newcommand{\jfwangcode}{\author[0000-0002-9586-093X]{\jfwang}}

\newcommand{\lqiao}{L. Qiao (乔亮)}
\newcommand{\lqiaocode}{\author[0000-0002-4411-2229]{\lqiao}}

\newcommand{\lianemail}{lllian\_hit@163.com}
\newcommand{\lllian}{L.-L. Lian (连乐乐)}
\newcommand{\llliancode}{\author[0000-0002-1097-1084]{\lllian}}
\newcommand{\jinhua}{Jinhua University of Vocational
Technology, Jinhua, Zhejiang, 321000, People’s Republic of China}
\newcommand{\jinhualian}{\jinhua; \lianemail}

\newcommand{\sswu}{S.-S. Wu (吴双双)}
\newcommand{\sswucode}{\author[0000-0002-5776-455X]{\sswu}}

\newcommand{\swang}{S. Wang (王姝)}
\newcommand{\swangcode}{\author[0000-0002-3051-0744]{\swang}}

\newcommand{\xnwang}{X.-N. Wang (王晓男)}
\newcommand{\xnwangcode}{\author[0000-0001-7358-6442]{\xnwang}}

\newcommand{\yszhong}{Y.-S. Zhong (仲雨水)}
\newcommand{\yszhongcode}{\author[0009-0002-7024-7386]{\yszhong}}

\newcommand{\fjkong}{F.-J. Kong (孔凡婧)}
\newcommand{\fjkongcode}{\author[0000-0001-7617-8268]{\fjkong}}
\newcommand{\ncwu}{School of Electronic Engineering, North China University
of Water Resources and Electric Power, Zhengzhou 450046, Peopleʼs Republic 
of China}

\newcommand{\aucorres}{\altaffiliation{Author of correspondence.}}
\newcommand{\hit}{School of
Science, Harbin Institute of
Technology, Shenzhen, 518055,
China}
\newcommand{\szlab}{Shenzhen Key Laboratory of Numerical Prediction for Space Storm,
Harbin Institute of Technology, Shenzhen, 518055, People's Republic of China}

\newcommand{\hitqin}{\hit; \qinemail}
\newcommand{\szlabqin}{\szlab; \qinemail}
\newcommand{\hitwangyang}{\hit; \wangyangemail}
\newcommand{\szlabwangyang}{\szlab; \wangyangemail}


\newcommand\aastex{AAS\TeX}
\newcommand\latex{La\TeX}

\newcommand{\ttsp}{\hspace{0.25em}}
\newcommand{\tsp}{\hspace{0.5em}}
\newcommand{\wsp}{\hspace{1em}}
\newcommand{\vdag}{(v)^\dagger}
\newcommand{\dee}{\mathrm{d}}
\newcommand{\txt}[1]{\mathrm{#1}}


\newcommand{\asr}{Adv. Space Res.}
\newcommand{\ag}{Ann. Geophys.}
\newcommand{\apr}{Appl. Phys. Res.}
\newcommand{\cpam}{Comm. Pure Appl. Math.}
\newcommand{\cpc}{Comput. Phys. Commun.}
\newcommand{\dossr}{DoSSR}
\newcommand{\jasa}{J. American Statistical Association}
\newcommand{\jastp}{J. Atmos. Sol.-Terr. Phys.}
\newcommand{\jcomph}{J. Comput. Phys.}
\newcommand{\joss}{J. Open Source Software}
\newcommand{\jpcs}{J. Phys.: Conf. Ser.}
\newcommand{\jpg}{J. Phys. G: Nucl. Part. Phys.}
\newcommand{\jplp}{J. Plasma Phys.}
\newcommand{\jswsc}{J. Space Weather Space Clim.}
\newcommand{\lrsp}{Living Rev. Sol. Phys.}
\newcommand{\phpl}{Phys. Plasmas}
\newcommand{\phr}{Phys. Rev.}
\newcommand{\rgsp}{Rev. Geophys. Space Phys.}
\newcommand{\rpp}{Reports Progress Phys.}
\newcommand{\sci}{Science}
\newcommand{\spwea}{Space Weather}
\newcommand{\zg}{Z. Geophys.}
\newcommand{\nnsfc}[1]
{
This work was supported by the grants \mylist{#1}.\hspace{-0.5em}
}

\newcommand{\szstp}[1] 
{
This work was supported by the Shenzhen Science and Technology Program under 
Grant \mylist{#1}.\hspace{-0.5em}
}

\newcommand{\nkrdpc}[1]
{
The authors would like to acknowledge the support of National
Key Research and Development Program of China \mylist{#1}.\hspace{-0.5em}
}

\newcommand{\szkllp}[1]
{
This work was supported by
Shenzhen Key Laboratory Launching Project (\mylist{#1}).\hspace{-0.5em}
}

\newcommand{\sprpcas}[1]
{
This work was supported
by the Strategic Priority Research Program of Chinese Academy of Sciences, 
Grant No. \mylist{#1}.\hspace{-0.5em}
}

\newcommand{\QinNNSFCelectrons}{NNSFC\ 42074206}
\newcommand{\QinNNSFCouter}{NNSFC\ 42374190}
\newcommand{\QinSZdisaster}{No.\ JCYJ20210324132812029}
\newcommand{\WangjfNNSFC}{NNSFC\ 42374189}
\newcommand{\GuoNNSFC}{NNSFC\ 42150105}
\newcommand{\GuoNKRDPC}{No.2021YFA0718600}
\newcommand{\ShenNKRDPC}{No.2022YFA1604600}
\newcommand{\FengLAB}{ZDSYS20210702140800001}
\newcommand{\FengCAS}{XDB\ 41000000}

\newcommand{\scTianjin}{The work was carried out at National Supercomputer 
Center in Tianjin, and the calculations were performed on TianHe-1 (A).
\hspace{0.5em}}
\newcommand{\ace}{ACE}
\newcommand{\acesc}{ACE Science Center}
\newcommand{\epam}{EPAM}

\newcommand{\stereo}{STEREO}
\newcommand{\stereosc}{STEREO Science Centers}

\newcommand{\goes}{GOES}
\newcommand{\xrs}{XRS}

\newcommand{\soho}{SOHO}
\newcommand{\celias}{CELIAS}

\newcommand{\halphaflare}{``H-alpha Flare” dataset}

\newcommand{\soon}{USAF Solar Observing Optical Network (SOON)}

\newcommand{\noaa}{NOAA}
\newcommand{\ngdc}{NOAA National Geophysical Data Center (NGDC)}

\newcommand{\cmedata}{CME}
\newcommand{\cmelist}{website \url{https://cdaw.gsfc.nasa.gov/CME_list/}}

\newcommand{\cdaweb}{Database for CDAweb (https://cdaweb.gsfc.nasa.gov/)}
\newcommand{\scinstr}[2]{#1/#2}

\newcommand{\dataprv}[2]{the #1 data are provided by the #2}
\newcommand{\dataprovideby}[1]{We acknowledge that \mylist{#1}.\hspace{0.5em}}
\newcommand{\dataprovide}[1]{We thank the \mylist{#1} for providing the data 
used in this paper.\hspace{0.5em}}
\newcommand{\dataprovavail}[3]{The #1 are prepared using data provided by the 
#2 and made available through the #3.\hspace{0.5em}}
\newcommand{\referee}{The authors thank the anonymous referees for their 
valuable comments.\hspace{0.5em}}

\arraycolsep 0pt

\title{
A Cosmic Ray Acceleration Mechanism Based on Background Flow Velocity 
Inhomogeneities Yielding Power-Law Spectra
}

\correspondingauthor{G. Qin}
\email{\qinemail}

\jfwangcode
\affiliation{\hitqin}

\gqincode
\aucorres
\affiliation{\hitqin}
\affiliation{\szlab}

\begin{abstract}
In modern astrophysics, 
one significant challenge‌ lies in understanding ‌the acceleration 
of cosmic ray particles ‌
along with‌ their power law mechanism‌.
In this article, 
momentum transport generated by the combined effects of 
pitch-angle diffusion
and 
Background Flow Velocity Inhomogeneities 
(BFVIs) is proposed to obtain a 
cosmic rays acceleration mechanism, 
starting from 
the well-known focusing equation describing particle diffusion and acceleration. 
The inhomogeneities of background flow velocity 
is ubiquitous in astrophysical environment.
The isotropic distribution function 
equation of charged energetic particles is derived, 
and its solution is obtained, demonstrating the form of 
momentum power laws of cosmic rays.
In addition, if it is assumed that cosmic rays penetrate compressible plasma waves
or turbulence, for quasi-steady states, the spectral index $\delta$ 
of 
the momentum power law spectrum of cosmic rays
is found to be in the range $[-5, -3]$,
which includes the observed power law indices of 
galactic cosmic rays. 
The results obtained in this article demonstrate that the mechanism proposed 
in this article, 
along with 
shock acceleration, may also contribute to the acceleration 
of galactic cosmic rays. 
Furthermore, when momentum convection effect
and higher-order momentum derivative
terms are considered, 
the indices of power laws should be 
smaller than $-5$. This may explain 
the power laws of 
solar energetic particle events. 

\end{abstract}


\section{INTRODUCTION}
The momentum transport, i.e., particle acceleration, of cosmic rays
is a ubiquitous phenomenon across the Universe,
and this scientific topic continues to attract considerable
attention from numerical, observational, and theoretical perspectives 
\citep{Ruffolo1999, Schlickeiser2002, Shalchi2009, MartinEA2012,
QinEA2013, Zank2014, QinEA2018, Shalchi2020, YangEA2023}. 
Initially, \cite{Fermi1949} pointed out that
plasma turbulence is 
an important agent for 
particle acceleration; and the second-order Fermi acceleration 
by magnetized plasma turbulence has been explored 
by many authors in a variety of 
astrophysical environments
\citep{SchlickeiserEA1998, Chandran2003, PetrosianEA2004}. Later,  
many scientists have discovered that
large populations of particles are accelerated by shock waves in plasmas 
\citep{MartinEA2012}. In addition, magnetic reconnection 
may be responsible for some particle acceleration events,
offering a potential mechanism for the observed phenomena
\citep{RaymondEA2012, CargillEA2012}.
Additionally, the spatial non-uniformity of mean 
magnetic fields leads to the 
adiabatic focusing effect with
acceleration of charged particles, which has been extensively researched 
\citep{SchlickeiserEA2007, SchlickeiserShalchi2008, SchlickeiserJenko2010, WangEA2021}.
For the suprathermal tail 
of the solar wind,
the pump acceleration mechanism 
has been proposed 
\citep{FiskEA2008, FiskEA2010, 
FiskGloeckler2012},
which is generated by perpendicular 
diffusion as well as 
compressive waves of plasmas.  

When charged particles  
interact with
the turbulent magnetic fields,
penetrating magnetized
plasmas,
the irreversible 
stochastic transport processes 
occur, such as gyrophase and  
pitch-angle scattering,
spatial perpendicular 
and parallel diffusion, and so forth
\citep{Schlickeiser2002, 
Shalchi2009, Zank2014, Shalchi2020}. 
Research
on this problem remains 
a highly 
significant topic, encompassing 
plasma physics, astrophysics,
and interplanetary physics. 
The diffusion coefficient 
$D_{\mu\mu}$ must be utilised
in particle kinetic 
equations 
to describe the 
stochastic changes in
particle pitch-angle 
\citep{Shalchi2005, Shalchi2009, QinandShalchi2009, QinandShalchi2014}. 
Initially, \cite{Jokipii1966}
proposed a theory
for determining the
coefficient $D_{\mu\mu}$,
which is well-known in plasma physics
and referred to as the 
quasi-linear theory (QLT). 
However, QLT proves to be 
inaccurate
and provides inadequate 
description of 
pitch-angle scattering 
in plasma environments
\citep{Shalchi2009}. 
In order to address this problem, 
the second-order quasi-linear theory (SOQLT) 
has been proposed 
and further explored
\citep{Shalchi2005, ShalchiEA2009}.
When plasma turbulence is not 
overly weak,
the isotropic form of the pitch-angle
diffusion coefficient, 
$D_{\mu\mu}=D(1-\mu^2)$
with the constant parameter $D$,
is frequently utilised 
\citep{Shalchi2006, ShalchiEA2009, 
Shalchi2012,
LitvinenkoNoble2016, 
LasuikShalchi2017, 
Shalchi2018, 
WangEA2018, 
WangEA2019, WangEA2020}.

It is possible that 
the two significant scientific 
topics of astrophysics, namely
acceleration of cosmic rays and spatial diffusion, i.e., the parallel, 
perpendicular, pitch-angle, and gyrotropic diffusion, may be interrelated.  
\cite{ChoEA2006} suggested that
the pitch-angle diffusion might
be associated with
particle momentum
transport. Subsequently, 
some scientists \citep{FiskEA2008, FiskEA2010, FiskGloeckler2012}
developed the pump acceleration 
mechanism, which arises from  
the combined effect of 
perpendicular diffusion 
and compressional waves,
commencing with the 
Parker equation. 
The well-known focusing equation,
which encompasses 
more physical information
than the Parker equation
describing the transport 
of energetic particles in
background flow plasmas 
\citep{QinEA2005, QinEA2006, ZhangEA2009, DrogeEA2010, ZuoEA2011, WangEA2012}, 
has found extensive utilization in related
astrophysics and space science research
\citep{ZhangEA2009, QinEA2013, ZuoEA2013, WangEA2014, ZhangEA2019,
QinandQiEA2020, WangEA2024}. 
This transport equation  
actually includes
almost all essential physical 
processes,
such as perpendicular diffusion
\citep{Skilling1971, ZhangEA2009},
as well as spatial and 
pitch-angle  
convection scattering
\citep{ForbesEA2006,  
ZhangandZhao2017,
WijsenEA2019, ZhangEA2019, 
BianEA2019, BianEA2020}.
In addition, 
the stochastic acceleration
of charged particles 
resulting from turbulent electric 
fields
\citep{Schlickeiser2002}, 
as well as the adiabatic 
focusing effect, are also incorporated
\citep{Roelof1969, Earl1976, Kunstmann1979, BeeckEA1986, BieberEA1990, DanosEA2013, ShalchiEA2011, ShalchiDanos2013,
Litvinenko2012a, Litvinenko2012b, 
WangEA2018, WangEA2019, WangEA2020, 
WangEA2024}. 

For galactic cosmic rays of 
relativistic energies,
the momentum power laws with
the spectral index $-5.1\sim -4.7$
have been found 
\citep{Jokipii2001, ChangEA2008}.
For solar energetic particles, 
on the other hand, the 
single and double momentum
(or kinetic energy flux) 
power laws have
been reported 
\citep{LiuEA2022, YangEA2023}, 
with the spectral indices
in a wide range as high as $-3$,
and as low as $-9$. 

In this article, 
we invesitgate the momentum 
transport caused by 
pitch-angle diffusion and  
Background Flow Velocity Inhomogeneities (BFVIs),
to derive the powe laws of cosmic rays spectra 
from the focusing equation. 
This article is organised as follows. In Section 
\ref{The focusing equation
	and isotropic distribution function 
	equation}, 
the focusing equation
is introduced, and the 
isotropic distribution function 
equation, which incorporates 
the terms $T_1$ and $T_2$,
as well as the anisotropic distribution
function, are derived. 
In Section \ref{Evaluating T1 and T2}, 
the terms $T_1$ and $T_2$
are evaluated, and then the derivation of 
the governing equation 
for the isotropic distribution function with momentum 
transport and its solution, 
are presented in Section 
\ref{The solution of isotropic distribution function}.
In Section \ref{Discussion},
the physical meanings of 
the momentum transport terms, 
obtained in the preceding
sections, are explored, 
followed by 
the investigation into 
the momentum power laws 
of these terms.
The results are presented 
in Section 
\ref{SUMMARY AND CONCLUSION}.

\section{The focusing equation
and isotropic distribution function 
equation}
\label{The focusing equation
and isotropic distribution 
function equation}

\subsection{The focusing equation}
\label{The focusing equation}

The focusing equation, 
describing the charged particle
transport in magnetized plasmas,
is widely used  
by physicists in space and plasma 
physics
\citep{Skilling1971, Schlickeiser2002,
QinEA2005, QinEA2006, ZhangEA2009,
DrogeEA2010, ZuoEA2011, WangEA2012,
QinEA2013, WangEA2014, ZhangEA2019,
QinandQiEA2020, WangEA2024}. 
The version of the focusing equation,
satisfying the particle
conservation law,  
is presented as follows \citep{WangEA2024}
\begin{eqnarray}
\frac{\partial{f}}{\partial{t}}
=&&\nabla\cdot\left(\kappa_\bot\cdot 
\nabla f\right)
-\nabla\cdot \left[\left(v\mu 
\hat{\bm{b}}+\bm{V}\right)f\right]
+\frac{\partial{}}{\partial{\mu}}
\left(D_{\mu\mu}\frac{\partial{f}}
{\partial{\mu}}\right)
\nonumber\\
&&
+\frac{1}{p^2}\frac{\partial{}}
{\partial{p}}
\Bigg\{p^3\Bigg[\frac{1-\mu^2}
{2}\left(\nabla\cdot \bm{V}-\hat{\bm{b}}\hat{\bm{b}}
:\nabla \bm{V}\right)
+\mu^2\hat{\bm{b}}
\hat{\bm{b}}:\nabla \bm{V}
\Bigg]f\Bigg\}
\nonumber\\
&&
+\frac{\partial{}}{\partial{\mu}}
\Bigg\{
\frac{1-\mu^2}{2}\Bigg[-\frac{v}{L}-
\mu\left(\nabla\cdot \bm{V}-3\hat{\bm{b}}\hat{\bm{b}}
:\nabla \bm{V}\right)\Bigg]f
\Bigg\}.
\label{equation satisfying particle 
number conservation}
\end{eqnarray}
Here, $f=f(x, y, z, \mu, p, t)$ 
is the charged energetic particle
distribution function,
$t$ is time, 
$x, y, z$ are spatial coordinates, 
$\nabla$ is the
three dimensional nabla operator,
$\kappa_\bot$ is the 
perpendicular 
diffusion coefficient,
$v$ is the speed of particle,
$\mu$ is pitch-angle cosine, 
$\hat{\bm{b}}$ is the unit vector
along the mean magnetic field,
$p$ is the radial component of 
particle momentum, 
$D_{\mu\mu}(\mu)$ is diffusion 
coefficient of 
pitch-angle, $L$ is the 
adiabaic focusing
characteristic length,
and $\bm{V}$ is the background flow
velocity, respectively. 
In this article, the study focuses on
momentum transport
in an even background magnetic
field, with  
the adiabatic focusing effect
disregarded. Thus,
the focusing 
equation is simplified to
\begin{eqnarray}
\frac{\partial{f}}{\partial{t}}
=&&\nabla\cdot\left(\kappa_\bot\cdot \nabla f\right)
-v\mu\frac{\partial{f}}{\partial{z}}
-\nabla\cdot \left(\bm{V}f\right)
+\frac{\partial{}}{\partial{\mu}}
\left(D_{\mu\mu}\frac{\partial{f}}{\partial{\mu}}\right)
\nonumber\\
&&
+\frac{1}{p^2}\frac{\partial{}}{\partial{p}}
\Bigg\{p^3\Bigg[\frac{1-\mu^2}{2}
\nabla\cdot \bm{V}
+\frac{3\mu^2-1}{2}\frac{\partial{V_z}}
{\partial{z}}
\Bigg]f\Bigg\}
+\frac{\partial{}}{\partial{\mu}}
\Bigg\{
\frac{1-\mu^2}{2}\Bigg[-
\mu\left(\nabla\cdot \bm{V}-
3\frac{\partial{V_z}}{\partial{z}}
\right)\Bigg]f
\Bigg\},
\label{starting point}	
\end{eqnarray}  
which serves as the 
starting point of
our research in this article.

\subsection{The isotropic distribution 
function equation with anisotropic 
distribution function} 
\label{The isotropic distribution 
function equation with anisotropic 
distribution function}

The distribution function 
$f(x, y, z,\mu,p,t)$
of charged energetic particles 
can be decomposed into 
an isotropic part 
$F(x, y, z,p,t)$
and 
an anisotropic part 
$g(x, y, z,\mu,p,t)$
\citep{BeeckEA1986, BieberEA1990, 
Shalchi2009a, ArtmannEA2011, LitvinenkoSchlickeiser2013, LitvinenkoEA2015,
LitvinenkoNoble2016, WangEA2017, WangEA2018, WangEA2019, WangEA2020, WangEA2023, WangEA2024}
\begin{eqnarray}
f(x, y, z,\mu,p,t)
=F(x, y, z,p,t)+g(x, y, z,\mu,p,t). 
\label{f=F+g}
\end{eqnarray}
With the latter formula,
Equation (\ref{starting point}) 
becomes
\begin{eqnarray}
\frac{\partial{F}}{\partial{t}}
&&+\frac{\partial{g}}{\partial{t}}
=\nabla\cdot\left(\kappa_\bot\cdot \nabla F\right)
+\nabla\cdot\left(\kappa_\bot\cdot \nabla g\right)
-v\mu\frac{\partial{F}}{\partial{z}}
-v\mu\frac{\partial{g}}{\partial{z}}
-\nabla\cdot \left(\bm{V}F\right)
-\nabla\cdot \left(\bm{V}g\right)
+\frac{\partial{}}{\partial{\mu}}
\left(D_{\mu\mu}\frac{\partial{g}}{\partial{\mu}}\right)
\nonumber\\
&&
+\frac{1}{p^2}\frac{\partial{}}{\partial{p}}
\Bigg\{p^3\Bigg[\frac{1-\mu^2}{2}
\nabla\cdot \bm{V}
+\frac{3\mu^2-1}{2}\frac{\partial{V_z}}
{\partial{z}}
\Bigg]F\Bigg\}
+\frac{1}{p^2}\frac{\partial{}}{\partial{p}}
\Bigg\{p^3\Bigg[\frac{1-\mu^2}{2}
\nabla\cdot \bm{V}
+\frac{3\mu^2-1}{2}\frac{\partial{V_z}}
{\partial{z}}
\Bigg]g\Bigg\}
\nonumber\\
&&
+\frac{\partial{}}{\partial{\mu}}
\Bigg\{
\frac{1-\mu^2}{2}\Bigg[-
\mu\left(\nabla\cdot \bm{V}-
3\frac{\partial{V_z}}{\partial{z}}
\right)\Bigg]F
\Bigg\}
+\frac{\partial{}}{\partial{\mu}}
\Bigg\{
\frac{1-\mu^2}{2}\Bigg[-
\mu\left(\nabla\cdot \bm{V}-
3\frac{\partial{V_z}}{\partial{z}}
\right)\Bigg]g
\Bigg\}.	
\label{F and g equation}
\end{eqnarray}
By integrating Equation 
(\ref{F and g equation})
with respect to the 
pitch-angle cosine $\mu$
over the range from 
from $-1$ to $1$,
the isotropic 
distribution
function equation 
is derived 
\begin{eqnarray}
\frac{\partial{F}}{\partial{t}}	
=&&
\nabla\cdot\left(\kappa_\bot\cdot \nabla F\right)
-\nabla \cdot\left(\bm{V} F\right)
+\frac{1}{3}
\left(\nabla\cdot\bm{V} \right)
\frac{1}{p^2}
\frac{\partial{}}{\partial{p}}
\left(
 p^3F\right)
+T_1+T_2
\label{F with T}	
\end{eqnarray}
with 
\begin{eqnarray}
&&T_1=\frac{1}{p^2}\frac{\partial{}}{\partial{p}}
\Bigg[
\frac{1}{2}\int_{-1}^1\dee\mu
\frac{1-\mu^2}{2}g
p^3\nabla\cdot \bm{V}
+\frac{1}{2}\int_{-1}^1\dee\mu
\frac{3\mu^2-1}{2}g
p^3\frac{\partial{V_z}}
{\partial{z}}
\Bigg],
\label{T1}\\
&&T_2=-\frac{v}{2}\frac{\partial{}}
{\partial{z}}\int_{-1}^{1}\dee 
\mu\mu g,
\label{T2}	
\end{eqnarray}
It is evident that Equation 
(\ref{F with T}) encompasses
the anisotropic distribution 
function $g(x, y, z,\mu,p,t)$. 
Previous research indicates that
the spatial parallel diffusion term, 
$\kappa_\parallel \partial^2{F}
/\partial{z^2}$, 
can be derived from   
the term $T_2$ in Equation
(\ref{F with T}) 
\citep{WangEA2018, WangEA2019,
WangEA2020, WangEA2024}.
Once the parallel and perpendicular 
diffusion terms have been merged,
the lowest-order form 
of Equation 
(\ref{F with T}) is presented 
as follows
\begin{eqnarray}
\frac{\partial{F}}{\partial{t}}	
=&&
\nabla\cdot\left(\kappa
\cdot \nabla F\right)
-\nabla \cdot\left(\bm{V} F\right)
+
\frac{1}{3}
\left(\nabla\cdot\bm{V} \right)
\frac{\partial{F}}
{\partial{\ln p}},
\label{Parker equation}		
\end{eqnarray}
which is the well-known Parker equation.
Here, $\kappa$ is spatial diffusion 
tensor. 

This article solely explores 
the momentum transport
resultant from
the combined effects 
of pitch-angle diffusion 
and BFVIs. 
The terms expressing
momentum transport, 
containing the anisotropic
distribution function 
$g(x, y, z,\mu,p,t)$, 
are to be encompassed 
within $T_1$ and 
$T_2$
\citep{SchlickeiserJenko2010}. Therefore, 
in the following subsection, 
the specific formula
of $g(x, y, z,\mu,p,t)$
will be derived. 

\subsection{The anisotropic distribution
function}
\label{The anisotropic distribution
function of charged particles}

Upon integrating Equation 
(\ref{F and g equation}) 
with respect to 
pitch-angle cosine
from $-1$ to $\mu$, we obtain 
\begin{eqnarray}
\frac{\partial{F}}{\partial{t}}
&&(\mu+1)
+\int_{-1}^{\mu}\dee\mu
\frac{\partial{g}}{\partial{t}}
=\nabla\cdot\left(\kappa_\bot\cdot \nabla F\right)(\mu+1)
+\nabla\cdot\left(\kappa_\bot\cdot \nabla \int_{-1}^{\mu}\dee\mu g\right)
-v\frac{\mu^2-1}{2}\frac{\partial{F}}{\partial{z}}
-v\frac{\partial{}}{\partial{z}}
\int_{-1}^{\mu}\dee\mu\mu g
\nonumber\\
&&
-\nabla\cdot \left(\bm{V}F\right)(\mu+1)
-\nabla\cdot \left(\bm{V}
\int_{-1}^{\mu}\dee\mu g\right)
+D_{\mu\mu}\frac{\partial{g}}{\partial{\mu}}
+\frac{1}{p^2}\frac{\partial{}}{\partial{p}}
\Bigg\{p^3\Bigg[
\int_{-1}^{\mu}\dee\mu\frac{1-\mu^2}{2}
\nabla\cdot \bm{V}
+\int_{-1}^{\mu}\dee\mu
\frac{3\mu^2-1}{2}\frac{\partial{V_z}}
{\partial{z}}
\Bigg]F\Bigg\}
\nonumber\\
&&
+\frac{1}{p^2}\frac{\partial{}}{\partial{p}}
\Bigg[
\int_{-1}^{\mu}\dee\mu\frac{1-\mu^2}{2}
gp^3\nabla\cdot \bm{V}
+\int_{-1}^{\mu}\dee\mu
\frac{3\mu^2-1}{2}g
p^3\frac{\partial{V_z}}
{\partial{z}}
\Bigg]
\nonumber\\
&&
+
\frac{1-\mu^2}{2}\Bigg[-
\mu\left(\nabla\cdot \bm{V}-
3\frac{\partial{V_z}}{\partial{z}}
\right)\Bigg]F
+
\frac{1-\mu^2}{2}\Bigg[-
\mu\left(\nabla\cdot \bm{V}-
3\frac{\partial{V_z}}{\partial{z}}
\right)\Bigg]g. 
\end{eqnarray}
The latter equation can be rewritten as
\begin{eqnarray}
\frac{\partial{g}}
{\partial{\mu}}
=\frac{1}{D_{\mu\mu}}\Phi(\mu, p, t)
\label{g/mu}
\end{eqnarray}
with
\begin{eqnarray}
\Phi(\mu, p, t)
=&&-\nabla\cdot\left(\kappa_\bot\cdot \nabla F\right)(\mu+1)
-\int_{-1}^{\mu}\dee\mu
\nabla\cdot\left(\kappa_\bot\cdot \nabla  g\right)
+v\frac{\mu^2-1}{2}\frac{\partial{F}}{\partial{z}}
+v
\int_{-1}^{\mu}\dee\mu\mu \frac{\partial{g}}{\partial{z}}
\nonumber\\
&&
+F\left(\nabla\cdot \bm{V}\right)(\mu+1)
+(\mu+1)\bm{V}\cdot \nabla F
+\int_{-1}^{\mu}\dee\mu g
\left(\nabla\cdot \bm{V}\right)
+\bm{V}\cdot
\int_{-1}^{\mu}\dee\mu \nabla g
\nonumber\\
&&
-\frac{1}{p^2}\frac{\partial{}}{\partial{p}}
\Bigg\{p^3\Bigg[
\int_{-1}^{\mu}\dee\mu\frac{1-\mu^2}{2}
\nabla\cdot \bm{V}
+\int_{-1}^{\mu}\dee\mu
\frac{3\mu^2-1}{2}\frac{\partial{V_z}}
{\partial{z}}
\Bigg]F\Bigg\}
\nonumber\\
&&
-\frac{1}{p^2}\frac{\partial{}}{\partial{p}}
\Bigg[
\int_{-1}^{\mu}\dee\mu\frac{1-\mu^2}{2}
gp^3\nabla\cdot \bm{V}
+\int_{-1}^{\mu}\dee\mu
\frac{3\mu^2-1}{2}g
p^3\frac{\partial{V_z}}
{\partial{z}}
\Bigg]
\nonumber\\
&&
-\frac{1-\mu^2}{2}\Bigg[-
\mu\left(\nabla\cdot \bm{V}-
3\frac{\partial{V_z}}{\partial{z}}
\right)\Bigg]F
-\frac{1-\mu^2}{2}\Bigg[-
\mu\left(\nabla\cdot \bm{V}-
3\frac{\partial{V_z}}{\partial{z}}
\right)\Bigg]g
\nonumber\\
&&
+\frac{\partial{F}}{\partial{t}}
(\mu+1)
+\int_{-1}^{\mu}\dee\mu
\frac{\partial{g}}{\partial{t}}. 
\label{Phi}
\end{eqnarray}
Integrating Equation (\ref{g/mu})
over $\mu$ from $-1$ to $\mu$ yields
\begin{eqnarray}
g(x, y, z, \mu, p, t)
=C+\int_{-1}^{\mu}\dee\mu
\frac{1}{D_{\mu\mu}}\Phi(\mu)
\label{g with C}
\end{eqnarray}
with 
\begin{eqnarray}
C=g(x, y, z, \mu=-1, p, t). 
\end{eqnarray}
To integrate Equation 
(\ref{g with C})
with respect to $\mu$ from $-1$ to $1$,
one finds 
\begin{eqnarray}
	0=2C+\int_{-1}^{1}\dee\mu
	\int_{-1}^{\mu}\dee\mu
	\frac{1}{D_{\mu\mu}}\Phi(\mu)
	=2C+
	\int_{-1}^{1}\dee\mu
	\frac{1-\mu}{D_{\mu\mu}}\Phi(\mu). 	
\end{eqnarray}
Here, the integration by parts
is used in the second term on the 
right-hand side of the equals sign.
To proceed, 
the constant $C$ can be obtained
from the latter formula as follows
\begin{eqnarray}
	C=-\frac{1}{2}
	\int_{-1}^{1}\dee\mu
	\frac{1-\mu}{D_{\mu\mu}}\Phi(\mu). 	
\end{eqnarray}
Inserting the latter formula 
into Equation (\ref{g with C})
yields the anisotropic distribution 
function
\begin{eqnarray}
g(x, y, z,\mu,p,t)=\int_{-1}^{\mu}\dee \mu \frac{1}{D_{\mu\mu}}\Phi(\mu)
-\frac{1}{2}
\int_{-1}^{1}\dee \mu 
\frac{1-\mu}{D_{\mu\mu}}\Phi(\mu). 
\label{g}
\end{eqnarray}
By employing the latter formula,
the terms $T_1$ and $T_2$ can be
evaluated. 

\section{Evaluating 
$T_1$ and $T_2$}
\label{Evaluating T1 and T2}

\subsection{The term $T_1$}

Using the anisotropic 
function $g(x, y, z,\mu,p,t)$ 
(\ref{g}), we find  
\begin{eqnarray}
T_1=\frac{1}{p^2}\frac{\partial{}}{\partial{p}}
\Bigg[
\frac{1}{2}\int_{-1}^1\dee\mu
\frac{1-\mu^2}{2}
p^3\nabla\cdot \bm{V}
+\frac{1}{2}\int_{-1}^1\dee\mu
\frac{3\mu^2-1}{2}
p^3\frac{\partial{V_z}}
{\partial{z}}
\Bigg]g
=\sum_{n=1}^{16}Z_n.	
\end{eqnarray} 
Here, the formulas for
$Z_1\sim Z_{16}$ are provided 
in Appendix 
\ref{The formilas of Zn}.

\subsubsection{Evaluating $Z_1\sim
Z_{16}$}
\label{Evaluating Z1-Z16}
Numerous studies indicates that 
a minimal ratio of the perpendicular
to parallel diffusion coefficients
of energetic particles
\citep{Palmer1982, JokipiiEA1995, Ferrando1997, BurgerEA2000, FerreiraEA2001}. 
Thus, 
within this article, 
the perpendicular 
diffusion effects are 
disregarded, 
and respective terms are set 
to zero. 
Therefore,
we can have
\begin{eqnarray}
&&Z_1=-\frac{1}{p^2}\frac{\partial{}}{\partial{p}}
\Bigg[
\frac{1}{2}\int_{-1}^1\dee\mu
\frac{1-\mu^2}{2}
p^3\nabla\cdot \bm{V}
+\frac{1}{2}\int_{-1}^1\dee\mu
\frac{3\mu^2-1}{2}
p^3\frac{\partial{V_z}}
{\partial{z}}
\Bigg]
\nonumber\\
&&
\times
\Bigg(\int_{-1}^{\mu}\dee \mu \frac{1}{D_{\mu\mu}}
-\frac{1}{2}
\int_{-1}^{1}\dee \mu 
\frac{1-\mu}{D_{\mu\mu}}\Bigg)(\mu+1)
\nabla\cdot\left(\kappa_\bot\cdot \nabla F\right)=0,
\nonumber\\
&&
Z_2=-\frac{1}{p^2}\frac{\partial{}}{\partial{p}}
\Bigg[
\frac{1}{2}\int_{-1}^1\dee\mu
\frac{1-\mu^2}{2}
p^3\nabla\cdot \bm{V}
+\frac{1}{2}\int_{-1}^1\dee\mu
\frac{3\mu^2-1}{2}
p^3\frac{\partial{V_z}}
{\partial{z}}
\Bigg]
\nonumber\\
&&
\times
\Bigg(\int_{-1}^{\mu}\dee \mu \frac{1}{D_{\mu\mu}}
-\frac{1}{2}
\int_{-1}^{1}\dee \mu 
\frac{1-\mu}{D_{\mu\mu}}\Bigg)
\int_{-1}^{\mu}\dee\mu
\nabla\cdot\left(\kappa_\bot\cdot \nabla g\right)=0.
\end{eqnarray}
Due to the power laws 
of isotropic distribution function
in quasi-steady state will be 
investigated in 
Section 
\ref{Discussion},
the time effect 
of distribution function, i.e.,
$\partial{F}/\partial{t}$
and $\partial{g}/\partial{t}$,
are henceforth disregarded 
in this article. 
Thus, we can obtain
\begin{eqnarray}
&&
Z_{15}=\frac{1}{p^2}\frac{\partial{}}{\partial{p}}
\Bigg[
\frac{1}{2}\int_{-1}^1\dee\mu
\frac{1-\mu^2}{2}
p^3\nabla\cdot \bm{V}
+\frac{1}{2}\int_{-1}^1\dee\mu
\frac{3\mu^2-1}{2}
p^3\frac{\partial{V_z}}
{\partial{z}}
\Bigg]
\nonumber\\
&&
\times
\Bigg(\int_{-1}^{\mu}\dee \mu \frac{1}{D_{\mu\mu}}
-\frac{1}{2}
\int_{-1}^{1}\dee \mu 
\frac{1-\mu}{D_{\mu\mu}}\Bigg)
\frac{\partial{F}}{\partial{t}}
(\mu+1)=0,
\nonumber\\
&&
Z_{16}=\frac{1}{p^2}\frac{\partial{}}{\partial{p}}
\Bigg[
\frac{1}{2}\int_{-1}^1\dee\mu
\frac{1-\mu^2}{2}
p^3\nabla\cdot \bm{V}
+\frac{1}{2}\int_{-1}^1\dee\mu
\frac{3\mu^2-1}{2}
p^3\frac{\partial{V_z}}
{\partial{z}}
\Bigg]
\nonumber\\
&&
\times
\Bigg(\int_{-1}^{\mu}\dee \mu \frac{1}{D_{\mu\mu}}
-\frac{1}{2}
\int_{-1}^{1}\dee \mu 
\frac{1-\mu}{D_{\mu\mu}}\Bigg)
\int_{-1}^{\mu}\dee\mu
\frac{\partial{g}}
{\partial{t}}=0. 	
\end{eqnarray}
To continue, we evaluate the third term
of $Z_n$
\begin{eqnarray}
Z_3=&&\frac{1}{p^2}\frac{\partial{}}{\partial{p}}
\Bigg[
\frac{1}{2}\int_{-1}^1\dee\mu
\frac{1-\mu^2}{2}
p^3\nabla\cdot \bm{V}
+\frac{1}{2}\int_{-1}^1\dee\mu
\frac{3\mu^2-1}{2}
p^3\frac{\partial{V_z}}
{\partial{z}}
\Bigg]
\nonumber\\
&&
\times
\Bigg(\int_{-1}^{\mu}\dee \mu \frac{1}{D_{\mu\mu}}
-\frac{1}{2}
\int_{-1}^{1}\dee \mu 
\frac{1-\mu}{D_{\mu\mu}}\Bigg)
v\frac{\mu^2-1}{2}
\frac{\partial{F}}{\partial{z}},	
\end{eqnarray}  
which contains the pitch-angle 
diffusion coefficient 
$D_{\mu\mu}$. 
If turbulence is not too weak,
the pitch-angle diffusion coefficient,
$D_{\mu\mu}$, approximately conforms to
the isotropic scattering form
\citep{Shalchi2006, ShalchiEA2009}
\begin{eqnarray} 
D_{\mu\mu}=D(1-\mu^2).
\label{Dmumu}
\end{eqnarray} 
Here, $D$ is a constant. 
With the latter formula, 
the formula of $Z_3$ becomes
\begin{eqnarray}
Z_3
=-\frac{1}{4D}
\frac{1}{p^2}\frac{\partial{}}{\partial{p}}p^3
\Bigg[
\int_{-1}^1\dee\mu
\frac{1-\mu^2}{2}\mu
\nabla\cdot \bm{V}
+\int_{-1}^1\dee\mu
\frac{3\mu^2-1}{2}\mu
\frac{\partial{V_z}}
{\partial{z}}
\Bigg]
v
\frac{\partial{F}}{\partial{z}}=0. 
\end{eqnarray}
To proceed, 
inserting the anisotropic 
distribution function 
$g(x, y, z,\mu,p,t)$ into
the formula of $Z_4$ yields
\begin{eqnarray}
Z_4&&=\frac{1}{p^2}\frac{\partial{}}{\partial{p}}
\Bigg[
\frac{1}{2}\int_{-1}^1\dee\mu
\frac{1-\mu^2}{2}
p^3\nabla\cdot \bm{V}
+\frac{1}{2}\int_{-1}^1\dee\mu
\frac{3\mu^2-1}{2}
p^3\frac{\partial{V_z}}
{\partial{z}}
\Bigg]
\nonumber\\
&&
\times
\Bigg(\int_{-1}^{\mu}\dee \mu \frac{1}{D_{\mu\mu}}
-\frac{1}{2}
\int_{-1}^{1}\dee \mu 
\frac{1-\mu}{D_{\mu\mu}}\Bigg)
v\frac{\partial{}}{\partial{z}}
\int_{-1}^{\mu}\dee\mu\mu g
\nonumber\\
&&
=\sum_{n=1}^{12}Z_4(n).
\end{eqnarray}
The specific formulas of $Z_4(n)$
are listed in Appendix 
\ref{Evaluating Zn}. 
Obviously, some formulas, e.g., 
$Z_4(2)$, $Z_4(5)$, etc, 
again contain the anisotropic 
distribution function 
$g(x, y, z,\mu,p,t)$. 
In this article, 
the assumption regarding 
the gentle spatial 
change of background flow velocity
is used, and 
the second and higher order spatial
gradients of background flow
velocity components
and the divergence 
of background flow velocity
are disregarded. 
In addition, the third and higher
power of spatial
gradients of background flow velocity 
components and the divergence
of background flow velocity
are also ignored. Moreover,
we only consider
the 
first and second derivative terms
of isotropic distribution function,
and ignore the terms with
the derivatives of higher order. 
Based on the above assumptions,
the formulas of 
$Z_4(1)\sim Z_4(12)$ are 
evaluated and 
listed in Appendix
\ref{Evaluating Zn}, 
allowing $Z_4$ to be
subsequently determined 
\begin{eqnarray}
Z_4&&=\sum_{n=1}^{12}Z_4(n)
\nonumber\\
&&
=
\Bigg[
\nabla\cdot \bm{V}
-3
\frac{\partial{V_z}}
{\partial{z}}
\Bigg]
\frac{1}{180D^2}
\frac{1}{m^2}
\frac{1}{p^2}\frac{\partial{}}{\partial{p}}p^5
\frac{\partial^2{F}}{\partial{z^2}}
\nonumber\\
&&
+\Bigg[
-2
\frac{\partial{V_z}}
{\partial{z}}
\left(\nabla\cdot \bm{V}\right)
+3
\left(
\frac{\partial{V_z}}
{\partial{z}}\right)^2
\Bigg]
\frac{1}{72D^2m}
\frac{1}{p^2}\frac{\partial{}}{\partial{p}}
\Bigg\{
p^4
\frac{\partial{F}}{\partial{z}}
\Bigg\}
\nonumber\\
&&
+
\Bigg[
-V_z
\left(\nabla\cdot \bm{V}\right)
+3V_z
\frac{\partial{V_z}}
{\partial{z}}
\Bigg]
\frac{1}{72D^2m}
\frac{1}{p^2}\frac{\partial{}}{\partial{p}}
\Bigg(
p^4\frac{\partial^2{F}}
{\partial{z^2}}
\Bigg)
\nonumber\\
&&
+
\Bigg[
\left(\nabla\cdot \bm{V}\right)^2
-4
\nabla\cdot \bm{V}
\frac{\partial{V_z}}
{\partial{z}}
\Bigg]
\frac{1}{216mD^2}
\frac{1}{p^2}
\frac{\partial{}}{\partial{p}}
p^5
\frac{\partial{}}{\partial{p}}
\frac{\partial{F}}{\partial{z}}. 
\end{eqnarray} 
Similarly, the terms $Z_5, Z_6,
\cdots, Z_{14}$  are also
evaluated, 
and the results are 
listed in Appendix
\ref{Evaluating Zn}. 

\subsubsection{Evaluating $T_1$}
\label{Evaluating T1}

With the formulas
listed in  Appendix
\ref{Evaluating Zn}, the term 
$T_1$ becomes
\begin{eqnarray}
T_1&&=\sum_{n=1}^{16}Z_n
=A_0+A_{zp2}+A_{zp3}
\label{T1 with A}	
\end{eqnarray}
with
\begin{eqnarray}
&&A_0=
\frac{1}{90D}
\Bigg[
2
(\nabla\cdot \bm{V})^2
-7
\frac{\partial{V_z}}{\partial{z}}
\nabla\cdot \bm{V}
+3
\left(\frac{\partial{V_z}}
{\partial{z}}\right)^2
\Bigg]
\frac{1}{p^2}\frac{\partial{}}{\partial{p}}
\left(p^4\frac{\partial{F}}{\partial{p}}\right), 
\label{A0}\\
&&A_{zp2}=
\Bigg[
-2
\frac{\partial{V_z}}
{\partial{z}}
\left(\nabla\cdot \bm{V}\right)
+3
\left(
\frac{\partial{V_z}}
{\partial{z}}\right)^2
\Bigg]
\frac{1}{72D^2m}
\frac{1}{p^2}\frac{\partial{}}{\partial{p}}
\Bigg(
p^4
\frac{\partial{F}}{\partial{z}}
\Bigg)
\nonumber\\
&&
+
\Bigg[
\left(\nabla\cdot \bm{V}\right)^2
-4
\nabla\cdot \bm{V}
\frac{\partial{V_z}}
{\partial{z}}
\Bigg]
\frac{1}{216mD^2}
\frac{1}{p^2}
\frac{\partial{}}{\partial{p}}
\Bigg(
p^5
\frac{\partial{}}{\partial{p}}
\frac{\partial{F}}{\partial{z}}
\Bigg)
\nonumber\\
&&
+
\Bigg[
3
V_z\frac{\partial{V_z}}
{\partial{z}}
-
V_z\nabla\cdot \bm{V}
\Bigg]
\frac{1}{18D}
\frac{1}{p^2}\frac{\partial{}}{\partial{p}}
\left(
p^3
\frac{\partial{F}}{\partial{z}}
\right)
\nonumber\\
&&
+
\Bigg[
-5
\left(\nabla\cdot \bm{V}\right)^2
+
21
\frac{\partial{V_z}}
{\partial{z}}
\left(\nabla\cdot \bm{V}\right)
\Bigg]V_z
\frac{1}{216D^2}
\frac{1}{p^2}
\frac{\partial{}}{\partial{p}}
\Bigg(
p^3
\frac{\partial{F}}{\partial{z}}
\Bigg)
\nonumber\\
&&
+
\Bigg[
-2
\left(\nabla\cdot \bm{V}\right)^2
+7
\left(\nabla\cdot \bm{V}\right)
\left(\frac{\partial{V_z}}
{\partial{z}}\right)
-3
\left(\frac{\partial{V_z}}
{\partial{z}}\right)^2
\Bigg]
\frac{1}{p^2}
\frac{\partial{}}{\partial{p}}
\Bigg(
p^4
\frac{\partial{}}{\partial{p}}
\frac{\partial{F}}{\partial{z}}
\Bigg), 
\label{Az1}
\\
&&A_{zp3}=
\Bigg[
\nabla\cdot \bm{V}
-3
\frac{\partial{V_z}}
{\partial{z}}
\Bigg]
\frac{1}{180D^2}
\frac{1}{m^2}
\frac{1}{p^2}
\frac{\partial{}}{\partial{p}}
\Bigg(
p^5
\frac{\partial^2{F}}{\partial{z^2}}
\Bigg)
\nonumber\\
&&
+
\Bigg[
-V_z
\left(\nabla\cdot \bm{V}\right)
+3V_z
\frac{\partial{V_z}}
{\partial{z}}
\Bigg]
\frac{1}{72D^2m}
\frac{1}{p^2}\frac{\partial{}}{\partial{p}}
\Bigg(
p^4\frac{\partial^2{F}}
{\partial{z^2}}
\Bigg)
\nonumber\\
&&
+
\Bigg[
-
\left(\nabla\cdot \bm{V}\right)
+
3
\frac{\partial{V_z}}
{\partial{z}}
\Bigg]V_z^2
\frac{1}{72D^2}
\frac{1}{p^2}
\frac{\partial{}}{\partial{p}}
\Bigg(
p^3
\frac{\partial^2{F}}
{\partial{z^2}}
\Bigg).	
\label{Az2}
\end{eqnarray}
Here, $A_0$ is the momentum 
transport process that 
arises from  
the combined effects of 
pitch-angle diffusion and BFVIs, 
and it is new.
The terms $A_{zp2}$ and $A_{zp3}$
represent the cross-terms
between spatial and momentum 
derivatives. 
Another new momentum transport process
can be derived from $A_{zp2}$ and 
$A_{zp3}$ by applying 
the following methods.

According to previous research
\citep{WangEA2018, WangEA2019, WangEA2020, WangEA2024},
the formula for the
second-order 
spatial derivative
along the mean magnetic field,
$\partial^2{F}/\partial{z^2}$,
can be derived from the term $T_2$
in Equation (\ref{F with T}).
Thus,
with reference to 
Equation (\ref{F with T}), 
the governing equation for
the isotropic distribution function
$F(x, y, z, p, t)$ can be obtained,
which encompasses
the spatial convection term 
$\partial{F}/\partial{z}$, 
the spatial diffusion term
$\partial^2{F}/\partial{z^2}$,
the momentum convection term 
$\partial{F}/\partial{p}$,
and so on 
\begin{eqnarray}
\frac{\partial{F}}{\partial{t}}	
=&&
\nabla\cdot\left(\kappa_\bot\cdot \nabla F\right)
-\bm{V}_\bot \cdot\nabla_\bot F
-V_z\frac{\partial{F}}{\partial{z}}
+\frac{1}{3}
\left(
\nabla \cdot\bm{V}
\right)
p\frac{\partial{F}}{\partial{p}}
+\kappa_\parallel
\frac{\partial^2{F}}{\partial{z^2}}
+\cdots.
\label{F eq}
\end{eqnarray}
Here, 
the following formulas 
are used
\begin{eqnarray}
&&\nabla \cdot\left(\bm{V} F\right)
=F\nabla \cdot\bm{V}
+\bm{V}_\bot \cdot\nabla_\bot F
+V_z\frac{\partial{F}}{\partial{z}},
\\
&&
\frac{1}{3}\left(
\nabla \cdot\bm{V}
\right)
p\frac{\partial{F}}{\partial{p}}
=-F\left(\nabla \cdot\bm{V}\right)
+\frac{1}{3}\left(
\nabla \cdot\bm{V}
\right)
\frac{1}{p^2}
\frac{\partial{}}{\partial{p}}
\left(
p^3F
\right).		
\end{eqnarray}
By using Equation (\ref{F eq}),
the formula for the
first- and second-order derivative
terms are obtained
\begin{eqnarray} 
&&\frac{\partial{F}}{\partial{z}}
=\frac{1}{V_z}
\Bigg[
-\frac{\partial{F}}{\partial{t}}	
+\nabla\cdot\left(\kappa_\bot\cdot \nabla F\right)
-\bm{V}_\bot \cdot\nabla_\bot F
+\frac{1}{3}\left(
\nabla \cdot\bm{V}
\right)
p\frac{\partial{F}}{\partial{p}}
+\kappa_\parallel
\frac{\partial^2{F}}{\partial{z^2}}
+\cdots
\Bigg],
\label{pF/pz}
\\
&&\frac{\partial^2{F}}{\partial{z^2}}
=\frac{1}{\kappa_\parallel}
\Bigg[
\frac{\partial{F}}{\partial{t}}	
-\nabla\cdot\left(\kappa_\bot\cdot \nabla F\right)
+\bm{V}_\bot \cdot\nabla_\bot F
+V_z\frac{\partial{F}}{\partial{z}}
-\frac{1}{3}\left(
\nabla \cdot\bm{V}
\right)
p\frac{\partial{F}}{\partial{p}}
-\cdots
\Bigg].
\label{p2F/pz2}
\end{eqnarray}
To proceed, 
by inserting Equations     
(\ref{pF/pz}) and (\ref{p2F/pz2})
into Equations  (\ref{Az1})
and (\ref{Az2}), 
the momentum diffusion terms arise.
These terms are new and 
induced by
the combined effects of 
the spatial parallel diffusion 
and BFVIs. 
Thus,  
in order to obtain the new 
momentum transport terms, 
it is necessary 
to derive the governing 
equation for the isotropic 
distribution function.
Subsequently, 
we can derive 
the corresponding 
formulas for both the
first- and second-order derivatives.
The term $T_1$ has already been obtained
in this subsection,
and subsequently, 
the term $T_2$ needs to be deduced.

\subsection{Evaluating $T_2$}
With the anisotropic distribution
function $g(x, y, z, \mu, p, t)$
(See Equation 
(\ref{g})), the term $T_2$ becomes
\begin{eqnarray}
T_2=-\frac{v}{2}\frac{\partial{}}
{\partial{z}}\int_{-1}^{1}\dee \mu\mu g
=\sum_{n=1}^{12}Y_{n}. 
\label{T2 with Yn}
\end{eqnarray}
Here, the formulas for $Y_1,
Y_2, \cdots,$ and $Y_{12}$ 
are presented in Appendix 
\ref{Yn}, 
and their evaluations are 
provided in Appendix
\ref{Evaluating Yn}, respectively. 
Inserting the formulas in 
Appendix
\ref{Evaluating Yn} into
Equation (\ref{T2 with Yn}) yields
\begin{eqnarray}
T_2
=B_0+B_{zp2}+B_{zp3}
\label{T2 with B}		
\end{eqnarray}
with
\begin{eqnarray}
&&B_0=\kappa_{\parallel}
\frac{\partial^2{F}}
{\partial{z^2}}
\\
&&B_{zp2}=\Bigg[-
\frac{5v}{72D^2}
\left(\nabla\cdot \bm{V}\right)^2
-\frac{v}{8D^2}
\left(\nabla\cdot \bm{V}\right)
\frac{\partial{V_z}}{\partial{z}}
\Bigg]
\frac{1}{p^2}
\frac{\partial{}}{\partial{p}}
\Bigg(
p^3
\frac{\partial{F}}{\partial{z}}
\Bigg)
\nonumber\\
&&
+
\Bigg[
\frac{v}{40D^2}
\left(\nabla\cdot \bm{V}\right)^2
+
\frac{v}{360D^2}
\nabla\cdot \bm{V}
\left(\frac{\partial{V_z}}
{\partial{z}}\right)
+
\frac{v}{60D^2}
\left(\frac{\partial{V_z}}
{\partial{z}}\right)^2
\Bigg]
\frac{1}{p^2}\frac{\partial{}}{\partial{p}}
\left(p^4
\frac{\partial{}}{\partial{p}}
\frac{\partial{F}}
{\partial{z}}\right),
\label{Bz1}
\\
&&
B_{zp3}=
\frac{v^2}{9D^2}
\Bigg(
2
\nabla\cdot \bm{V}
-
\frac{\partial{V_z}}
{\partial{z}}
\Bigg)
\frac{1}{p^2}\frac{\partial{}}{\partial{p}}
\left(p^3
\frac{\partial^2{F}}
{\partial{z^2}}\right).	
\label{Bz2}
\end{eqnarray}
Here, $B_0$ is indicative of 
the spatial 
parallel diffusion, while
$B_{zp2}$ and $B_{zp3}$ contain 
the cross terms of the spatial and momentum derivatives. 
In addition, the conditions
$v\gg V$, 
and $v/\lambda\gg 
\partial{V_z}/\partial{z}$ are 
used. 

\subsection{Iterating and deriving the momentum transport terms}
\label{Iterating and deriving the momentum transport terms}

By inserting Equations 
(\ref{T1 with A})
and
(\ref{T2 with B})
into Equation 
(\ref{F with T}), we 
obtain  
\begin{eqnarray}
\frac{\partial{F}}{\partial{t}}	
=&&
\nabla\cdot\left(\kappa_\bot\cdot \nabla F\right)
-\bm{V}_\bot \cdot\nabla_\bot F
-V_z\frac{\partial{F}}{\partial{z}}
+\frac{1}{3}\left(
\nabla \cdot\bm{V}
\right)
p\frac{\partial{F}}{\partial{p}}
\nonumber\\
&&
+A_0+A_{zp2}+A_{zp3}
+B_0+B_{zp2}+B_{zp3}, 	
\label{F equation with A and B}	
\end{eqnarray}
which represents the 
governing eqution for 
the isotropic distribution function,
excluding any anisotropic distribution function. 
The third and higher-order 
spatial and momentum derivative
terms are ignored. 
From the latter equation,
the formulas for the
first- and second-order spatial 
derivatives along the mean magnetic
field can be derived, respectively
\begin{eqnarray} 
&&\frac{\partial{F}}{\partial{z}}
=\frac{1}{V_z}
\Bigg[
\frac{\partial{F}}{\partial{t}}	
+\nabla\cdot\left(\kappa_\bot\cdot \nabla F\right)
-\bm{V}_\bot \cdot\nabla_\bot F
+\frac{1}{3}
\left(\nabla \cdot\bm{V}\right)
p\frac{\partial{F}}{\partial{p}}
+\kappa_{\parallel}
\frac{\partial^2{F}}
{\partial{z^2}}
\nonumber\\
&&
+A_0+A_{zp2}+A_{zp3}
+B_{zp2}+B_{zp3}\Bigg],
\label{pF/pz-s}
\\
&&\frac{\partial^2{F}}{\partial{z^2}}
=\frac{1}{\kappa_\parallel}
\Bigg[
\frac{\partial{F}}{\partial{t}}	
-\nabla\cdot\left(\kappa_\bot\cdot \nabla F\right)
+\bm{V}_\bot \cdot\nabla_\bot F
+V_z\frac{\partial{F}}{\partial{z}}
-\frac{1}{3}
\left(\nabla \cdot\bm{V}\right)
p\frac{\partial{F}}{\partial{p}}
\nonumber\\
&&
-A_0-A_{zp2}-A_{zp3}
-B_{zp2}-B_{zp3}
\Bigg].
\label{p2F/pz2-s}
\end{eqnarray}
Inserting Equations 
(\ref{pF/pz-s})
and 
(\ref{p2F/pz2-s})
into formulas 
(\ref{Az1}) and (\ref{Az2}) yields
\begin{eqnarray}
&&A_{zp2}=
\Bigg[
3
\frac{\partial{V_z}}
{\partial{z}}
-
\nabla\cdot \bm{V}
\Bigg]
\frac{1}{18D}
\frac{1}{p^2}\frac{\partial{}}{\partial{p}}
p^3
\Bigg[\frac{1}{3}
\left(\nabla \cdot\bm{V}\right)
p\frac{\partial{F}}{\partial{p}}
+\kappa_{\parallel}
\frac{\partial^2{F}}
{\partial{z^2}}
\Bigg],
\label{Az1-1}
\\ 
&&A_{zp3}=
\Bigg[
\nabla\cdot \bm{V}
-3
\frac{\partial{V_z}}
{\partial{z}}
\Bigg]
\frac{1}{180D^2}
\frac{1}{m^2}
\frac{1}{p^2}
\frac{\partial{}}{\partial{p}}
p^5
\frac{1}{\kappa_\parallel}
\Bigg[
V_z\frac{\partial{F}}{\partial{z}}
-\frac{1}{3}
\left(\nabla \cdot\bm{V}\right)
p\frac{\partial{F}}{\partial{p}}
\Bigg]
\nonumber\\
&&
+
\Bigg[
-V_z
\left(\nabla\cdot \bm{V}\right)
+3V_z
\frac{\partial{V_z}}
{\partial{z}}
\Bigg]
\frac{1}{72D^2m}
\frac{1}{p^2}\frac{\partial{}}{\partial{p}}
p^4\frac{1}{\kappa_\parallel}
\Bigg[
V_z\frac{\partial{F}}{\partial{z}}
-\frac{1}{3}
\left(\nabla \cdot\bm{V}\right)
p\frac{\partial{F}}{\partial{p}}
\Bigg]
\nonumber\\
&&
+
\Bigg[
-
\left(\nabla\cdot \bm{V}\right)
+
3
\frac{\partial{V_z}}
{\partial{z}}
\Bigg]V_z^2
\frac{1}{72D^2}
\frac{1}{p^2}
\frac{\partial{}}{\partial{p}}
p^3
\frac{1}{\kappa_\parallel}
\Bigg[
V_z\frac{\partial{F}}{\partial{z}}
-\frac{1}{3}
\left(\nabla \cdot\bm{V}\right)
p\frac{\partial{F}}{\partial{p}}
\Bigg].
\label{Az2-1}				
\end{eqnarray}
Here, 
the same assumptions and
settings as those in subsection
\ref{Evaluating Z1-Z16}
are used. 
In the latter equations,
each formula comprises 
one momentum transport term
and one cross-term.
Hereinafter, 
our exploration will be solely
confined to 
the momentum transport,
as it constitutes the primary focus 
of this article,
and only momentum derivative terms
are expored.
Thus, Equations 
(\ref{Az1-1}) and (\ref{Az2-1}) 
are simplified respectively as
\begin{eqnarray}
&&A_{zp2}=
\Bigg[
3
\frac{\partial{V_z}}
{\partial{z}}
\left(\nabla \cdot\bm{V}\right)
-
\left(\nabla \cdot\bm{V}\right)^2
\Bigg]
\frac{1}{54D}
\frac{1}{p^2}\frac{\partial{}}{\partial{p}}
\left(
p^4\frac{\partial{F}}{\partial{p}}
\right),
\label{Az1-2}
\\ 
&&A_{zp3}=
\Bigg[
-\left(\nabla \cdot\bm{V}\right)^2
+3
\frac{\partial{V_z}}
{\partial{z}}
\left(\nabla \cdot\bm{V}\right)
\Bigg]
\frac{v^2}{540D^2\kappa_\parallel}
\frac{1}{p^4}
\frac{\partial{}}{\partial{p}}
\left(
p^6
\frac{\partial{F}}{\partial{p}}
\right)
\nonumber\\
&&
+\Bigg[
\left(\nabla\cdot \bm{V}\right)^2
-3
\frac{\partial{V_z}}
{\partial{z}}
\left(\nabla \cdot\bm{V}\right)
\Bigg]
\frac{vV_z}{216D^2\kappa_\parallel}
\frac{1}{p^3}\frac{\partial{}}{\partial{p}}
\left(
p^5
\frac{\partial{F}}{\partial{p}}
\right)
\nonumber\\
&&
+
\Bigg[
\left(\nabla\cdot \bm{V}\right)^2
-
3
\frac{\partial{V_z}}
{\partial{z}}
\left(\nabla \cdot\bm{V}\right)
\Bigg]
\frac{V_z^2}{216D^2\kappa_\parallel}
\frac{1}{p^2}
\frac{\partial{}}{\partial{p}}
\left(p^4
\frac{\partial{F}}{\partial{p}}
\right).
\label{Az2-2}					
\end{eqnarray} 
Similarly, 
with Equations 
(\ref{pF/pz-s}) and 
(\ref{p2F/pz2-s}),
we obtain 
from Equation (\ref{Bz1}) and 
(\ref{Bz2}) 
\begin{eqnarray}
&&B_{zp2}=0,
\label{Bz1-last}\\
&&B_{zp3}=
-
\frac{v^2}{27\kappa_\parallel D^2}
\Bigg(
2
\left(\nabla \cdot\bm{V}\right)^2
-
\frac{\partial{V_z}}
{\partial{z}}
\left(\nabla \cdot\bm{V}\right)
\Bigg)
\frac{1}{p^2}\frac{\partial{}}{\partial{p}}
\left(
p^4
\frac{\partial{F}}{\partial{p}}
\right).
\label{Bz2-last}
\end{eqnarray}

\subsection{The isotropic distribution
function equation excluding spatial derivative terms}
\label{The isotropic distribution
function equation excluding spatial derivative terms}

With Equations (\ref{A0}),
(\ref{Az1-2}), (\ref{Az2-2}),
(\ref{Bz1-last}), and 
(\ref{Bz2-last}),
the isotropic 
distribution function
equation, 
excluding spatial derivative
terms, is obtained
\begin{eqnarray}
\frac{\partial{F}}{\partial{t}}	
=&&
\nabla\cdot\left(\kappa\cdot \nabla F\right)
-\nabla\cdot\left(
\bm{V} F\right)
+\frac{\nabla \cdot\bm{V}}{3}
\frac{1}{p^2}
\frac{\partial{}}{\partial{p}}
\left(p^3F\right)
+S_1+S_2+S_3+S_4+S_5	
\label{F equation-last}		
\end{eqnarray}
with
\begin{eqnarray}
&&S_1=\Bigg[
(\nabla\cdot \bm{V})^2
-6
\frac{\partial{V_z}}{\partial{z}}
\nabla\cdot \bm{V}
+9
\left(\frac{\partial{V_z}}
{\partial{z}}\right)^2
\Bigg]
\frac{1}{270D}
\frac{1}{p^2}\frac{\partial{}}{\partial{p}}
\left(p^4\frac{\partial{F}}{\partial{p}}\right),
\label{S1}
\\
&&
S_2=\Bigg[
-\left(\nabla \cdot\bm{V}\right)^2
+3
\frac{\partial{V_z}}
{\partial{z}}
\left(\nabla \cdot\bm{V}\right)
\Bigg]
\frac{v^2}{540D^2\kappa_\parallel}
\frac{1}{p^4}
\frac{\partial{}}{\partial{p}}
\left(
p^6
\frac{\partial{F}}{\partial{p}}
\right),
\label{S2}
\\
&&
S_3=\Bigg[
\left(\nabla\cdot \bm{V}\right)^2
-3
\frac{\partial{V_z}}
{\partial{z}}
\left(\nabla \cdot\bm{V}\right)
\Bigg]
\frac{vV_z}{216D^2\kappa_\parallel}
\frac{1}{p^3}\frac{\partial{}}{\partial{p}}
\left(
p^5
\frac{\partial{F}}{\partial{p}}
\right),
\label{S3}
\\
&&
S_4=
\Bigg[
\left(\nabla\cdot \bm{V}\right)^2
-
3
\frac{\partial{V_z}}
{\partial{z}}
\left(\nabla \cdot\bm{V}\right)
\Bigg]
\frac{V_z^2}{216D^2\kappa_\parallel}
\frac{1}{p^2}
\frac{\partial{}}{\partial{p}}
\left(p^4
\frac{\partial{F}}{\partial{p}}
\right),
\label{S4}
\\
&&
S_5=
\Bigg(
-2
\left(\nabla \cdot\bm{V}\right)^2
+
\frac{\partial{V_z}}
{\partial{z}}
\left(\nabla \cdot\bm{V}\right)
\Bigg)
\frac{v^2}{27\kappa_\parallel D^2}
\frac{1}{p^2}\frac{\partial{}}{\partial{p}}
\left(
p^4
\frac{\partial{F}}{\partial{p}}
\right).
\label{S5}
\end{eqnarray}
In Equation 
(\ref{F equation-last}),
the momentum transport processes
are represented by the 
terms $S_1$, $S_2$, $\cdots$,
$S_5$. 
Among them, 
the term $S_1$ arises from
the combination of pitch-angle
diffusion and BFVIs.
The remaining terms, 
$S_2$, $S_2$, $S_3$, 
$S_4$, and $S_5$,
pertain to
the combined effects
of parallel diffusion, 
and BFVIs. 

When charged particles move 
in magnetized plasmas that possess 
a large-scale magnetic field, 
pitch-angle scattering of 
these particles
results in parallel diffusion
\citep{Jokipii1966, Shalchi2009},
utimately leading to 
the well-known formula
\begin{eqnarray}
\kappa_\parallel=\frac{v^2}{8}
\int_{-1}^{1}\dee\mu
\frac{\left(1-\mu^2\right)^2}
{D_{\mu\mu}}.
\end{eqnarray}
By using Equation (\ref{Dmumu}),
the latter formula is simplified 
as follows
\begin{eqnarray}
\kappa_\parallel=\frac{v^2}{6D}.
\label{relation of k and D}
\end{eqnarray} 
Inserting the latter formula into 
Equations (\ref{S2})-(\ref{S5}) gives
\begin{eqnarray}
&&
S_2=\Bigg[
-\left(\nabla \cdot\bm{V}\right)^2
+3
\frac{\partial{V_z}}
{\partial{z}}
\left(\nabla \cdot\bm{V}\right)
\Bigg]
\frac{1}{90D}
\frac{1}{p^4}
\frac{\partial{}}{\partial{p}}
\left(
p^6
\frac{\partial{F}}{\partial{p}}
\right),
\label{S2 without kzz}
\\
&&
S_3=\Bigg[
\left(\nabla\cdot \bm{V}\right)^2
-3
\frac{\partial{V_z}}
{\partial{z}}
\left(\nabla \cdot\bm{V}\right)
\Bigg]\frac{V_z}{v}
\frac{1}{36D}
\frac{1}{p^3}\frac{\partial{}}{\partial{p}}
\left(
p^5
\frac{\partial{F}}{\partial{p}}
\right),
\label{S3 without kzz}
\\
&&
S_4=
\Bigg[
\left(\nabla\cdot \bm{V}\right)^2
-
3
\frac{\partial{V_z}}
{\partial{z}}
\left(\nabla \cdot\bm{V}\right)
\Bigg]
\frac{V_z^2}{v^2}
\frac{1}{36D}
\frac{1}{p^2}
\frac{\partial{}}{\partial{p}}
\left(p^4
\frac{\partial{F}}{\partial{p}}
\right),
\label{S4 without kzz}
\\
&&
S_5=
\Bigg(
-2
\left(\nabla \cdot\bm{V}\right)^2
+
\frac{\partial{V_z}}
{\partial{z}}
\left(\nabla \cdot\bm{V}\right)
\Bigg)
\frac{2}{9 D}
\frac{1}{p^2}\frac{\partial{}}{\partial{p}}
\left(
p^4
\frac{\partial{F}}{\partial{p}}
\right).
\end{eqnarray} 
As parallel diffusion emanates
from pitch-angle scattering,
the generation mechanism underlying
the momentum transport processes,
denoted by 
$S_1$ through $S_5$, ought to
identical. 
Merging $S_1$ and $S_5$
yields
\begin{eqnarray}  
S^{(1)}=S_1+S_5
=
\Bigg[
9
\left(\frac{\partial{V_z}}
{\partial{z}}\right)^2
-119
\left(\nabla\cdot \bm{V}\right)^2
+
54
\left(\nabla\cdot \bm{V}\right)
\frac{\partial{V_z}}
{\partial{z}}
\Bigg]
\frac{1}{270D}
\frac{1}{p^2}\frac{\partial{}}{\partial{p}}
\left(p^4\frac{\partial{F}}{\partial{p}}\right).
\label{SI}	
\end{eqnarray} 
Considering Equations 
(\ref{F equation-last}),
(\ref{S2 without kzz}),
(\ref{S3 without kzz}),
(\ref{S4 without kzz}),
and (\ref{SI}), we obtain
\begin{eqnarray}
\frac{\partial{F}}{\partial{t}}	
=&&
\nabla\cdot\left(\kappa\cdot \nabla F\right)
-\nabla\cdot\left(
\bm{V} F\right)
+\frac{\nabla \cdot\bm{V}}{3}
\frac{1}{p^2}
\frac{\partial{}}{\partial{p}}
\left(p^3F\right)
+S^{(1)}+S_2+S_3+S_4. 	
\end{eqnarray}
Because $v\gg V$, the terms 
$S_3$ and $S_4$ can be neglected,
the isotropic distribution 
function becomes
\begin{eqnarray}
\frac{\partial{F}}{\partial{t}}	
&&=
\nabla\cdot\left(\kappa\cdot \nabla F\right)
-\bm{V}\cdot\nabla F
+\frac{1}{3}
\left(\nabla \cdot\bm{V}\right)
p
\frac{\partial{F}}{\partial{p}}
+S^{(1)}+S^{(2)} 	
\label{F equation}	
\end{eqnarray}
with $S^{(2)}=S_2$. 
Here, the momentum 
diffusion terms $S^{(1)}$
and $S^{(2)}$ are new, 
whereby the momentum transport 
is generated by the combination
of pitch-angle scattering and
the combined effects of pitch-angle diffusion and BFVIs.

In this article,
we only delve into the momentum
transport, whereby 
the spatial transport terms
are omitted from consideration. 
Thus, Equation
(\ref{F equation})
becomes
\begin{eqnarray}
\frac{\partial{F}}{\partial{t}}	
&&=
\kappa_p
p
\frac{\partial{F}}{\partial{p}}
+
\kappa_{pp}^{(1)}
\frac{1}{p^2}\frac{\partial{}}{\partial{p}}
\left(p^4\frac{\partial{F}}{\partial{p}}\right)
+\kappa_{pp}^{(2)}
\frac{1}{p^4}\frac{\partial{}}{\partial{p}}
\left(
p^6
\frac{\partial{F}}{\partial{p}}
\right)	
\label{F equation without spatial transport}	
\end{eqnarray}
with
\begin{eqnarray}
&&\kappa_p=\frac{1}{3}
\left(\nabla \cdot\bm{V}\right),
\label{kp1}
\\
&&\kappa_{pp}^{(1)}=
\frac{1}{270D}
\Bigg[
9
\left(\frac{\partial{V_z}}
{\partial{z}}\right)^2
-119
\left(\nabla\cdot \bm{V}\right)^2
+
54
\left(\nabla\cdot \bm{V}\right)
\frac{\partial{V_z}}
{\partial{z}}
\Bigg],
\label{kpp1}\\
&&\kappa_{pp}^{(2)}
=
\frac{1}{270D}
\Bigg[-3
\left(\nabla\cdot\bm{V}
\right)^2
+9\left(\nabla\cdot\bm{V}\right)
\frac{\partial{V_z}}
{\partial{z}}
\Bigg].	
\label{kpp2}	
\end{eqnarray}
Here, the third and higher order
momentum derivative terms 
are ignored.

\section{The solution of isotropic distribution function equation}
\label{The solution of isotropic distribution function}

In this section, we solve
the isotropic distribution equation,
subsequently deriving its solution.  
By using the method of separation 
of variables,
we are able to express $F=T(t)V(p)$, and insert
it into Equation
(\ref{F equation without spatial transport})
\begin{eqnarray}
V\frac{\dee{T}}{\dee{t}}	
&&=
T\kappa_p
p
\frac{\dee{V}}{\dee{p}}
+
T\kappa_{pp}^{(1)}
\frac{1}{p^2}\frac{\dee{}}{\dee{p}}
\left(p^4\frac{\dee{V}}{\dee{p}}\right)
+T\kappa_{pp}^{(2)}
\frac{1}{p^4}\frac{\dee{}}{\dee{p}}
\left(
p^6
\frac{\dee{V}}{\dee{p}}
\right).		
\end{eqnarray}
Upon dividing both sides of the latter equation
by $TV$, we find
\begin{eqnarray}
\frac{1}{T}\frac{\dee{T}}{\dee{t}}	
&&=
\frac{1}{V}\left[\kappa_p
p
\frac{\dee{V}}{\dee{p}}
+
\kappa_{pp}^{(1)}
\frac{1}{p^2}\frac{\dee{}}{\dee{p}}
\left(p^4\frac{\dee{V}}{\dee{p}}\right)
+\kappa_{pp}^{(2)}
\frac{1}{p^4}\frac{\dee{}}{\dee{p}}
\left(
p^6
\frac{\dee{V}}{\dee{p}}
\right)
\right].
\end{eqnarray}
Given that the left-hand side of 
the latter equation is solely 
related to 
time $t$, whilst the right-hand 
side depends solely on 
the variable $p$,
we can equate both sides to 
a constant, denoted as $\lambda$,
\begin{eqnarray}
\frac{1}{T}\frac{\dee{T}}{\dee{t}}	
&&=
\frac{1}{V}\left[\kappa_p
p
\frac{\dee{V}}{\dee{p}}
+
\kappa_{pp}^{(1)}
\frac{1}{p^2}\frac{\dee{}}{\dee{p}}
\left(p^4\frac{\dee{V}}{\dee{p}}\right)
+\kappa_{pp}^{(2)}
\frac{1}{p^4}\frac{\dee{}}{\dee{p}}
\left(
p^6
\frac{\dee{V}}{\dee{p}}
\right)
\right]=-\lambda
\label{left=right=-lambda}	
\end{eqnarray} 
Here, $\lambda>0$.  
From Equation 
(\ref{left=right=-lambda}),
the equations can be derived
\begin{eqnarray}
&&\frac{\dee{T}}{\dee{t}}=-\lambda 
T,\\
&&
\kappa_p
p
\frac{\dee{V}}{\dee{p}}
+
\kappa_{pp}^{(1)}
\frac{1}{p^2}\frac{\dee{}}{\dee{p}}
\left(p^4\frac{\dee{V}}{\dee{p}}\right)
+\kappa_{pp}^{(2)}
\frac{1}{p^4}\frac{\dee{}}{\dee{p}}
\left(
p^6
\frac{\dee{V}}{\dee{p}}
\right)=-\lambda V.
\end{eqnarray}
The solution to the first equation 
can be easily obtained 
\begin{eqnarray}
T=T_0e^{-\lambda t}.
\end{eqnarray}
The second equation is known as
the Euler equation, 
and it can be solved as
\begin{eqnarray}
V(p)=c_1p^{r_1}+c_2p^{r_2}
\end{eqnarray}
with
\begin{eqnarray}
r_{1,2}=\frac{1}{2}
\left[
-\frac{\kappa_p+3\kappa_{pp}^{(1)}
+5\kappa_{pp}^{(2)}}{\kappa_{pp}^{(1)}
+\kappa_{pp}^{(2)}}
\pm
\sqrt{
\left(
\frac{\kappa_p+3\kappa_{pp}^{(1)}
	+5\kappa_{pp}^{(2)}}
{\kappa_{pp}^{(1)}
	+\kappa_{pp}^{(2)}}
\right)^2
-\frac{4\lambda}
{\kappa_{pp}^{(1)}
	+\kappa_{pp}^{(2)}}
}
\right].
\label{r1,2 with convection and time}
\end{eqnarray}
Thus, the solution to Equation
(\ref{F equation without spatial transport}) can be obtained
\begin{eqnarray}
F(p, t)=T_0
\left(
c_1p^{r_1}+c_2p^{r_2}
\right)
e^{-\lambda t}
=\left(
c_1'p^{r_1}+c_2'p^{r_2}
\right)
e^{-\lambda t}.
\end{eqnarray}
It is clear that 
the particle distribution
demonstrates momentum power laws.

For the quasi-steady state, i.e.,
$\lambda=0$, the indices are given
as follows
\begin{eqnarray}
&&r_1=0,\\
&&r_2=-\frac{\kappa_p+3\kappa_{pp}^{(1)}
	+5\kappa_{pp}^{(2)}}
{\kappa_{pp}^{(1)}
	+\kappa_{pp}^{(2)}},
\end{eqnarray}
and the corresponding solution 
formula is
\begin{eqnarray}
F(p, t)=c_1'+c_2'p^{r_2}.	
\end{eqnarray}
The latter formula represents 
a single power law, 
with a spectral index denoted 
as $r_2$. 

If the following formula holds
\begin{eqnarray}
\lambda
=\frac{1}{4}
\frac{\left(
	\kappa_p+3\kappa_{pp}^{(1)}
	+5\kappa_{pp}^{(2)}
	\right)^2}{\kappa_{pp}^{(1)}
	+\kappa_{pp}^{(2)}},	
\end{eqnarray}
the indices becomes
\begin{eqnarray}
r_1=r_2=r=-
\frac{1}{2}
\frac{\kappa_p+3\kappa_{pp}^{(1)}
	+5\kappa_{pp}^{(2)}}
{\kappa_{pp}^{(1)}
	+\kappa_{pp}^{(2)}},
\end{eqnarray}
and accordingly the solution can be 
found
\begin{eqnarray}
F(p, t)=\left(
c_1'+c_2'\ln p
\right)p^{r}
e^{-\lambda t}.	
\end{eqnarray}
If $r_1\ne r_2$ and
$r_1\ne0$ and $r_2\ne0$,  
the solution is 
\begin{eqnarray}
F(p, t)=\left(
c_1'p^{r_1}+c_2'p^{r_2}
	\right)
	e^{-\lambda t},
\end{eqnarray}
which constitutes 
a double power law.

\section{Discussion}
\label{Discussion}

In the above section, 
the isotropic distribution function
equation for charged energetic 
particles
has been derived, with its solution
obtained. In this section,
the physical meanings 
of momentum transport terms and 
solution are
investigated.

\subsection{Momentum transport for incompressible plasmas}
\label{For incompressible 
plasmas}

In the case of 
incompressible plasmas,
the formula $\nabla \cdot\bm{V}
=0$ is satisfied. Thus, 
Equation 
(\ref{F equation without spatial transport}) becomes
\begin{eqnarray}
\frac{\partial{F}}{\partial{t}}	
=
\left(\frac{\partial{V_z}}
{\partial{z}}\right)^2
\frac{1}{30D}
\frac{1}{p^2}\frac{\partial{}}{\partial{p}}
\left(p^4\frac{\partial{F}}{\partial{p}}\right).
\label{F equation for incompressible plasma}			
\end{eqnarray}
The latter equation suggests 
that, for incompressible plasmas,
the momentum transport term 
still exists due to the combined effects of  
the pitch-angle scattering
and BFVIs. 
The solution of Equation
(\ref{F equation for incompressible plasma}) is given by
\begin{eqnarray}
F(p, t)=T_0
\left(
c_1p^{r_1}+c_2p^{r_2}
\right)
e^{-\lambda t}
=\left(
c_1'p^{r_1}+c_2'p^{r_2}
\right)
e^{-\lambda t}.
\end{eqnarray}
with
\begin{eqnarray}
r_{1,2}=\frac{1}{2}
\left[
-3
\pm
\sqrt{
9
	-\frac{4\lambda}
	{\kappa_{pp}^I
	}
}
\right].
\end{eqnarray}
For quasi-steady state, i.e.,
$\lambda=0$, it can be found that  
$r_1=0$ and $r_2=-3$. 
The solution of
Equation 
(\ref{F equation for incompressible plasma}) is
\begin{eqnarray}
F(p, t)
=
c_1'+c_2'p^{-3}.
\end{eqnarray}
The latter formula demonstrates
that momentum distribution of 
particles follows a single power law
with a spectral index $-3$.
In addition, 
when the following relation holds
\begin{eqnarray}
\lambda
=\frac{9}{4}\kappa_{pp}^I,
\end{eqnarray}
the index becomes $r_1=r_2=r=-3/2$,
and the solution can be derived 
\begin{eqnarray}
F(p, t)
=\left(
c_1'+c_2'\ln p
\right)p^{r}
e^{-\lambda t}.	
\end{eqnarray}
If $r_1\ne r_2$ and $r_1\ne0$
and $r_2\ne0$,
the solution is 
\begin{eqnarray}
F(p, t)
=\left(
c_1'p^{r_1}+c_2'p^{r_2}
\right)
e^{-\lambda t},
\end{eqnarray}
which is a double power law.

\subsection{The case. $\partial{V_z}/\partial{z}=0$}

When $\partial{V_z}/\partial{z}
=0$, Equation 
(\ref{F equation without spatial transport})
becomes 
\begin{eqnarray}
\frac{\partial{F}}{\partial{t}}	
&&=
\kappa_p
p
\frac{\partial{F}}{\partial{p}}
+
\kappa_{pp}^{(1)}
\frac{1}{p^2}\frac{\partial{}}{\partial{p}}
\left(p^4\frac{\partial{F}}{\partial{p}}\right)
+\kappa_{pp}^{(2)}
\frac{1}{p^4}\frac{\partial{}}{\partial{p}}
\left(
p^6
\frac{\partial{F}}{\partial{p}}
\right)	
\label{F equation for two dimensional space}		
\end{eqnarray}
with
\begin{eqnarray}
&&\kappa_p
=\frac{1}{3}
\left(\frac{\partial{V_x}}{\partial{x}}
+
\frac{\partial{V_y}}{\partial{y}}\right),
\\
&&\kappa_{pp}^{(1)}
=\frac{1}{270D}
\Bigg[
-119\left(\frac{\partial{V_x}}{\partial{x}}
+
\frac{\partial{V_y}}{\partial{y}}\right)^2
\Bigg],
\\
&&\kappa_{pp}^{(2)}
=\frac{1}{270D}
\Bigg[-3
\left(\frac{\partial{V_x}}{\partial{x}}
+
\frac{\partial{V_y}}{\partial{y}}\right)^2
\Bigg].
\end{eqnarray}
With the exception of 
the coefficients,
Equation 
(\ref{F equation for two dimensional space})
is exactly the same in form
as Equation
(\ref{F equation without spatial transport}). 
As a result, the solution of 
Equation 
(\ref{F equation for two dimensional space})
also exhibits 
the momentum power laws, 
with the exception of 
different indices.

\subsection{Momentum transport for charged particles
in compressible plasma waves}
\label{Momentum transport equation with compressible waves}

When particles 
penetrate compressible plasma waves or  turbulence, after performing the 
averaging operation,
all terms  $(\partial{V_z}
/\partial{z})^{k}$ and $(\nabla\cdot \bm{V})^{k}$
with odd number $k$
are zero.
However,
the influence of the terms with an even power 
ought to be taken into account.
Therefore, 
Equation
(\ref{F equation without spatial transport})
simplifies to
\begin{eqnarray}
\frac{\partial{F}}{\partial{t}}	
&&=
\kappa_{pp}^{(1)}
\frac{1}{p^2}\frac{\partial{}}{\partial{p}}
\left(p^4\frac{\partial{F}}{\partial{p}}\right)
+\kappa_{pp}^{(2)}
\frac{1}{p^4}\frac{\partial{}}{\partial{p}}
\left(
p^6
\frac{\partial{F}}{\partial{p}}
\right)	
\label{F equation for compressible waves}		
\end{eqnarray}
The solution of Equation 
(\ref{F equation for compressible waves}) can be derived 
for $r_1\ne r_2$
\begin{eqnarray} 
F(p, t)=T_0
\left(
c_1p^{r_1}+c_2p^{r_2}
\right)
e^{-\lambda t}
=\left(
c_1'p^{r_1}+c_2'p^{r_2}
\right)
e^{-\lambda t}
\end{eqnarray}
with the indices
\begin{eqnarray} 
r_{1,2}=\frac{1}{2}
\left[
-\frac{3\kappa_{pp}^{(1)}
	+5\kappa_{pp}^{(2)}}
{\kappa_{pp}^{(1)}
	+\kappa_{pp}^{(2)}}
\pm
\sqrt{
	\left(
	\frac{3\kappa_{pp}^{(1)}
		+5\kappa_{pp}^{(2)}}
	{\kappa_{pp}^{(1)}
		+\kappa_{pp}^{(2)}}
	\right)^2
	-\frac{4\lambda}
	{\kappa_{pp}^{(1)}
		+\kappa_{pp}^{(2)}}
}
\right].
\end{eqnarray}
When $r_1=r_2$, the index can be 
derives as follows
\begin{eqnarray} 
r_1=r_2=r=-\frac{1}{2}
\frac{3\kappa_{pp}^{(1)}
	+5\kappa_{pp}^{(2)}}
{\kappa_{pp}^{(1)}
	+\kappa_{pp}^{(2)}},
\end{eqnarray}
and the corresponding solution is
\begin{eqnarray} 
F(p, t)=T_0
\left(
c_1p^{r_1}+c_2p^{r_2}
\right)
e^{-\lambda t}
=\left(
c_1'+c_2'\ln p
\right)p^{r}
e^{-\lambda t}. 	
\end{eqnarray}

\subsubsection{The solution for quasi-steady state}
\label{The solution for quasi-steady state}

For quasi-steady state, 
the indices of momentum power law
becomes
\begin{eqnarray}
	&&r_1=0,\\
	&&r_2=-\frac{3\kappa_{pp}^{(1)}
		+5\kappa_{pp}^{(2)}}{\kappa_{pp}^{(1)}
		+\kappa_{pp}^{(2)}}
	=-\frac{3
		+5\kappa_{pp}^{(2)}/\kappa_{pp}^{(1)}}
	{1+
		\kappa_{pp}^{(2)}/\kappa_{pp}^{(1)}}.
\end{eqnarray}
Thus, the solution can be found
\begin{eqnarray}
	F(p, t)
	=
	c_1'+c_2'p^{r_2}. 
	\label{solution of two term equation}
\end{eqnarray}
which indicates that
the spectral index of
momentum spectrum 
is determined by
the ratio of $\kappa_{pp}^{(2)}$
to $\kappa_{pp}^{(1)}$. 

To proceed, 
by using the formula
\begin{eqnarray} 
	&&
	\alpha
	=
	\frac{\partial{V_x}/\partial{x}}
	{\partial{V_z}/\partial{z}},
	\\
	&&
	\beta=
	\frac{\partial{V_y}/\partial{y}}
	{\partial{V_z}/\partial{z}},		
\end{eqnarray}
the coefficients, $\kappa_{pp}^{(1)}$
and $\kappa_{pp}^{(2)}$, can be 
rewritten as
\begin{eqnarray} 
	&&\kappa_{pp}^I=
	\frac{1}{270D}
	\left(\frac{\partial{V_z}}
	{\partial{z}}\right)^2h_1,\\
	&&\kappa_{pp}^{II}
	=
	\frac{1}{270D}
	\left(\frac{\partial{V_z}}
	{\partial{z}}\right)^2
	h_2		
\end{eqnarray}	
with
\begin{eqnarray} 	
	&&h_1=-119\xi^2+54\xi+9,
	\label{h1}
	\\
	&&h_2=-3\xi^2+9\xi.	
	\label{h2}		
\end{eqnarray}	
Here, the variable $\xi$ is
defined as 
\begin{eqnarray} 	
	\xi=\alpha+\beta+1
	=\frac{\nabla \cdot \bm{V}}{\partial{V_z}/\partial{z}},
\end{eqnarray}
which denotes the compressible ratio
of plasmas, and its value range 
spans $(-\infty, \infty)$. 
We set 
\begin{eqnarray}
	\eta=\frac{\kappa_{pp}^{(2)}}
	{\kappa_{pp}^{(1)}}
	=\frac{h_2}{h_1},
	\label{eta definition}
\end{eqnarray}
which becomes with Equations
(\ref{h1}) and (\ref{h2})
\begin{eqnarray}
	\eta=\frac{h_2}{h_1}
	=\frac{-3\xi^2+9\xi}
	{-119\xi^2+54\xi+9}.	
\end{eqnarray}
Given that $\kappa_{pp}^{(1)}$
and $\kappa_{pp}^{(2)}$
are 
greater than zero, 
subsequently 
$h_1$ and $h_2$ are also
both equal to or 
greater than zero
\begin{eqnarray}  
	&&-3\xi^2+9\xi\ge0,
	\label{xi range of h1}\\
	&&-119\xi^2+54\xi+9>0.
	\label{xi range of h2}
\end{eqnarray}
From formula (\ref{xi range of h1}), we have 
\begin{eqnarray} 
	0\le \xi \le 3,
	\label{xi range-1}
\end{eqnarray}
and from formula
(\ref{xi range of h2}),
we obtain
\begin{eqnarray} 
	-0.13\le \xi \le 0.58.
	\label{xi range-2}
\end{eqnarray}
The intersection of 
inequalities 
(\ref{xi range-1}) and
(\ref{xi range-2}) can be found
\begin{eqnarray} 
	0\le \xi \le 0.58.
	\label{intersection xi}	
\end{eqnarray}
For the latter inequality,
the range of variable $\eta$ 
can be derived
\begin{eqnarray} 
	0\le \eta \le \infty.
	\label{eta range}	
\end{eqnarray}
Applying Equations
(\ref{delta formula}) and
(\ref{eta definition}) yields
\begin{eqnarray}
\delta=r_2=-\frac{3
+5\eta}{1+\eta}=-5+\frac{2法}{1+\eta}.
\label{delta formula}		
\end{eqnarray}
With inequality 
(\ref{eta range}),
the range of the variable $\delta$
can be found
\begin{eqnarray} 
	-5\le \delta \le -3.
	\label{eta range-last}		
\end{eqnarray}
The results presented 
in this subsection 
may be applicable to
the investigation of 
galatic cosmic rays.
According to the 
conversion formula in Appendix 
\ref{Conversion formulas},
the observation data of 
galactic cosmic rays 
exhibit a momentum 
power law ranging from
$-5.1$ to 
$-4.7$, which is approximately
within the interval 
$-5\le \delta \le -3$. 
When $\delta=-4.7$ and $\delta=-5$, 
we can find that 
$\xi\approx 0.5746$ and 
$\xi\approx 0.5834$, 
respectively.

\subsubsection{The single power law with logarithm form}
\label{The single power law with logarithm form}

If the following relation holds
\begin{eqnarray}
\left(
\frac{3\kappa_{pp}^{(1)}
	+5\kappa_{pp}^{(2)}}
{\kappa_{pp}^{(1)}
	+\kappa_{pp}^{(2)}}
\right)^2
=\frac{4\lambda}
{\kappa_{pp}^{(1)}
	+\kappa_{pp}^{(2)}},
\end{eqnarray}
the solution of Equation 
(\ref{F equation for compressible waves}) becomes
\begin{eqnarray}
F(p, t)
=\left(
c_1'+c_2'\ln p
\right)p^{r}
e^{-\lambda t}.	
\end{eqnarray}
with 
\begin{eqnarray}
r_1=r_2=r=-\frac{1}{2}
\frac{3
+5\kappa_{pp}^{(2)}/\kappa_{pp}^{(1)}}
{1+\kappa_{pp}^{(2)}/\kappa_{pp}^{(1)}}.
\end{eqnarray}
The latter formula denotes that 
the index of power law
is determined by the relation
$\kappa_{pp}^{(2)}/\kappa_{pp}^{(1)}$.

\subsubsection{The double power law}
\label{The double power law}

When the indices satisfy the 
relations $r_1\ne r_2$, $r_1\ne0$, 
and $r_2\ne0$, the solution
of Equation 
(\ref{F equation for compressible waves}) is
\begin{eqnarray}
	F(p, t)=T_0
	\left(
	c_1p^{r_1}+c_2p^{r_2}
	\right)
	e^{-\lambda t}
	=\left(
	c_1'p^{r_1}+c_2'p^{r_2}
	\right)
	e^{-\lambda t}
	\label{double power law for
	compressible waves}
\end{eqnarray}
with
\begin{eqnarray}
	r_{1,2}=\frac{1}{2}
	\left[
	-\frac{3\kappa_{pp}^{(1)}
		+5\kappa_{pp}^{(2)}}{\kappa_{pp}^{(1)}
		+\kappa_{pp}^{(2)}}
	\pm
	\sqrt{
		\left(
		\frac{3\kappa_{pp}^{(1)}
			+5\kappa_{pp}^{(2)}}{\kappa_{pp}^{(1)}
			+\kappa_{pp}^{(2)}}
		\right)^2
		-\frac{4\lambda}
		{\kappa_{pp}^{(1)}
			+\kappa_{pp}^{(2)}}
	}
	\right].	
\end{eqnarray}
The distribution function
(\ref{double power law for
compressible waves}) 
demonstrates a double power law
with spectral indices $r_1$
and $r_2$.

\subsection{The case 
	$S^{(2)}\gg S^{(1)}$}

When $\xi\approx 0.8$, 
it is observed that $h_1\approx 0$, 
which subsequently results in the 
retention of only $S^{(2)}$. 
For this case, 
Equation 
(\ref{F equation for compressible waves})
becomes
\begin{eqnarray}
\frac{\partial{F}}{\partial{t}}	
=
\kappa_{pp}^{(2)}
\frac{1}{p^4}\frac{\partial{}}{\partial{p}}
\left(
p^6
\frac{\partial{F}}{\partial{p}}
\right).
\label{F equation for SII}						
\end{eqnarray}
The solution of the
latter equation is 
\begin{eqnarray}
F(p, t)
=
\left(c_1'p^{r_1}+c_2'p^{r_2}\right)
e^{-\lambda t}	 
\label{solution for equation with S2}
\end{eqnarray}
with
\begin{eqnarray}
r_{1,2}=\frac{1}{2}
\left(-5\pm\sqrt{
25-4\lambda/\kappa_{pp}^{(2)}}
\right).
\end{eqnarray}
Equation 
(\ref{solution for equation with S2}) 
denotes 
that the momentum distribution
obeys a double power law.
For the quasi-steady state,
$r_1=0$ and $r_2=-5$, and 
the corresponding solution
becomes
\begin{eqnarray}
F(p, t)=
c_1'+c_2'p^{-5},
\end{eqnarray}  
which denotes a single powe law
with the spectral index $-5$. 
If the relation 
$\lambda=25\kappa_{pp}^{(2)}/4$,
the indices can be found 
$r_1=r_2=r=-5/2$. 
Correspondingly, the solution
is 
\begin{eqnarray}
F(p, t)
=
\left(c_1'+c_2'\ln p\right)
p^{-5/2}
e^{-25\kappa_{pp}^{II}t/4}. 	
\end{eqnarray}

\subsection{The case 
$S^{I}\gg S^{II}$}

When $\xi$ is very 
close to zero, we find that 
$S^{II}\approx0$, and Equation 
(\ref{F equation for compressible waves})
is simplified as
\begin{eqnarray}
	\frac{\partial{F}}{\partial{t}}	
	=
	\kappa_{pp}^{(1)}
	\frac{1}{p^2}\frac{\partial{}}{\partial{p}}
	\left(p^4\frac{\partial{F}}{\partial{p}}\right).
	\label{F equation with SI}					
\end{eqnarray}
The solution of the latter equation
can be easily derive as follows
\begin{eqnarray}
	F(p, t)
	=
	\left(c_1'p^{r_1}+c_2'p^{r_2}\right)
	e^{-\lambda t}	 
	\label{solution for equation with S1}
\end{eqnarray}
with
\begin{eqnarray}
	r_{1,2}=\frac{1}{2}
	\left(-3\pm\sqrt{
		9-4\lambda/\kappa_{pp}^{(1)}}
	\right).
\end{eqnarray}
If $\lambda=0$, the formulas
$r_1=0$ and $r_2=-3$ holds.
The solution demonstrates a 
single power law as follows
\begin{eqnarray}
F(p, t)
=c_1'+c_2'p^{-3}.
\end{eqnarray}
When $\lambda=9\kappa_{pp}^{(1)}/4$,
the solution becomes
\begin{eqnarray}
F(p, t)
=
\left(c_1'+c_2'\ln p\right)
p^{-3/2}
e^{-9\kappa_{pp}^{(1)}t/4}. 		
\end{eqnarray}

\subsection{Discussion on the 
power law of solar energetic particles}
\label{Discussion on the 
power law of solar energetic particles}

In the case of 
solar energetic electrons 
and protons, the convection effect
needs to be considered. 
Hence, the results, derived in
Section \ref{The solution of isotropic distribution function}, 
might potentially 
describe solar 
energetic particle events.
Given that the momentum
convection effect is
significantly geater 
than diffusion,
the index formula 
(\ref{r1,2 with convection and time}) can be simplified as follows
\begin{eqnarray}
r_{1,2}=\frac{1}{2}
\left[
-\frac{\kappa_p}
{\kappa_{pp}^{(1)}
	+\kappa_{pp}^{(2)}}
\pm
\sqrt{
\left(
\frac{\kappa_p}{\kappa_{pp}^{(1)}
+\kappa_{pp}^{(2)}}
\right)^2
-\frac{4\lambda}
{\kappa_{pp}^{(1)}
+\kappa_{pp}^{(2)}}
}
\right],	
\end{eqnarray}
which denotes that the indices
are typically less than
the indices of galactic cosmic rays,
$-5$.
In addition, the isotropic 
distribution function equation
pertaining to 
momentum transport can be expressed 
as follows
\begin{eqnarray}  
\frac{\partial{F}}{\partial{t}}	
&&=
\kappa_p
p
\frac{\partial{F}}{\partial{p}}
+
\kappa_{pp}^{(1)}
\frac{1}{p^2}\frac{\partial{}}{\partial{p}}
\left(p^4\frac{\partial{F}}{\partial{p}}\right)
+\kappa_{pp}^{(2)}
\frac{1}{p^4}\frac{\partial{}}{\partial{p}}
\left(
p^6
\frac{\partial{F}}{\partial{p}}
\right)+\cdots
\nonumber\\
&&
=S_p+S_{pp}^{(1)}+S_{pp}^{(2)}
+\cdots+S_{n-1}+S_{n}+\cdots,
\label{momentum transport equation
with infinite terms}
\end{eqnarray}
where the right-hand side
of the latter equation
is a series of function terms.
The ratio the $n$th $T_n$ and 
$(n-1)$th $T_{n-1}$ is
\begin{eqnarray} 
\frac{T_n}{T_{n-1}}=
\frac{\frac{1}{D^{2n}}(
\nabla\cdot\bm{V})^{2n}}
{\frac{1}{D^{2n-1}}(
\nabla\cdot\bm{V})^{2n-1}} 
=\frac{\nabla\cdot\bm{V}}{D}
\sim \frac{VT_{\mu}}{L}.
\end{eqnarray}
Here, $V$ is the characteristic
scale of background flow velocity,
$L$ is the characteristic length
of spatial gradient of background flow
velocity, and $T_{\mu}$
is the characteristic time scale
of pithc-angle scattering.
For galactic cosmic rays,
it is noted that 
the pitch-angle scattering 
tends to be isotropic,
and the characteristic time 
$T_{\mu}$ is notably small.
Thus, it follows that 
$VT_{\mu}/L\ll 1$, or equivalently,
that is, $T_n\ll T_{n-1}$,
indicating that the right-hand
side of Equation 
(\ref{momentum transport equation
with infinite terms}) 
converges rapidly. 
Thus, it is likely that 
only the retained 
second-order momentum
derivative terms can
accurately describe 
the momentum transport and 
power law properties of
galactic socmic rays.
However, for solar energetic 
particle events, the pitch-angle 
scattering is far from reaching
the isotropic state, and the 
characteristic time $T_{\mu}$
of $D_{\mu\mu}$ should not be 
considered as a
small value. Thus, the ratio
of  $VT_{\mu}$ to $L$ 
is not too small.
In this case, the right-hand side
of Equation 
(\ref{momentum transport equation
with infinite terms}) should 
retain additional terms,
which contain higher-order 
momentum derivatives. In Appendix
\ref{The power laws for momentum transport equation up to the third-order
momentum derivatives},
we derive a third-order momentum
derivative term, and find 
it satisfies a momentum power law
with a spectral index $-6$.
Similarly, we deduce a fourth-order
momentum derivative term,
which obeys a power law with a spectral index $-7$. Due to
length and complexity of 
deriving all third and 
fourth-order momentum 
derivative terms, we only  
derive this special third and 
fourth-order momentum derivative
term, specifically. However, 
from the two special cases,
we can draw 
the conclusion that
higher-order momentum derivative
terms might obey higher spectral 
indices.

\subsection{The physical meanings 
of spatial gradients
of background flow velocity components 
and velocity divergence}
\label{The physical meanings 
of spatial gradients
of solar wind velocity components 
and velocity divergence}

The volume change of background flow
plasma elements  is indicated by the 
divergence operator, 
$\nabla\cdot\bm{V}$.
When 
$\nabla\cdot\bm{V}>0$,
the volume of the background flow 
plasma element
increases; whereas
when $\nabla\cdot\bm{V}<0$,
the background flow volume decreases. 
The condition $\nabla\cdot\bm{V}=0$
implies that
the volume of background flow element
remains unchanged. 
In addition, 
the spatial gradients 
$\partial{V_x}/\partial{x}$,
$\partial{V_y}/\partial{y}$,
and $\partial{V_z}/\partial{z}$
indicate that the  
tension or compression 
of plasma fluid elements 
along the x-, y-, and z-axis 
directions, respectively. 
Moreover, velocity divergence
\begin{eqnarray}
\nabla\cdot\bm{V}=
\frac{\partial{V_x}}{\partial{x}}
+\frac{\partial{V_y}}{\partial{y}}
+\frac{\partial{V_z}}{\partial{z}}
\end{eqnarray}	
is the sum of spatial gradients of
velocity components 
along the x-, y-, and z-axis 
directions. Thus, velocity divergence 
is also indicative of 
tension or compression 
of background flow plasma elements.
In summary, the spatial gradients
of background flow velocity  
components and velocity divergence 
serve as indicators of 
BFVIs.
Therefore, 
the momentum transport processes
$S^{I}$ and $S^{II}$
are induced by
the combined effects of
pitch-angle diffusion and BFVIs.

\section{Summary and conclusion}
\label{SUMMARY AND CONCLUSION}

In this article, 
with 
the well-known focusing equation, 
we demonstrate that 
the combined effects of 
pitch-angle diffusion and 
BFVIs
can induce new momentum transport
processes. 
The inhomogeneities of 
background flow velocity 
is ubiquitous in 
astrophysical environment.
By using the method of separation 
of variables, 
the solution to
the isotropic distribution function 
equation of energetic particles is obtained, which 
reveals the form
of momentum power laws.

Firstly, we study the 
isotropic distribution function 
in the 
incompressible plasma cases. 
In three conditions,
we find power laws with different 
spectral indices. 
Among them, 
in the quasi-steady state,
the distribution function 
of charged energetic particles 
follows
the momentum spectrum with 
a spectral index of $-3$.

Secondly, 
we explore the 
isotropic distribution function 
in the 
compressible plasma cases. 
In the condition of
compressible plasma waves or turbulence, we assume that 
all terms  $(\partial{V_z}
/\partial{z})^{k}$ and $(\nabla\cdot \bm{V})^{k}$
with odd number $k$ are ignorable. 
For the quasi-steady state, momentum distribution of 
energetic particles obeys the power laws, with the spectral index 
$\delta$ determined
by the background flow velocity
divergence and the 
spatial gradients of its components.  
The spectral index $\delta$ 
is further shown 
to range in $[-5, -3]$.
Furthermore, 
the observation results 
demonstrate that the 
galactic cosmic rays 
obey the momentum power law,
with spectral index ranging from
$-5.1$ to $-4.7$,
which approximately falls 
within the interval $[-5, -3]$.  
Therefore,  
it is 
possible that, in addition to shock acceleration, the mechanism proposed in 
this article may 
also contributes to 
galactic cosmic rays.

For solar energetic particle events, 
the momentum
convection effect need to be 
considered. In addition,
because 
the charcateristic time scale
of the pitch-angle diffusion
is not very small, 
in order to accurately describe 
solar energetic particle events,
the momentum 
transport equation might need 
to retain 
third-order or even higher-order 
terms. According to the 
investigation
in this article, the indices of 
momentum power laws should be smaller than $-5$, which qualitatively 
agrees with 
the observations.  
    
It is noted that the well-known
theory, 
the pump acceleration mechanism, 
is presented,
in order to explain 
suprathermal particles
in solar wind \citep{FiskEA2008, 
	FiskEA2010, FiskGloeckler2012}. 
In this theory,
the momentum transport
is caused by the 
combined effects of 
perpendicular diffusion as well as
compression and expansion of plasma
waves, and 
the momentum power law with 
a spectral index of
$-5$ is presented. 
It is possible that 
the other irreversible 
processes,
e.g., gyrophase scattering,
drift diffusion and so forth,
as well as BFVIs,  
might also produce particle 
momentum transport processes. 
We will expore these various physical 
processes in the future. 
Additionally, the 
formulas of pitch-angle diffusion
coefficient $D_{\mu\mu}$ should 
take different forms 
depending on the
particle energy ranges,
the magnetic turbulence 
strength, and so on. 
For these cases, the parameter $\xi$
exhibits different 
values across various scenarios. 
Thus,
the influence of specific forms 
of $D_{\mu\mu}$ on momentum transport
should be investigated, and 
the corresponding acceleration of 
energetic particles are obtained. 
The results presented in this article 
can also be used 
to explore momentum transport
of energetic particles 
in other scenarios, 
i.e.,
solar flares and coronal mass 
ejections at the Sun considering
the time effect from $\lambda\ne0$ 
in the 
distribution function. \\

\begin{acknowledgments}
\nnsfc{\WangjfNNSFC, \QinNNSFCouter, \QinNNSFCelectrons, \GuoNNSFC}
\szstp{\QinSZdisaster}
\nkrdpc{\GuoNKRDPC, \ShenNKRDPC}
\szkllp{\FengLAB}
\sprpcas{\FengCAS}
\referee
\end{acknowledgments}

\renewcommand{\theequation}{\Alph{section}-\arabic{equation}}
\setcounter{equation}{0}  
\begin{appendices}
\section{The formilas of $Z_n$} 
\label{The formilas of Zn}

The formulas of $Z_n$,
where $n$ ranges from  
$1$ to $16$, are listed
as follows
\begin{eqnarray}
&&Z_1=-\frac{1}{p^2}\frac{\partial{}}{\partial{p}}
\Bigg[
\frac{1}{2}\int_{-1}^1\dee\mu
\frac{1-\mu^2}{2}
p^3\nabla\cdot \bm{V}
+\frac{1}{2}\int_{-1}^1\dee\mu
\frac{3\mu^2-1}{2}
p^3\frac{\partial{V_z}}
{\partial{z}}
\Bigg]
\nonumber\\
&&
\times
\Bigg(\int_{-1}^{\mu}\dee \mu \frac{1}{D_{\mu\mu}}
-\frac{1}{2}
\int_{-1}^{1}\dee \mu 
\frac{1-\mu}{D_{\mu\mu}}\Bigg)(\mu+1)
\nabla\cdot\left(\kappa_\bot\cdot \nabla F\right),
\label{Z1}
\\
&&
Z_2=-\frac{1}{p^2}\frac{\partial{}}{\partial{p}}
\Bigg[
\frac{1}{2}\int_{-1}^1\dee\mu
\frac{1-\mu^2}{2}
p^3\nabla\cdot \bm{V}
+\frac{1}{2}\int_{-1}^1\dee\mu
\frac{3\mu^2-1}{2}
p^3\frac{\partial{V_z}}
{\partial{z}}
\Bigg]
\nonumber\\
&&
\times
\Bigg(\int_{-1}^{\mu}\dee \mu \frac{1}{D_{\mu\mu}}
-\frac{1}{2}
\int_{-1}^{1}\dee \mu 
\frac{1-\mu}{D_{\mu\mu}}\Bigg)
\int_{-1}^{\mu}\dee\mu
\nabla\cdot\left(\kappa_\bot\cdot \nabla g\right),
\\
&&
Z_3=\frac{1}{p^2}\frac{\partial{}}{\partial{p}}
\Bigg[
\frac{1}{2}\int_{-1}^1\dee\mu
\frac{1-\mu^2}{2}
p^3\nabla\cdot \bm{V}
+\frac{1}{2}\int_{-1}^1\dee\mu
\frac{3\mu^2-1}{2}
p^3\frac{\partial{V_z}}
{\partial{z}}
\Bigg]
\nonumber\\
&&
\times
\Bigg(\int_{-1}^{\mu}\dee \mu \frac{1}{D_{\mu\mu}}
-\frac{1}{2}
\int_{-1}^{1}\dee \mu 
\frac{1-\mu}{D_{\mu\mu}}\Bigg)
v\frac{\mu^2-1}{2}
\frac{\partial{F}}{\partial{z}},
\\
&&
Z_4=\frac{1}{p^2}\frac{\partial{}}{\partial{p}}
\Bigg[
\frac{1}{2}\int_{-1}^1\dee\mu
\frac{1-\mu^2}{2}
p^3\nabla\cdot \bm{V}
+\frac{1}{2}\int_{-1}^1\dee\mu
\frac{3\mu^2-1}{2}
p^3\frac{\partial{V_z}}
{\partial{z}}
\Bigg]
\nonumber\\
&&
\times
\Bigg(\int_{-1}^{\mu}\dee \mu \frac{1}{D_{\mu\mu}}
-\frac{1}{2}
\int_{-1}^{1}\dee \mu 
\frac{1-\mu}{D_{\mu\mu}}\Bigg)
v\frac{\partial{}}{\partial{z}}
\int_{-1}^{\mu}\dee\mu\mu g,
\\
&&
Z_5=\frac{1}{p^2}\frac{\partial{}}{\partial{p}}
\Bigg[
\frac{1}{2}\int_{-1}^1\dee\mu
\frac{1-\mu^2}{2}
p^3\nabla\cdot \bm{V}
+\frac{1}{2}\int_{-1}^1\dee\mu
\frac{3\mu^2-1}{2}
p^3\frac{\partial{V_z}}
{\partial{z}}
\Bigg]
\nonumber\\
&&
\times
\Bigg(\int_{-1}^{\mu}\dee \mu \frac{1}{D_{\mu\mu}}
-\frac{1}{2}
\int_{-1}^{1}\dee \mu 
\frac{1-\mu}{D_{\mu\mu}}\Bigg)
F\left(\nabla\cdot \bm{V}\right)(\mu+1),
\\
&&
Z_6=\frac{1}{p^2}\frac{\partial{}}{\partial{p}}
\Bigg[
\frac{1}{2}\int_{-1}^1\dee\mu
\frac{1-\mu^2}{2}
p^3\nabla\cdot \bm{V}
+\frac{1}{2}\int_{-1}^1\dee\mu
\frac{3\mu^2-1}{2}
p^3\frac{\partial{V_z}}
{\partial{z}}
\Bigg]
\nonumber\\
&&
\times
\Bigg(\int_{-1}^{\mu}\dee \mu \frac{1}{D_{\mu\mu}}
-\frac{1}{2}
\int_{-1}^{1}\dee \mu 
\frac{1-\mu}{D_{\mu\mu}}\Bigg)
\big(\bm{V}_\perp\cdot\nabla_\perp \big)F(\mu+1),
\\
&&
Z_7=\frac{1}{p^2}\frac{\partial{}}{\partial{p}}
\Bigg[
\frac{1}{2}\int_{-1}^1\dee\mu
\frac{1-\mu^2}{2}
p^3\nabla\cdot \bm{V}
+\frac{1}{2}\int_{-1}^1\dee\mu
\frac{3\mu^2-1}{2}
p^3\frac{\partial{V_z}}
{\partial{z}}
\Bigg]
\nonumber\\
&&
\times
\Bigg(\int_{-1}^{\mu}\dee \mu \frac{1}{D_{\mu\mu}}
-\frac{1}{2}
\int_{-1}^{1}\dee \mu 
\frac{1-\mu}{D_{\mu\mu}}\Bigg)
V_z\frac{\partial{F}}{\partial{z}}
(\mu+1),
\\
&&
Z_8=\frac{1}{p^2}\frac{\partial{}}{\partial{p}}
\Bigg[
\frac{1}{2}\int_{-1}^1\dee\mu
\frac{1-\mu^2}{2}
p^3\nabla\cdot \bm{V}
+\frac{1}{2}\int_{-1}^1\dee\mu
\frac{3\mu^2-1}{2}
p^3\frac{\partial{V_z}}
{\partial{z}}
\Bigg]
\nonumber\\
&&
\times
\Bigg(\int_{-1}^{\mu}\dee \mu \frac{1}{D_{\mu\mu}}
-\frac{1}{2}
\int_{-1}^{1}\dee \mu 
\frac{1-\mu}{D_{\mu\mu}}\Bigg)
\int_{-1}^{\mu}\dee\mu
g\left(\nabla\cdot \bm{V}\right),
\\
&&
Z_9=\frac{1}{p^2}\frac{\partial{}}{\partial{p}}
\Bigg[
\frac{1}{2}\int_{-1}^1\dee\mu
\frac{1-\mu^2}{2}
p^3\nabla\cdot \bm{V}
+\frac{1}{2}\int_{-1}^1\dee\mu
\frac{3\mu^2-1}{2}
p^3\frac{\partial{V_z}}
{\partial{z}}
\Bigg]
\nonumber\\
&&
\times
\Bigg(\int_{-1}^{\mu}\dee \mu \frac{1}{D_{\mu\mu}}
-\frac{1}{2}
\int_{-1}^{1}\dee \mu 
\frac{1-\mu}{D_{\mu\mu}}\Bigg)
\left(\bm{V}_\perp\cdot\nabla_\perp \right)
\int_{-1}^{\mu}\dee\mu g,
\\
&&
Z_{10}=\frac{1}{p^2}\frac{\partial{}}{\partial{p}}
\Bigg[
\frac{1}{2}\int_{-1}^1\dee\mu
\frac{1-\mu^2}{2}
p^3\nabla\cdot \bm{V}
+\frac{1}{2}\int_{-1}^1\dee\mu
\frac{3\mu^2-1}{2}
p^3\frac{\partial{V_z}}
{\partial{z}}
\Bigg]
\nonumber\\
&&
\times
\Bigg(\int_{-1}^{\mu}\dee \mu \frac{1}{D_{\mu\mu}}
-\frac{1}{2}
\int_{-1}^{1}\dee \mu 
\frac{1-\mu}{D_{\mu\mu}}\Bigg)
V_z\frac{\partial{}}{\partial{z}}
\int_{-1}^{\mu}\dee\mu g,
\\
&&
Z_{11}=-\frac{1}{p^2}\frac{\partial{}}{\partial{p}}
\Bigg[
\frac{1}{2}\int_{-1}^1\dee\mu
\frac{1-\mu^2}{2}
p^3\nabla\cdot \bm{V}
+\frac{1}{2}\int_{-1}^1\dee\mu
\frac{3\mu^2-1}{2}
p^3\frac{\partial{V_z}}
{\partial{z}}
\Bigg]
\nonumber\\
&&
\times
\Bigg(\int_{-1}^{\mu}\dee \mu \frac{1}{D_{\mu\mu}}
-\frac{1}{2}
\int_{-1}^{1}\dee \mu 
\frac{1-\mu}{D_{\mu\mu}}\Bigg)
\frac{1}{p^2}\frac{\partial{}}{\partial{p}}
\Bigg\{p^3\Bigg[\frac{3\mu-\mu^3+2}{6}
\nabla\cdot \bm{V}
+\frac{\mu^3-\mu}{2}\frac{\partial{V_z}}
{\partial{z}}
\Bigg]F\Bigg\},
\\
&&
Z_{12}=-\frac{1}{p^2}\frac{\partial{}}{\partial{p}}
\Bigg[
\frac{1}{2}\int_{-1}^1\dee\mu
\frac{1-\mu^2}{2}
p^3\nabla\cdot \bm{V}
+\frac{1}{2}\int_{-1}^1\dee\mu
\frac{3\mu^2-1}{2}
p^3\frac{\partial{V_z}}
{\partial{z}}
\Bigg]
\nonumber\\
&&
\times
\Bigg(\int_{-1}^{\mu}\dee \mu \frac{1}{D_{\mu\mu}}
-\frac{1}{2}
\int_{-1}^{1}\dee \mu 
\frac{1-\mu}{D_{\mu\mu}}\Bigg)
\frac{1}{p^2}\frac{\partial{}}{\partial{p}}
\Bigg\{p^3\int_{-1}^{\mu}\dee\mu
\Bigg[\frac{1-\mu^2}{2}
\nabla\cdot \bm{V}
+\frac{3\mu^2-1}{2}\frac{\partial{V_z}}
{\partial{z}}
\Bigg]g\Bigg\},
\\
&&
Z_{13}=\frac{1}{p^2}\frac{\partial{}}{\partial{p}}
\Bigg[
\frac{1}{2}\int_{-1}^1\dee\mu
\frac{1-\mu^2}{2}
p^3\nabla\cdot \bm{V}
+\frac{1}{2}\int_{-1}^1\dee\mu
\frac{3\mu^2-1}{2}
p^3\frac{\partial{V_z}}
{\partial{z}}
\Bigg]
\nonumber\\
&&
\times
\Bigg(\int_{-1}^{\mu}\dee \mu \frac{1}{D_{\mu\mu}}
-\frac{1}{2}
\int_{-1}^{1}\dee \mu 
\frac{1-\mu}{D_{\mu\mu}}\Bigg)
\frac{1-\mu^2}{2}
\mu\left(\nabla\cdot \bm{V}-
3\frac{\partial{V_z}}{\partial{z}}
\right)F,
\\
&&
Z_{14}=\frac{1}{p^2}\frac{\partial{}}{\partial{p}}
\Bigg[
\frac{1}{2}\int_{-1}^1\dee\mu
\frac{1-\mu^2}{2}
p^3\nabla\cdot \bm{V}
+\frac{1}{2}\int_{-1}^1\dee\mu
\frac{3\mu^2-1}{2}
p^3\frac{\partial{V_z}}
{\partial{z}}
\Bigg]
\nonumber\\
&&
\times
\Bigg(\int_{-1}^{\mu}\dee \mu \frac{1}{D_{\mu\mu}}
-\frac{1}{2}
\int_{-1}^{1}\dee \mu 
\frac{1-\mu}{D_{\mu\mu}}\Bigg)
\frac{1-\mu^2}{2}
\mu\left(\nabla\cdot \bm{V}-
3\frac{\partial{V_z}}{\partial{z}}
\right)g,
\\
&&
Z_{15}=\frac{1}{p^2}\frac{\partial{}}{\partial{p}}
\Bigg[
\frac{1}{2}\int_{-1}^1\dee\mu
\frac{1-\mu^2}{2}
p^3\nabla\cdot \bm{V}
+\frac{1}{2}\int_{-1}^1\dee\mu
\frac{3\mu^2-1}{2}
p^3\frac{\partial{V_z}}
{\partial{z}}
\Bigg]
\nonumber\\
&&
\times
\Bigg(\int_{-1}^{\mu}\dee \mu \frac{1}{D_{\mu\mu}}
-\frac{1}{2}
\int_{-1}^{1}\dee \mu 
\frac{1-\mu}{D_{\mu\mu}}\Bigg)
\frac{\partial{F}}{\partial{t}}(\mu+1),
\\
&&
Z_{16}=\frac{1}{p^2}\frac{\partial{}}{\partial{p}}
\Bigg[
\frac{1}{2}\int_{-1}^1\dee\mu
\frac{1-\mu^2}{2}
p^3\nabla\cdot \bm{V}
+\frac{1}{2}\int_{-1}^1\dee\mu
\frac{3\mu^2-1}{2}
p^3\frac{\partial{V_z}}
{\partial{z}}
\Bigg]
\nonumber\\
&&
\times
\Bigg(\int_{-1}^{\mu}\dee \mu \frac{1}{D_{\mu\mu}}
-\frac{1}{2}
\int_{-1}^{1}\dee \mu 
\frac{1-\mu}{D_{\mu\mu}}\Bigg)
\int_{-1}^{\mu}\dee\mu
\frac{\partial{g}}{\partial{t}}	
\label{Z16}.
\end{eqnarray}

\section{Evaluating $Z_n$} 
\label{Evaluating Zn}

Considering 
the requirements and assumptions
outlined
in the main text, 
and using the same evaluation methods 
in subection \ref{Evaluating Z1-Z16},
we now proceed to evaluate Equations 
(\ref{Z1})-(\ref{Z16}) as
\begin{eqnarray}
Z_1=0, Z_2=0, Z_3=0.  	
\end{eqnarray} 
For term $Z_4$, 
by inserting the formula of $g$ into 
it, we find  
\begin{eqnarray}
Z_4&&=\sum_{n=1}^{12}Z_4(n)	
\end{eqnarray}
with
\begin{eqnarray}
&&Z_4(1)=\frac{1}{2m}
\frac{1}{p^2}\frac{\partial{}}{\partial{p}}p^4
\Bigg[
\int_{-1}^1\dee\mu
\frac{1-\mu^2}{2}
\nabla\cdot \bm{V}
+\int_{-1}^1\dee\mu
\frac{3\mu^2-1}{2}
\frac{\partial{V_z}}
{\partial{z}}
\Bigg]
\nonumber\\
&&
\times
\Bigg(\int_{-1}^{\mu}\dee \mu \frac{1}{D_{\mu\mu}}
-\frac{1}{2}
\int_{-1}^{1}\dee \mu 
\frac{1-\mu}{D_{\mu\mu}}\Bigg)
\int_{-1}^{\mu}\dee\mu\mu 
\Bigg(\int_{-1}^{\mu}\dee \mu \frac{1}{D_{\mu\mu}}
-\frac{1}{2}
\int_{-1}^{1}\dee \mu 
\frac{1-\mu}{D_{\mu\mu}}\Bigg)
\frac{\partial{}}{\partial{z}}
v\frac{\mu^2-1}{2}
\frac{\partial{F}}{\partial{z}},
\\
&&
Z_4(2)=
\frac{1}{2m}
\frac{1}{p^2}\frac{\partial{}}{\partial{p}}p^4
\Bigg[
\int_{-1}^1\dee\mu
\frac{1-\mu^2}{2}
\nabla\cdot \bm{V}
+\int_{-1}^1\dee\mu
\frac{3\mu^2-1}{2}
\frac{\partial{V_z}}
{\partial{z}}
\Bigg]
\nonumber\\
&&
\times
\Bigg(\int_{-1}^{\mu}\dee \mu \frac{1}{D_{\mu\mu}}
-\frac{1}{2}
\int_{-1}^{1}\dee \mu 
\frac{1-\mu}{D_{\mu\mu}}\Bigg)
\int_{-1}^{\mu}\dee\mu\mu 
\Bigg(\int_{-1}^{\mu}\dee \mu \frac{1}{D_{\mu\mu}}
-\frac{1}{2}
\int_{-1}^{1}\dee \mu 
\frac{1-\mu}{D_{\mu\mu}}\Bigg)
\frac{\partial{}}{\partial{z}}
v\frac{\partial{}}{\partial{z}}
\int_{-1}^{\mu}\dee\mu\mu g,
\label{Z4(2)}
\\
&&
Z_4(3)=
\frac{1}{2m}
\frac{1}{p^2}\frac{\partial{}}{\partial{p}}p^4
\Bigg[
\int_{-1}^1\dee\mu
\frac{1-\mu^2}{2}
\nabla\cdot \bm{V}
+\int_{-1}^1\dee\mu
\frac{3\mu^2-1}{2}
\frac{\partial{V_z}}
{\partial{z}}
\Bigg]
\nonumber\\
&&
\times
\Bigg(\int_{-1}^{\mu}\dee \mu \frac{1}{D_{\mu\mu}}
-\frac{1}{2}
\int_{-1}^{1}\dee \mu 
\frac{1-\mu}{D_{\mu\mu}}\Bigg)
\int_{-1}^{\mu}\dee\mu\mu 
\Bigg(\int_{-1}^{\mu}\dee \mu \frac{1}{D_{\mu\mu}}
-\frac{1}{2}
\int_{-1}^{1}\dee \mu 
\frac{1-\mu}{D_{\mu\mu}}\Bigg)
\frac{\partial{}}{\partial{z}}
F\left(\nabla\cdot \bm{V}\right)(\mu+1),
\\
&&
Z_4(4)=
\frac{1}{2m}
\frac{1}{p^2}\frac{\partial{}}{\partial{p}}p^4
\Bigg[
\int_{-1}^1\dee\mu
\frac{1-\mu^2}{2}
\nabla\cdot \bm{V}
+\int_{-1}^1\dee\mu
\frac{3\mu^2-1}{2}
\frac{\partial{V_z}}
{\partial{z}}
\Bigg]
\nonumber\\
&&
\times
\Bigg(\int_{-1}^{\mu}\dee \mu \frac{1}{D_{\mu\mu}}
-\frac{1}{2}
\int_{-1}^{1}\dee \mu 
\frac{1-\mu}{D_{\mu\mu}}\Bigg)
\int_{-1}^{\mu}\dee\mu\mu 
\Bigg(\int_{-1}^{\mu}\dee \mu \frac{1}{D_{\mu\mu}}
-\frac{1}{2}
\int_{-1}^{1}\dee \mu 
\frac{1-\mu}{D_{\mu\mu}}\Bigg)
\frac{\partial{}}{\partial{z}}
V_z\frac{\partial{F}}{\partial{z}}
(\mu+1),
\\
&&
Z_4(5)=
\frac{1}{2m}
\frac{1}{p^2}\frac{\partial{}}{\partial{p}}p^4
\Bigg[
\int_{-1}^1\dee\mu
\frac{1-\mu^2}{2}
\nabla\cdot \bm{V}
+\int_{-1}^1\dee\mu
\frac{3\mu^2-1}{2}
\frac{\partial{V_z}}
{\partial{z}}
\Bigg]
\nonumber\\
&&
\times
\Bigg(\int_{-1}^{\mu}\dee \mu \frac{1}{D_{\mu\mu}}
-\frac{1}{2}
\int_{-1}^{1}\dee \mu 
\frac{1-\mu}{D_{\mu\mu}}\Bigg)
\int_{-1}^{\mu}\dee\mu\mu 
\Bigg(\int_{-1}^{\mu}\dee \mu \frac{1}{D_{\mu\mu}}
-\frac{1}{2}
\int_{-1}^{1}\dee \mu 
\frac{1-\mu}{D_{\mu\mu}}\Bigg)
\frac{\partial{}}{\partial{z}}
\int_{-1}^{\mu}\dee\mu
g\left(\nabla\cdot \bm{V}\right),
\\
&&
Z_4(6)=
\frac{1}{2m}
\frac{1}{p^2}\frac{\partial{}}{\partial{p}}p^4
\Bigg[
\int_{-1}^1\dee\mu
\frac{1-\mu^2}{2}
\nabla\cdot \bm{V}
+\int_{-1}^1\dee\mu
\frac{3\mu^2-1}{2}
\frac{\partial{V_z}}
{\partial{z}}
\Bigg]
\nonumber\\
&&
\times
\Bigg(\int_{-1}^{\mu}\dee \mu \frac{1}{D_{\mu\mu}}
-\frac{1}{2}
\int_{-1}^{1}\dee \mu 
\frac{1-\mu}{D_{\mu\mu}}\Bigg)
\int_{-1}^{\mu}\dee\mu\mu 
\Bigg(\int_{-1}^{\mu}\dee \mu \frac{1}{D_{\mu\mu}}
-\frac{1}{2}
\int_{-1}^{1}\dee \mu 
\frac{1-\mu}{D_{\mu\mu}}\Bigg)
\frac{\partial{}}{\partial{z}}
V_z\frac{\partial{}}{\partial{z}}
\int_{-1}^{\mu}\dee\mu g,
\\
&&
Z_4(7)=-
\frac{1}{2m}
\frac{1}{p^2}\frac{\partial{}}{\partial{p}}p^4
\Bigg[
\int_{-1}^1\dee\mu
\frac{1-\mu^2}{2}
\nabla\cdot \bm{V}
+\int_{-1}^1\dee\mu
\frac{3\mu^2-1}{2}
\frac{\partial{V_z}}
{\partial{z}}
\Bigg]
\nonumber\\
&&
\times
\Bigg(\int_{-1}^{\mu}\dee \mu \frac{1}{D_{\mu\mu}}
-\frac{1}{2}
\int_{-1}^{1}\dee \mu 
\frac{1-\mu}{D_{\mu\mu}}\Bigg)
\int_{-1}^{\mu}\dee\mu\mu 
\Bigg(\int_{-1}^{\mu}\dee \mu \frac{1}{D_{\mu\mu}}
-\frac{1}{2}
\int_{-1}^{1}\dee \mu 
\frac{1-\mu}{D_{\mu\mu}}\Bigg)
\frac{\partial{}}{\partial{z}}
\frac{1}{p^2}\frac{\partial{}}{\partial{p}}
\Bigg\{p^3\Bigg[\frac{3\mu-\mu^3+2}{6}
\nabla\cdot \bm{V}
\nonumber\\
&&	
+\frac{\mu^3-\mu}{2}\frac{\partial{V_z}}
{\partial{z}}
\Bigg]F\Bigg\},
\\
&&
Z_4(8)=-
\frac{1}{2m}
\frac{1}{p^2}\frac{\partial{}}{\partial{p}}p^4
\Bigg[
\int_{-1}^1\dee\mu
\frac{1-\mu^2}{2}
\nabla\cdot \bm{V}
+\int_{-1}^1\dee\mu
\frac{3\mu^2-1}{2}
\frac{\partial{V_z}}
{\partial{z}}
\Bigg]
\nonumber\\
&&
\times
\Bigg(\int_{-1}^{\mu}\dee \mu \frac{1}{D_{\mu\mu}}
-\frac{1}{2}
\int_{-1}^{1}\dee \mu 
\frac{1-\mu}{D_{\mu\mu}}\Bigg)
\int_{-1}^{\mu}\dee\mu\mu 
\Bigg(\int_{-1}^{\mu}\dee \mu \frac{1}{D_{\mu\mu}}
-\frac{1}{2}
\int_{-1}^{1}\dee \mu 
\frac{1-\mu}{D_{\mu\mu}}\Bigg)
\frac{\partial{}}{\partial{z}}
\frac{1}{p^2}\frac{\partial{}}{\partial{p}}
\Bigg\{p^3\int_{-1}^{\mu}\dee\mu
\Bigg[\frac{1-\mu^2}{2}
\nabla\cdot \bm{V}
\nonumber\\
&&	
+\frac{3\mu^2-1}{2}\frac{\partial{V_z}}
{\partial{z}}
\Bigg]g\Bigg\},
\\
&&
Z_4(9)=
\frac{1}{2m}
\frac{1}{p^2}\frac{\partial{}}{\partial{p}}p^4
\Bigg[
\int_{-1}^1\dee\mu
\frac{1-\mu^2}{2}
\nabla\cdot \bm{V}
+\int_{-1}^1\dee\mu
\frac{3\mu^2-1}{2}
\frac{\partial{V_z}}
{\partial{z}}
\Bigg]
\nonumber\\
&&
\times
\Bigg(\int_{-1}^{\mu}\dee \mu \frac{1}{D_{\mu\mu}}
-\frac{1}{2}
\int_{-1}^{1}\dee \mu 
\frac{1-\mu}{D_{\mu\mu}}\Bigg)
\int_{-1}^{\mu}\dee\mu\mu 
\Bigg(\int_{-1}^{\mu}\dee \mu \frac{1}{D_{\mu\mu}}
-\frac{1}{2}
\int_{-1}^{1}\dee \mu 
\frac{1-\mu}{D_{\mu\mu}}\Bigg)
\frac{\partial{}}{\partial{z}}
\frac{1-\mu^2}{2}
\mu\left(\nabla\cdot \bm{V}-
3\frac{\partial{V_z}}{\partial{z}}
\right)F,
\\
&&
Z_4(10)=
\frac{1}{2m}
\frac{1}{p^2}\frac{\partial{}}{\partial{p}}p^4
\Bigg[
\int_{-1}^1\dee\mu
\frac{1-\mu^2}{2}
\nabla\cdot \bm{V}
+\int_{-1}^1\dee\mu
\frac{3\mu^2-1}{2}
\frac{\partial{V_z}}
{\partial{z}}
\Bigg]
\nonumber\\
&&
\times
\Bigg(\int_{-1}^{\mu}\dee \mu \frac{1}{D_{\mu\mu}}
-\frac{1}{2}
\int_{-1}^{1}\dee \mu 
\frac{1-\mu}{D_{\mu\mu}}\Bigg)
\int_{-1}^{\mu}\dee\mu\mu 
\Bigg(\int_{-1}^{\mu}\dee \mu \frac{1}{D_{\mu\mu}}
-\frac{1}{2}
\int_{-1}^{1}\dee \mu 
\frac{1-\mu}{D_{\mu\mu}}\Bigg)
\frac{\partial{}}{\partial{z}}
\frac{1-\mu^2}{2}
\mu\left(\nabla\cdot \bm{V}-
3\frac{\partial{V_z}}{\partial{z}}
\right)g,
\\
&&
Z_4(11)=
\frac{1}{2m}
\frac{1}{p^2}\frac{\partial{}}{\partial{p}}p^4
\Bigg[
\int_{-1}^1\dee\mu
\frac{1-\mu^2}{2}
\nabla\cdot \bm{V}
+\int_{-1}^1\dee\mu
\frac{3\mu^2-1}{2}
\frac{\partial{V_z}}
{\partial{z}}
\Bigg]
\nonumber\\
&&
\times
\Bigg(\int_{-1}^{\mu}\dee \mu \frac{1}{D_{\mu\mu}}
-\frac{1}{2}
\int_{-1}^{1}\dee \mu 
\frac{1-\mu}{D_{\mu\mu}}\Bigg)
\int_{-1}^{\mu}\dee\mu\mu 
\Bigg(\int_{-1}^{\mu}\dee \mu \frac{1}{D_{\mu\mu}}
-\frac{1}{2}
\int_{-1}^{1}\dee \mu 
\frac{1-\mu}{D_{\mu\mu}}\Bigg)
\frac{\partial{}}{\partial{z}}
\frac{\partial{F}}{\partial{t}}(\mu+1),
\\
&&
Z_4(12)=
\frac{1}{2m}
\frac{1}{p^2}\frac{\partial{}}{\partial{p}}p^4
\Bigg[
\int_{-1}^1\dee\mu
\frac{1-\mu^2}{2}
\nabla\cdot \bm{V}
+\int_{-1}^1\dee\mu
\frac{3\mu^2-1}{2}
\frac{\partial{V_z}}
{\partial{z}}
\Bigg]
\nonumber\\
&&
\times
\Bigg(\int_{-1}^{\mu}\dee \mu \frac{1}{D_{\mu\mu}}
-\frac{1}{2}
\int_{-1}^{1}\dee \mu 
\frac{1-\mu}{D_{\mu\mu}}\Bigg)
\int_{-1}^{\mu}\dee\mu\mu 
\Bigg(\int_{-1}^{\mu}\dee \mu \frac{1}{D_{\mu\mu}}
-\frac{1}{2}
\int_{-1}^{1}\dee \mu 
\frac{1-\mu}{D_{\mu\mu}}\Bigg)
\frac{\partial{}}{\partial{z}}
\int_{-1}^{\mu}\dee\mu
\frac{\partial{g}}{\partial{t}}.	
\end{eqnarray} 

The latter formulas can be evaluated as
follows
\begin{eqnarray}
&&Z_4(1)=
\Bigg[
\nabla\cdot \bm{V}
-3
\frac{\partial{V_z}}
{\partial{z}}
\Bigg]
\frac{1}{180D^2}
\frac{1}{m^2}
\frac{1}{p^2}\frac{\partial{}}{\partial{p}}p^5
\frac{\partial^2{F}}{\partial{z^2}}, 
\nonumber\\
&&
Z_4(2)=0,
\nonumber\\
&&
Z_4(3)=	
\Bigg[
-
\left(\nabla\cdot \bm{V}\right)^2
+3
\frac{\partial{V_z}}
{\partial{z}}
\left(\nabla\cdot \bm{V}\right)
\Bigg]
\frac{1}{72D^2m}
\frac{1}{p^2}\frac{\partial{}}{\partial{p}}
\Bigg\{
p^4
\frac{\partial{F}}{\partial{z}}
\Bigg\},
\nonumber\\
&&
Z_4(4)=
\Bigg[
-
\left(\nabla\cdot \bm{V}\right)
\frac{\partial{V_z}}{\partial{z}}
+3
\frac{\partial{V_z}}
{\partial{z}}
\frac{\partial{V_z}}{\partial{z}}
\Bigg]
\frac{1}{72D^2m}
\frac{1}{p^2}\frac{\partial{}}{\partial{p}}
\Bigg(
p^4
\frac{\partial{F}}{\partial{z}}
\Bigg)
\nonumber\\
&&
+
\Bigg[
-V_z
\left(\nabla\cdot \bm{V}\right)
+3V_z
\frac{\partial{V_z}}
{\partial{z}}
\Bigg]
\frac{1}{72D^2m}
\frac{1}{p^2}\frac{\partial{}}{\partial{p}}
\Bigg(
p^4\frac{\partial^2{F}}
{\partial{z^2}}
\Bigg),
\nonumber\\
&&
Z_4(5)=0,
Z_4(6)=0,
\nonumber\\
&&
Z_4(7)=
\frac{1}{216mD^2}
\frac{1}{p^2}\frac{\partial{}}{\partial{p}}p^4
\Bigg[
(\nabla\cdot \bm{V})^2
-
4
(\nabla\cdot \bm{V})
\frac{\partial{V_z}}
{\partial{z}}
\Bigg]
\frac{1}{p^2}\frac{\partial{}}{\partial{p}}
\Bigg\{p^3
\frac{\partial{F}}{\partial{z}}\Bigg\},
\nonumber\\
&&
Z_4(8)=0,
Z_4(9)=0,
Z_4(10)=0,
Z_4(11)=0,
Z_4(12)=0.
\end{eqnarray} 
With the latter formulas of 
$Z_{4}(1)\sim Z_4(12)$, we obtain
\begin{eqnarray}
Z_4&&=
\Bigg[
\nabla\cdot \bm{V}
-3
\frac{\partial{V_z}}
{\partial{z}}
\Bigg]
\frac{1}{180D^2}
\frac{1}{m^2}
\frac{1}{p^2}\frac{\partial{}}{\partial{p}}p^5
\frac{\partial^2{F}}{\partial{z^2}}
\nonumber\\
&&
+\Bigg[
-2
\frac{\partial{V_z}}
{\partial{z}}
\left(\nabla\cdot \bm{V}\right)
+3
\left(
\frac{\partial{V_z}}
{\partial{z}}\right)^2
\Bigg]
\frac{1}{72D^2m}
\frac{1}{p^2}\frac{\partial{}}{\partial{p}}
\Bigg\{
p^4
\frac{\partial{F}}{\partial{z}}
\Bigg\}
\nonumber\\
&&
+
\Bigg[
-V_z
\left(\nabla\cdot \bm{V}\right)
+3V_z
\frac{\partial{V_z}}
{\partial{z}}
\Bigg]
\frac{1}{72D^2m}
\frac{1}{p^2}\frac{\partial{}}{\partial{p}}
\Bigg(
p^4\frac{\partial^2{F}}
{\partial{z^2}}
\Bigg)
\nonumber\\
&&
+
\Bigg[
\left(\nabla\cdot \bm{V}\right)^2
-4
\nabla\cdot \bm{V}
\frac{\partial{V_z}}
{\partial{z}}
\Bigg]
\frac{1}{216mD^2}
\frac{1}{p^2}
\frac{\partial{}}{\partial{p}}
p^5
\frac{\partial{}}{\partial{p}}
\frac{\partial{F}}{\partial{z}}. 	
\end{eqnarray} 
To proceed, we can find
\begin{eqnarray}
&&Z_5
=\frac{1}{p^2}\frac{\partial{}}{\partial{p}}\left(p^3F\right)
\frac{1}{18D}\Bigg[-
\left(\nabla\cdot \bm{V}\right)^2
+3
\frac{\partial{V_z}}
{\partial{z}}\left(\nabla\cdot \bm{V}\right)
\Bigg],
\nonumber\\
&&
Z_6=0,
\nonumber\\
&&
Z_7
=
\Bigg[
3
V_z\frac{\partial{V_z}}
{\partial{z}}
-
V_z\nabla\cdot \bm{V}
\Bigg]
\frac{1}{18D}
\frac{1}{p^2}\frac{\partial{}}{\partial{p}}
\left(
p^3
\frac{\partial{F}}{\partial{z}}
\right),
\nonumber\\
&&
Z_8=0, Z_9=0. 
\end{eqnarray}
With the anisotropic distribution
function, we have
\begin{eqnarray}
Z_{10}=\sum_{n=1}^{12}Z_{10}(n)
\end{eqnarray}
with
\begin{eqnarray}
&&Z_{10}(1)=\frac{1}{p^2}\frac{\partial{}}{\partial{p}}
\Bigg[
\frac{1}{2}\int_{-1}^1\dee\mu
\frac{1-\mu^2}{2}
p^3\nabla\cdot \bm{V}
+\frac{1}{2}\int_{-1}^1\dee\mu
\frac{3\mu^2-1}{2}
p^3\frac{\partial{V_z}}
{\partial{z}}
\Bigg]
\nonumber\\
&&
\times
\Bigg(\int_{-1}^{\mu}\dee \mu \frac{1}{D_{\mu\mu}}
-\frac{1}{2}
\int_{-1}^{1}\dee \mu 
\frac{1-\mu}{D_{\mu\mu}}\Bigg)
V_z
\int_{-1}^{\mu}\dee\mu 
\Bigg(\int_{-1}^{\mu}\dee \mu \frac{1}{D_{\mu\mu}}
-\frac{1}{2}
\int_{-1}^{1}\dee \mu 
\frac{1-\mu}{D_{\mu\mu}}\Bigg)
\frac{\partial{}}{\partial{z}}
v\frac{\mu^2-1}{2}
\frac{\partial{F}}{\partial{z}},
\\
&&
Z_{10}(2)=
\frac{1}{p^2}\frac{\partial{}}{\partial{p}}
\Bigg[
\frac{1}{2}\int_{-1}^1\dee\mu
\frac{1-\mu^2}{2}
p^3\nabla\cdot \bm{V}
+\frac{1}{2}\int_{-1}^1\dee\mu
\frac{3\mu^2-1}{2}
p^3\frac{\partial{V_z}}
{\partial{z}}
\Bigg]
\nonumber\\
&&
\times
\Bigg(\int_{-1}^{\mu}\dee \mu \frac{1}{D_{\mu\mu}}
-\frac{1}{2}
\int_{-1}^{1}\dee \mu 
\frac{1-\mu}{D_{\mu\mu}}\Bigg)
V_z
\int_{-1}^{\mu}\dee\mu 
\Bigg(\int_{-1}^{\mu}\dee \mu \frac{1}{D_{\mu\mu}}
-\frac{1}{2}
\int_{-1}^{1}\dee \mu 
\frac{1-\mu}{D_{\mu\mu}}\Bigg)
\frac{\partial{}}{\partial{z}}
v\frac{\partial{}}{\partial{z}}
\int_{-1}^{\mu}\dee\mu\mu g,
\\
&&
Z_{10}(3)=
\frac{1}{p^2}\frac{\partial{}}{\partial{p}}
\Bigg[
\frac{1}{2}\int_{-1}^1\dee\mu
\frac{1-\mu^2}{2}
p^3\nabla\cdot \bm{V}
+\frac{1}{2}\int_{-1}^1\dee\mu
\frac{3\mu^2-1}{2}
p^3\frac{\partial{V_z}}
{\partial{z}}
\Bigg]
\nonumber\\
&&
\times
\Bigg(\int_{-1}^{\mu}\dee \mu \frac{1}{D_{\mu\mu}}
-\frac{1}{2}
\int_{-1}^{1}\dee \mu 
\frac{1-\mu}{D_{\mu\mu}}\Bigg)
V_z
\int_{-1}^{\mu}\dee\mu 
\Bigg(\int_{-1}^{\mu}\dee \mu \frac{1}{D_{\mu\mu}}
-\frac{1}{2}
\int_{-1}^{1}\dee \mu 
\frac{1-\mu}{D_{\mu\mu}}\Bigg)
\frac{\partial{}}{\partial{z}}
F\left(\nabla\cdot \bm{V}\right)(\mu+1),
\\
&&
Z_{10}(4)=
\frac{1}{p^2}\frac{\partial{}}{\partial{p}}
\Bigg[
\frac{1}{2}\int_{-1}^1\dee\mu
\frac{1-\mu^2}{2}
p^3\nabla\cdot \bm{V}
+\frac{1}{2}\int_{-1}^1\dee\mu
\frac{3\mu^2-1}{2}
p^3\frac{\partial{V_z}}
{\partial{z}}
\Bigg]
\nonumber\\
&&
\times
\Bigg(\int_{-1}^{\mu}\dee \mu \frac{1}{D_{\mu\mu}}
-\frac{1}{2}
\int_{-1}^{1}\dee \mu 
\frac{1-\mu}{D_{\mu\mu}}\Bigg)
V_z
\int_{-1}^{\mu}\dee\mu 
\Bigg(\int_{-1}^{\mu}\dee \mu \frac{1}{D_{\mu\mu}}
-\frac{1}{2}
\int_{-1}^{1}\dee \mu 
\frac{1-\mu}{D_{\mu\mu}}\Bigg)
\frac{\partial{}}{\partial{z}}
V_z\frac{\partial{F}}{\partial{z}}
(\mu+1),
\\
&&
Z_{10}(5)=
\frac{1}{p^2}\frac{\partial{}}{\partial{p}}
\Bigg[
\frac{1}{2}\int_{-1}^1\dee\mu
\frac{1-\mu^2}{2}
p^3\nabla\cdot \bm{V}
+\frac{1}{2}\int_{-1}^1\dee\mu
\frac{3\mu^2-1}{2}
p^3\frac{\partial{V_z}}
{\partial{z}}
\Bigg]
\nonumber\\
&&
\times
\Bigg(\int_{-1}^{\mu}\dee \mu \frac{1}{D_{\mu\mu}}
-\frac{1}{2}
\int_{-1}^{1}\dee \mu 
\frac{1-\mu}{D_{\mu\mu}}\Bigg)
V_z
\int_{-1}^{\mu}\dee\mu 
\Bigg(\int_{-1}^{\mu}\dee \mu \frac{1}{D_{\mu\mu}}
-\frac{1}{2}
\int_{-1}^{1}\dee \mu 
\frac{1-\mu}{D_{\mu\mu}}\Bigg)
\frac{\partial{}}{\partial{z}}
\int_{-1}^{\mu}\dee\mu
g\left(\nabla\cdot \bm{V}\right),
\\
&&
Z_{10}(6)=
\frac{1}{p^2}\frac{\partial{}}{\partial{p}}
\Bigg[
\frac{1}{2}\int_{-1}^1\dee\mu
\frac{1-\mu^2}{2}
p^3\nabla\cdot \bm{V}
+\frac{1}{2}\int_{-1}^1\dee\mu
\frac{3\mu^2-1}{2}
p^3\frac{\partial{V_z}}
{\partial{z}}
\Bigg]
\nonumber\\
&&
\times
\Bigg(\int_{-1}^{\mu}\dee \mu \frac{1}{D_{\mu\mu}}
-\frac{1}{2}
\int_{-1}^{1}\dee \mu 
\frac{1-\mu}{D_{\mu\mu}}\Bigg)
V_z
\int_{-1}^{\mu}\dee\mu 
\Bigg(\int_{-1}^{\mu}\dee \mu \frac{1}{D_{\mu\mu}}
-\frac{1}{2}
\int_{-1}^{1}\dee \mu 
\frac{1-\mu}{D_{\mu\mu}}\Bigg)
\frac{\partial{}}{\partial{z}}
V_z\frac{\partial{}}{\partial{z}}
\int_{-1}^{\mu}\dee\mu g,
\\
&&
Z_{10}(7)=-
\frac{1}{p^2}\frac{\partial{}}{\partial{p}}
\Bigg[
\frac{1}{2}\int_{-1}^1\dee\mu
\frac{1-\mu^2}{2}
p^3\nabla\cdot \bm{V}
+\frac{1}{2}\int_{-1}^1\dee\mu
\frac{3\mu^2-1}{2}
p^3\frac{\partial{V_z}}
{\partial{z}}
\Bigg]
\nonumber\\
&&
\times
\Bigg(\int_{-1}^{\mu}\dee \mu \frac{1}{D_{\mu\mu}}
-\frac{1}{2}
\int_{-1}^{1}\dee \mu 
\frac{1-\mu}{D_{\mu\mu}}\Bigg)
V_z
\int_{-1}^{\mu}\dee\mu 
\Bigg(\int_{-1}^{\mu}\dee \mu \frac{1}{D_{\mu\mu}}
-\frac{1}{2}
\int_{-1}^{1}\dee \mu 
\frac{1-\mu}{D_{\mu\mu}}\Bigg)
\frac{\partial{}}{\partial{z}}
\frac{1}{p^2}\frac{\partial{}}{\partial{p}}
\Bigg\{p^3\Bigg[\frac{3\mu-\mu^3+2}{6}
\nabla\cdot \bm{V}
\nonumber\\
&&	
+\frac{\mu^3-\mu}{2}\frac{\partial{V_z}}
{\partial{z}}
\Bigg]F\Bigg\},
\\
&&
Z_{10}(8)=-
\frac{1}{p^2}\frac{\partial{}}{\partial{p}}
\Bigg[
\frac{1}{2}\int_{-1}^1\dee\mu
\frac{1-\mu^2}{2}
p^3\nabla\cdot \bm{V}
+\frac{1}{2}\int_{-1}^1\dee\mu
\frac{3\mu^2-1}{2}
p^3\frac{\partial{V_z}}
{\partial{z}}
\Bigg]
\nonumber\\
&&
\times
\Bigg(\int_{-1}^{\mu}\dee \mu \frac{1}{D_{\mu\mu}}
-\frac{1}{2}
\int_{-1}^{1}\dee \mu 
\frac{1-\mu}{D_{\mu\mu}}\Bigg)
V_z
\int_{-1}^{\mu}\dee\mu 
\Bigg(\int_{-1}^{\mu}\dee \mu \frac{1}{D_{\mu\mu}}
-\frac{1}{2}
\int_{-1}^{1}\dee \mu 
\frac{1-\mu}{D_{\mu\mu}}\Bigg)
\nonumber\\
&&
\frac{\partial{}}{\partial{z}}
\frac{1}{p^2}\frac{\partial{}}{\partial{p}}
\Bigg\{p^3\int_{-1}^{\mu}\dee\mu
\Bigg[\frac{1-\mu^2}{2}
\nabla\cdot \bm{V}
+\frac{3\mu^2-1}{2}\frac{\partial{V_z}}
{\partial{z}}
\Bigg]g\Bigg\},
\\
&&
Z_{10}(9)=
\frac{1}{p^2}\frac{\partial{}}{\partial{p}}
\Bigg[
\frac{1}{2}\int_{-1}^1\dee\mu
\frac{1-\mu^2}{2}
p^3\nabla\cdot \bm{V}
+\frac{1}{2}\int_{-1}^1\dee\mu
\frac{3\mu^2-1}{2}
p^3\frac{\partial{V_z}}
{\partial{z}}
\Bigg]
\nonumber\\
&&
\times
\Bigg(\int_{-1}^{\mu}\dee \mu \frac{1}{D_{\mu\mu}}
-\frac{1}{2}
\int_{-1}^{1}\dee \mu 
\frac{1-\mu}{D_{\mu\mu}}\Bigg)
V_z
\int_{-1}^{\mu}\dee\mu 
\Bigg(\int_{-1}^{\mu}\dee \mu \frac{1}{D_{\mu\mu}}
-\frac{1}{2}
\int_{-1}^{1}\dee \mu 
\frac{1-\mu}{D_{\mu\mu}}\Bigg)
\frac{\partial{}}{\partial{z}}
\frac{1-\mu^2}{2}
\mu\left(\nabla\cdot \bm{V}-
3\frac{\partial{V_z}}{\partial{z}}
\right)F,
\\
&&
Z_{10}(10)=
\frac{1}{p^2}\frac{\partial{}}{\partial{p}}
\Bigg[
\frac{1}{2}\int_{-1}^1\dee\mu
\frac{1-\mu^2}{2}
p^3\nabla\cdot \bm{V}
+\frac{1}{2}\int_{-1}^1\dee\mu
\frac{3\mu^2-1}{2}
p^3\frac{\partial{V_z}}
{\partial{z}}
\Bigg]
\nonumber\\
&&
\times
\Bigg(\int_{-1}^{\mu}\dee \mu \frac{1}{D_{\mu\mu}}
-\frac{1}{2}
\int_{-1}^{1}\dee \mu 
\frac{1-\mu}{D_{\mu\mu}}\Bigg)
V_z
\int_{-1}^{\mu}\dee\mu 
\Bigg(\int_{-1}^{\mu}\dee \mu \frac{1}{D_{\mu\mu}}
-\frac{1}{2}
\int_{-1}^{1}\dee \mu 
\frac{1-\mu}{D_{\mu\mu}}\Bigg)
\frac{\partial{}}{\partial{z}}
\frac{1-\mu^2}{2}
\mu\left(\nabla\cdot \bm{V}-
3\frac{\partial{V_z}}{\partial{z}}
\right)g,
\\
&&
Z_{10}(11)=
\frac{1}{p^2}\frac{\partial{}}{\partial{p}}
\Bigg[
\frac{1}{2}\int_{-1}^1\dee\mu
\frac{1-\mu^2}{2}
p^3\nabla\cdot \bm{V}
+\frac{1}{2}\int_{-1}^1\dee\mu
\frac{3\mu^2-1}{2}
p^3\frac{\partial{V_z}}
{\partial{z}}
\Bigg]
\nonumber\\
&&
\times
\Bigg(\int_{-1}^{\mu}\dee \mu \frac{1}{D_{\mu\mu}}
-\frac{1}{2}
\int_{-1}^{1}\dee \mu 
\frac{1-\mu}{D_{\mu\mu}}\Bigg)
V_z
\int_{-1}^{\mu}\dee\mu 
\Bigg(\int_{-1}^{\mu}\dee \mu \frac{1}{D_{\mu\mu}}
-\frac{1}{2}
\int_{-1}^{1}\dee \mu 
\frac{1-\mu}{D_{\mu\mu}}\Bigg)
\frac{\partial{}}{\partial{z}}
\frac{\partial{F}}{\partial{t}}(\mu+1),
\\
&&
Z_{10}(12)=
\frac{1}{p^2}\frac{\partial{}}{\partial{p}}
\Bigg[
\frac{1}{2}\int_{-1}^1\dee\mu
\frac{1-\mu^2}{2}
p^3\nabla\cdot \bm{V}
+\frac{1}{2}\int_{-1}^1\dee\mu
\frac{3\mu^2-1}{2}
p^3\frac{\partial{V_z}}
{\partial{z}}
\Bigg]
\nonumber\\
&&
\times
\Bigg(\int_{-1}^{\mu}\dee \mu \frac{1}{D_{\mu\mu}}
-\frac{1}{2}
\int_{-1}^{1}\dee \mu 
\frac{1-\mu}{D_{\mu\mu}}\Bigg)
V_z
\int_{-1}^{\mu}\dee\mu 
\Bigg(\int_{-1}^{\mu}\dee \mu \frac{1}{D_{\mu\mu}}
-\frac{1}{2}
\int_{-1}^{1}\dee \mu 
\frac{1-\mu}{D_{\mu\mu}}\Bigg)
\frac{\partial{}}{\partial{z}}
\int_{-1}^{\mu}\dee\mu
\frac{\partial{g}}{\partial{t}}.	
\end{eqnarray}
Using the anisotropic distribution
function, we can find  
\begin{eqnarray}
&&Z_{10}(1)=0, Z_{10}(2)=0, 
\nonumber\\
&&
Z_{10}(3)
=
\frac{1}{72D^2}
\frac{1}{p^2}\frac{\partial{}}{\partial{p}}p^3
\Bigg[
-
\left(\nabla\cdot \bm{V}\right)^2
+
3
\frac{\partial{V_z}}
{\partial{z}}
\left(\nabla\cdot \bm{V}\right)
\Bigg]
V_z\frac{\partial{F}}{\partial{z}},
\nonumber\\
&&
Z_{10}(4)
=
\frac{1}{72D^2}
\frac{1}{p^2}\frac{\partial{}}{\partial{p}}p^3
\Bigg[
-
\left(\nabla\cdot \bm{V}\right)
\left(
\frac{\partial{V_z}}{\partial{z}}
\right)
+
3
\left(
\frac{\partial{V_z}}{\partial{z}}
\right)^2
\Bigg]
V_z
\frac{\partial{F}}{\partial{z}}
\nonumber\\
&&
+
\frac{1}{72D^2}
\frac{1}{p^2}\frac{\partial{}}{\partial{p}}p^3
\Bigg[
-
\left(\nabla\cdot \bm{V}\right)
+
3
\frac{\partial{V_z}}
{\partial{z}}
\Bigg]
V_z^2
\frac{\partial^2{F}}
{\partial{z^2}},
\nonumber\\
&&
Z_{10}(5)=0, Z_{10}(6)=0, 
\nonumber\\
&&
Z_{10}(7)
=-\frac{1}{540D^2}
\frac{1}{p^2}\frac{\partial{}}{\partial{p}}p^3
\Bigg[
2
\left(\nabla\cdot \bm{V}\right)^2
-7
\left(\nabla\cdot \bm{V}\right)
\left(\frac{\partial{V_z}}
{\partial{z}}\right)
+3
\left(\frac{\partial{V_z}}
{\partial{z}}\right)^2
\Bigg]
V_z
\frac{1}{p^2}\frac{\partial{}}{\partial{p}}
\Bigg\{p^3
\frac{\partial{F}}{\partial{z}}
\Bigg\},
\nonumber\\
&&
Z_{10}(8)=0, 
\nonumber\\
&&
Z_{10}(9)
=
\frac{1}{540D^2}
V_z
\Bigg[
\left(\nabla\cdot \bm{V}\right)^2
-
6
\frac{\partial{V_z}}
{\partial{z}}
\left(\nabla\cdot \bm{V}\right)
+9
\left(\frac{\partial{V_z}}
{\partial{z}}\right)^2
\Bigg]
\frac{1}{p^2}\frac{\partial{}}{\partial{p}}p^3
\frac{\partial{F}}{\partial{z}},
\nonumber\\
&&
Z_{10}(10)=0, Z_{10}(11)=0, 
Z_{10}(12)=0, 
\end{eqnarray}
With the latter equations,
we obtain
\begin{eqnarray}
Z_{10}=&&\Bigg[-5
\left(\nabla\cdot \bm{V}\right)^2
+
21
\frac{\partial{V_z}}
{\partial{z}}
\left(\nabla\cdot \bm{V}\right)
\Bigg]V_z
\frac{1}{216D^2}
\frac{1}{p^2}\frac{\partial{}}{\partial{p}}p^3
\frac{\partial{F}}{\partial{z}}
\nonumber\\
&&
+
\Bigg[
-
\left(\nabla\cdot \bm{V}\right)
+
3
\frac{\partial{V_z}}
{\partial{z}}
\Bigg]V_z^2
\frac{1}{72D^2}
\frac{1}{p^2}\frac{\partial{}}{\partial{p}}p^3
\frac{\partial^2{F}}
{\partial{z^2}}
\nonumber\\
&&
+
\Bigg[
-2
\left(\nabla\cdot \bm{V}\right)^2
+7
\left(\nabla\cdot \bm{V}\right)
\left(\frac{\partial{V_z}}
{\partial{z}}\right)
-3
\left(\frac{\partial{V_z}}
{\partial{z}}\right)^2
\Bigg]
\frac{1}{p^2}
\frac{\partial{}}{\partial{p}}
p^4
\frac{\partial{}}{\partial{p}}
\frac{\partial{F}}{\partial{z}}.	
\end{eqnarray}
To continue, we find
\begin{eqnarray}
&&Z_{11}
=\frac{1}{p^2}\frac{\partial{}}{\partial{p}}
\left[
p\frac{\partial{}}{\partial{p}}
\left(p^3F\right)\right]
\frac{1}{90D}
\Bigg[
2
(\nabla\cdot \bm{V})^2
-7
\frac{\partial{V_z}}{\partial{z}}
\nabla\cdot \bm{V}
+3
\left(\frac{\partial{V_z}}
{\partial{z}}\right)^2
\Bigg],
\nonumber\\
&&
Z_{12}=0, 	
\nonumber\\
&&
Z_{13}=
\frac{1}{p^2}\frac{\partial{}}{\partial{p}}\left(p^3F\right)
\frac{1}{90D}\Bigg[
-
\left(\nabla\cdot \bm{V}\right)
+3
\frac{\partial{V_z}}
{\partial{z}}
\Bigg]
\left(\nabla\cdot \bm{V}-
3\frac{\partial{V_z}}{\partial{z}}
\right),
\nonumber\\
&&
Z_{14}=0, Z_{15}=0, Z_{16}=0.
\end{eqnarray}

\setcounter{equation}{0}  

\section{The formulas of $Y_1\sim Y_{12}$}
\label{Yn}

The formulas of $Y_n$ are shown as follows
\begin{eqnarray}
&&Y_1=-\frac{v}{2}\frac{\partial{}}
{\partial{z}}\int_{-1}^{1}\dee \mu\mu \Bigg(\int_{-1}^{\mu}\dee \mu \frac{1}{D_{\mu\mu}}
-\frac{1}{2}
\int_{-1}^{1}\dee \mu 
\frac{1-\mu}{D_{\mu\mu}}\Bigg)
v\frac{\mu^2-1}{2}
\frac{\partial{F}}{\partial{z}},
\\
&&
Y_2=-\frac{v}{2}\frac{\partial{}}
{\partial{z}}\int_{-1}^{1}\dee \mu\mu \Bigg(\int_{-1}^{\mu}\dee \mu \frac{1}{D_{\mu\mu}}
-\frac{1}{2}
\int_{-1}^{1}\dee \mu 
\frac{1-\mu}{D_{\mu\mu}}\Bigg)
v\frac{\partial{}}{\partial{z}}
\int_{-1}^{\mu}\dee\mu\mu g,
\\
&&
Y_3=-\frac{v}{2}\frac{\partial{}}
{\partial{z}}\int_{-1}^{1}\dee \mu\mu \Bigg(\int_{-1}^{\mu}\dee \mu \frac{1}{D_{\mu\mu}}
-\frac{1}{2}
\int_{-1}^{1}\dee \mu 
\frac{1-\mu}{D_{\mu\mu}}\Bigg)
F\left(\nabla\cdot \bm{V}\right)(\mu+1),
\\
&&
Y_4=-\frac{v}{2}\frac{\partial{}}
{\partial{z}}\int_{-1}^{1}\dee \mu\mu \Bigg(\int_{-1}^{\mu}\dee \mu \frac{1}{D_{\mu\mu}}
-\frac{1}{2}
\int_{-1}^{1}\dee \mu 
\frac{1-\mu}{D_{\mu\mu}}\Bigg)
V_z\frac{\partial{F}}{\partial{z}}
(\mu+1),
\\
&&
Y_5=-\frac{v}{2}\frac{\partial{}}
{\partial{z}}\int_{-1}^{1}\dee \mu\mu \Bigg(\int_{-1}^{\mu}\dee \mu \frac{1}{D_{\mu\mu}}
-\frac{1}{2}
\int_{-1}^{1}\dee \mu 
\frac{1-\mu}{D_{\mu\mu}}\Bigg)
\int_{-1}^{\mu}\dee\mu
g\left(\nabla\cdot \bm{V}\right),
\\
&&
Y_6=-\frac{v}{2}\frac{\partial{}}
{\partial{z}}\int_{-1}^{1}\dee \mu\mu \Bigg(\int_{-1}^{\mu}\dee \mu \frac{1}{D_{\mu\mu}}
-\frac{1}{2}
\int_{-1}^{1}\dee \mu 
\frac{1-\mu}{D_{\mu\mu}}\Bigg)
V_z\frac{\partial{}}{\partial{z}}
\int_{-1}^{\mu}\dee\mu g,
\\
&&
Y_7=\frac{v}{2}\frac{\partial{}}
{\partial{z}}\int_{-1}^{1}\dee \mu\mu \Bigg(\int_{-1}^{\mu}\dee \mu \frac{1}{D_{\mu\mu}}
-\frac{1}{2}
\int_{-1}^{1}\dee \mu 
\frac{1-\mu}{D_{\mu\mu}}\Bigg)
\frac{1}{p^2}\frac{\partial{}}{\partial{p}}
\Bigg\{p^3\Bigg[\frac{3\mu-\mu^3+2}{6}
\nabla\cdot \bm{V}
+\frac{\mu^3-\mu}{2}\frac{\partial{V_z}}
{\partial{z}}
\Bigg]F\Bigg\},
\\
&&
Y_8=\frac{v}{2}\frac{\partial{}}
{\partial{z}}\int_{-1}^{1}\dee \mu\mu \Bigg(\int_{-1}^{\mu}\dee \mu \frac{1}{D_{\mu\mu}}
-\frac{1}{2}
\int_{-1}^{1}\dee \mu 
\frac{1-\mu}{D_{\mu\mu}}\Bigg)
\frac{1}{p^2}\frac{\partial{}}{\partial{p}}
\Bigg\{p^3\int_{-1}^{\mu}\dee\mu
\Bigg[\frac{1-\mu^2}{2}
\nabla\cdot \bm{V}
+\frac{3\mu^2-1}{2}\frac{\partial{V_z}}
{\partial{z}}
\Bigg]g\Bigg\},
\\
&&
Y_9=-\frac{v}{2}\frac{\partial{}}
{\partial{z}}\int_{-1}^{1}\dee \mu\mu \Bigg(\int_{-1}^{\mu}\dee \mu \frac{1}{D_{\mu\mu}}
-\frac{1}{2}
\int_{-1}^{1}\dee \mu 
\frac{1-\mu}{D_{\mu\mu}}\Bigg)
\frac{1-\mu^2}{2}
\mu\left(\nabla\cdot \bm{V}-
3\frac{\partial{V_z}}{\partial{z}}
\right)F,
\\
&&
Y_{10}=-\frac{v}{2}\frac{\partial{}}
{\partial{z}}\int_{-1}^{1}\dee \mu\mu \Bigg(\int_{-1}^{\mu}\dee \mu \frac{1}{D_{\mu\mu}}
-\frac{1}{2}
\int_{-1}^{1}\dee \mu 
\frac{1-\mu}{D_{\mu\mu}}\Bigg)
\frac{1-\mu^2}{2}
\mu\left(\nabla\cdot \bm{V}-
3\frac{\partial{V_z}}{\partial{z}}
\right)g,
\\
&&
Y_{11}=-\frac{v}{2}\frac{\partial{}}
{\partial{z}}\int_{-1}^{1}\dee \mu\mu \Bigg(\int_{-1}^{\mu}\dee \mu \frac{1}{D_{\mu\mu}}
-\frac{1}{2}
\int_{-1}^{1}\dee \mu 
\frac{1-\mu}{D_{\mu\mu}}\Bigg)
\frac{\partial{F}}{\partial{t}}(\mu+1),
\\
&&
Y_{12}=-\frac{v}{2}\frac{\partial{}}
{\partial{z}}\int_{-1}^{1}\dee \mu\mu \Bigg(\int_{-1}^{\mu}\dee \mu \frac{1}{D_{\mu\mu}}
-\frac{1}{2}
\int_{-1}^{1}\dee \mu 
\frac{1-\mu}{D_{\mu\mu}}\Bigg)
\int_{-1}^{\mu}\dee\mu
\frac{\partial{g}}{\partial{t}}.	
\end{eqnarray}

\setcounter{equation}{0}  

\section{Evaluating $Y_1\sim Y_{12}$}
\label{Evaluating Yn}

Using the formula $D_{\mu\mu}(\mu)=
D(1-\mu^2)$, we can have
\begin{eqnarray}
&&Y_1
=\kappa_\parallel
\frac{\partial^2{F}}{\partial{z^2}}. 	
\end{eqnarray}
With anisotropic distribution function,
we find
\begin{eqnarray}
Y_2=\sum_{n=1}^{12}Y_2(n)	
\end{eqnarray}
with
\begin{eqnarray}
&&Y_2(1)=-\frac{v^2}{2}
\int_{-1}^{1}\dee \mu\mu \Bigg(\int_{-1}^{\mu}\dee \mu \frac{1}{D_{\mu\mu}}
-\frac{1}{2}
\int_{-1}^{1}\dee \mu 
\frac{1-\mu}{D_{\mu\mu}}\Bigg)
\int_{-1}^{\mu}\dee\mu\mu 
\Bigg(\int_{-1}^{\mu}\dee \mu \frac{1}{D_{\mu\mu}}
-\frac{1}{2}
\int_{-1}^{1}\dee \mu 
\frac{1-\mu}{D_{\mu\mu}}\Bigg)
\nonumber\\
&&
\times
\frac{\partial{}}{\partial{z}}
\frac{\partial{}}{\partial{z}}
F\left(\nabla\cdot \bm{V}\right)(\mu+1),
\\
&&
Y_2(2)=-\frac{v^2}{2}
\int_{-1}^{1}\dee \mu\mu \Bigg(\int_{-1}^{\mu}\dee \mu \frac{1}{D_{\mu\mu}}
-\frac{1}{2}
\int_{-1}^{1}\dee \mu 
\frac{1-\mu}{D_{\mu\mu}}\Bigg)
\int_{-1}^{\mu}\dee\mu\mu 
\Bigg(\int_{-1}^{\mu}\dee \mu \frac{1}{D_{\mu\mu}}
-\frac{1}{2}
\int_{-1}^{1}\dee \mu 
\frac{1-\mu}{D_{\mu\mu}}\Bigg)
\nonumber\\
&&
\times
\frac{\partial{}}{\partial{z}}
\frac{\partial{}}{\partial{z}}
\int_{-1}^{\mu}\dee\mu
g\left(\nabla\cdot \bm{V}\right),
\\
&&
Y_2(3)=\frac{v^2}{2}
\int_{-1}^{1}\dee \mu\mu \Bigg(\int_{-1}^{\mu}\dee \mu \frac{1}{D_{\mu\mu}}
-\frac{1}{2}
\int_{-1}^{1}\dee \mu 
\frac{1-\mu}{D_{\mu\mu}}\Bigg)
\int_{-1}^{\mu}\dee\mu\mu 
\Bigg(\int_{-1}^{\mu}\dee \mu \frac{1}{D_{\mu\mu}}
-\frac{1}{2}
\int_{-1}^{1}\dee \mu 
\frac{1-\mu}{D_{\mu\mu}}\Bigg)
\nonumber\\
&&
\times
\frac{\partial{}}{\partial{z}}
\frac{\partial{}}{\partial{z}}
\frac{1}{p^2}\frac{\partial{}}{\partial{p}}
\Bigg\{p^3\Bigg[\frac{3\mu-\mu^3+2}{6}
\nabla\cdot \bm{V}
+\frac{\mu^3-\mu}{2}\frac{\partial{V_z}}
{\partial{z}}
\Bigg]F\Bigg\},
\\
&&
Y_2(4)=\frac{v^2}{2}
\int_{-1}^{1}\dee \mu\mu \Bigg(\int_{-1}^{\mu}\dee \mu \frac{1}{D_{\mu\mu}}
-\frac{1}{2}
\int_{-1}^{1}\dee \mu 
\frac{1-\mu}{D_{\mu\mu}}\Bigg)
\int_{-1}^{\mu}\dee\mu\mu 
\Bigg(\int_{-1}^{\mu}\dee \mu \frac{1}{D_{\mu\mu}}
-\frac{1}{2}
\int_{-1}^{1}\dee \mu 
\frac{1-\mu}{D_{\mu\mu}}\Bigg)
\nonumber\\
&&
\times
\frac{\partial{}}{\partial{z}}
\frac{\partial{}}{\partial{z}}
\frac{1}{p^2}\frac{\partial{}}{\partial{p}}
\Bigg\{p^3\int_{-1}^{\mu}\dee\mu
\Bigg[\frac{1-\mu^2}{2}
\nabla\cdot \bm{V}
+\frac{3\mu^2-1}{2}\frac{\partial{V_z}}
{\partial{z}}
\Bigg]g\Bigg\},
\\
&&
Y_2(5)=-\frac{v^2}{2}
\int_{-1}^{1}\dee \mu\mu \Bigg(\int_{-1}^{\mu}\dee \mu \frac{1}{D_{\mu\mu}}
-\frac{1}{2}
\int_{-1}^{1}\dee \mu 
\frac{1-\mu}{D_{\mu\mu}}\Bigg)
\int_{-1}^{\mu}\dee\mu\mu 
\Bigg(\int_{-1}^{\mu}\dee \mu \frac{1}{D_{\mu\mu}}
-\frac{1}{2}
\int_{-1}^{1}\dee \mu 
\frac{1-\mu}{D_{\mu\mu}}\Bigg)
\nonumber\\
&&
\times
\frac{\partial{}}{\partial{z}}
\frac{\partial{}}{\partial{z}}
\frac{1-\mu^2}{2}
\mu\left(\nabla\cdot \bm{V}-
3\frac{\partial{V_z}}{\partial{z}}
\right)F,
\\
&&
Y_2(6)=-\frac{v^2}{2}
\int_{-1}^{1}\dee \mu\mu \Bigg(\int_{-1}^{\mu}\dee \mu \frac{1}{D_{\mu\mu}}
-\frac{1}{2}
\int_{-1}^{1}\dee \mu 
\frac{1-\mu}{D_{\mu\mu}}\Bigg)
\int_{-1}^{\mu}\dee\mu\mu 
\Bigg(\int_{-1}^{\mu}\dee \mu \frac{1}{D_{\mu\mu}}
-\frac{1}{2}
\int_{-1}^{1}\dee \mu 
\frac{1-\mu}{D_{\mu\mu}}\Bigg)
\nonumber\\
&&
\times
\frac{\partial{}}{\partial{z}}
\frac{\partial{}}{\partial{z}}
\frac{1-\mu^2}{2}
\mu\left(\nabla\cdot \bm{V}-
3\frac{\partial{V_z}}{\partial{z}}
\right)g,
\\
&&
Y_2(7)=-\frac{v^2}{2}
\int_{-1}^{1}\dee \mu\mu \Bigg(\int_{-1}^{\mu}\dee \mu \frac{1}{D_{\mu\mu}}
-\frac{1}{2}
\int_{-1}^{1}\dee \mu 
\frac{1-\mu}{D_{\mu\mu}}\Bigg)
\int_{-1}^{\mu}\dee\mu\mu 
\Bigg(\int_{-1}^{\mu}\dee \mu \frac{1}{D_{\mu\mu}}
-\frac{1}{2}
\int_{-1}^{1}\dee \mu 
\frac{1-\mu}{D_{\mu\mu}}\Bigg)
\nonumber\\
&&
\times
\frac{\partial{}}{\partial{z}}
\frac{\partial{}}{\partial{z}}
\frac{\partial{F}}{\partial{t}}
(\mu+1),
\\
&&
Y_2(8)=-\frac{v^2}{2}
\int_{-1}^{1}\dee \mu\mu \Bigg(\int_{-1}^{\mu}\dee \mu \frac{1}{D_{\mu\mu}}
-\frac{1}{2}
\int_{-1}^{1}\dee \mu 
\frac{1-\mu}{D_{\mu\mu}}\Bigg)
\int_{-1}^{\mu}\dee\mu\mu 
\Bigg(\int_{-1}^{\mu}\dee \mu \frac{1}{D_{\mu\mu}}
-\frac{1}{2}
\int_{-1}^{1}\dee \mu 
\frac{1-\mu}{D_{\mu\mu}}\Bigg)
\nonumber\\
&&
\times
\frac{\partial{}}{\partial{z}}
\frac{\partial{}}{\partial{z}}
\int_{-1}^{\mu}\dee\mu
\frac{\partial{g}}{\partial{t}}.	
\end{eqnarray}
After performing 
straightforward algebraic 
operation, we have
\begin{eqnarray}
&&Y_2(1)
=
-\left(\nabla\cdot \bm{V}\right)
v^2
\frac{7}{36D^2}
\frac{\partial^2{F}}{\partial{z^2}},
\nonumber\\
&&
Y_2(2)=0, 	
\nonumber\\
&&	
Y_2(3)=
(\nabla\cdot \bm{V})
\frac{v^2}{12D^2}
\frac{1}{p^2}\frac{\partial{}}{\partial{p}}
p^3
\frac{\partial^2{F}}{\partial{z^2}},
\nonumber\\
&&	
Y_2(4)=0, 
\nonumber\\
&&
Y_2(5)
=\frac{v^2}{90D^2}
\left(\nabla\cdot \bm{V}-
3\frac{\partial{V_z}}{\partial{z}}
\right)\frac{\partial{}}{\partial{z}}
\frac{\partial{F}}{\partial{z}},
\nonumber\\
&&
Y_2(6)=0, Y_2(7)=0, Y_2(8)=0. 
\end{eqnarray}
Using the latter formulas gives
\begin{eqnarray}
Y_2&&=Y_2(1)+\cdots+Y_2(8)
\nonumber\\
&&
=-
\frac{v^2}{30D^2}
\left(11\nabla\cdot \bm{V}+
2\frac{\partial{V_z}}{\partial{z}}
\right)
\frac{\partial^2{F}}{\partial{z^2}}
+(\nabla\cdot \bm{V})
\frac{v^2}{12D^2}
\frac{1}{p^2}\frac{\partial{}}{\partial{p}}
p^3
\frac{\partial^2{F}}{\partial{z^2}}.	
\end{eqnarray}
To proceed, we have 
\begin{eqnarray}
&&Y_3	
=-
\left(\nabla\cdot \bm{V}\right)
\frac{v}{2D}
\frac{\partial{F}}
{\partial{z}},
\nonumber\\
&&
Y_4=
-
\frac{\partial{V_z}}{\partial{z}}
\frac{v}{2D}
\frac{\partial{F}}{\partial{z}}
-V_z\frac{v}{2D}
\frac{\partial^2{F}}{\partial{z^2}}. 	
\end{eqnarray}
In addition, we can find
\begin{eqnarray}
Y_5
=\sum_{n=1}^{12}Y_5(n)	
\end{eqnarray}
with
\begin{eqnarray}
&&Y_5(1)=-\left(\nabla\cdot \bm{V}\right)
\frac{v}{2}
\int_{-1}^{1}\dee \mu\mu \Bigg(\int_{-1}^{\mu}\dee \mu \frac{1}{D_{\mu\mu}}
-\frac{1}{2}
\int_{-1}^{1}\dee \mu 
\frac{1-\mu}{D_{\mu\mu}}\Bigg)
\int_{-1}^{\mu}\dee\mu
\Bigg(\int_{-1}^{\mu}\dee \mu \frac{1}{D_{\mu\mu}}
-\frac{1}{2}
\int_{-1}^{1}\dee \mu 
\frac{1-\mu}{D_{\mu\mu}}\Bigg)
\nonumber\\
&&
\times
\frac{\partial{}}{\partial{z}}
v\frac{\mu^2-1}{2}
\frac{\partial{F}}{\partial{z}},
\\
&&
Y_5(2)=-\left(\nabla\cdot \bm{V}\right)
\frac{v}{2}
\int_{-1}^{1}\dee \mu\mu \Bigg(\int_{-1}^{\mu}\dee \mu \frac{1}{D_{\mu\mu}}
-\frac{1}{2}
\int_{-1}^{1}\dee \mu 
\frac{1-\mu}{D_{\mu\mu}}\Bigg)
\int_{-1}^{\mu}\dee\mu
\Bigg(\int_{-1}^{\mu}\dee \mu \frac{1}{D_{\mu\mu}}
-\frac{1}{2}
\int_{-1}^{1}\dee \mu 
\frac{1-\mu}{D_{\mu\mu}}\Bigg)
\nonumber\\
&&
\times
\frac{\partial{}}{\partial{z}}
v\frac{\partial{}}{\partial{z}}
\int_{-1}^{\mu}\dee\mu\mu g,
\\
&&
Y_5(3)=-\left(\nabla\cdot \bm{V}\right)
\frac{v}{2}
\int_{-1}^{1}\dee \mu\mu \Bigg(\int_{-1}^{\mu}\dee \mu \frac{1}{D_{\mu\mu}}
-\frac{1}{2}
\int_{-1}^{1}\dee \mu 
\frac{1-\mu}{D_{\mu\mu}}\Bigg)
\int_{-1}^{\mu}\dee\mu
\Bigg(\int_{-1}^{\mu}\dee \mu \frac{1}{D_{\mu\mu}}
-\frac{1}{2}
\int_{-1}^{1}\dee \mu 
\frac{1-\mu}{D_{\mu\mu}}\Bigg)
\nonumber\\
&&
\times
\frac{\partial{}}{\partial{z}}
F\left(\nabla\cdot \bm{V}\right)(\mu+1),
\\
&&
Y_5(4)=-\left(\nabla\cdot \bm{V}\right)
\frac{v}{2}
\int_{-1}^{1}\dee \mu\mu \Bigg(\int_{-1}^{\mu}\dee \mu \frac{1}{D_{\mu\mu}}
-\frac{1}{2}
\int_{-1}^{1}\dee \mu 
\frac{1-\mu}{D_{\mu\mu}}\Bigg)
\int_{-1}^{\mu}\dee\mu
\Bigg(\int_{-1}^{\mu}\dee \mu \frac{1}{D_{\mu\mu}}
-\frac{1}{2}
\int_{-1}^{1}\dee \mu 
\frac{1-\mu}{D_{\mu\mu}}\Bigg)
\nonumber\\
&&
\times
\frac{\partial{}}{\partial{z}}
V_z\frac{\partial{F}}{\partial{z}}
(\mu+1),
\\
&&
Y_5(5)=-\left(\nabla\cdot \bm{V}\right)
\frac{v}{2}
\int_{-1}^{1}\dee \mu\mu \Bigg(\int_{-1}^{\mu}\dee \mu \frac{1}{D_{\mu\mu}}
-\frac{1}{2}
\int_{-1}^{1}\dee \mu 
\frac{1-\mu}{D_{\mu\mu}}\Bigg)
\int_{-1}^{\mu}\dee\mu
\Bigg(\int_{-1}^{\mu}\dee \mu \frac{1}{D_{\mu\mu}}
-\frac{1}{2}
\int_{-1}^{1}\dee \mu 
\frac{1-\mu}{D_{\mu\mu}}\Bigg)
\nonumber\\
&&
\times
\frac{\partial{}}{\partial{z}}
\int_{-1}^{\mu}\dee\mu
g\left(\nabla\cdot \bm{V}\right),
\\
&&
Y_5(6)=-\left(\nabla\cdot \bm{V}\right)
\frac{v}{2}
\int_{-1}^{1}\dee \mu\mu \Bigg(\int_{-1}^{\mu}\dee \mu \frac{1}{D_{\mu\mu}}
-\frac{1}{2}
\int_{-1}^{1}\dee \mu 
\frac{1-\mu}{D_{\mu\mu}}\Bigg)
\int_{-1}^{\mu}\dee\mu
\Bigg(\int_{-1}^{\mu}\dee \mu \frac{1}{D_{\mu\mu}}
-\frac{1}{2}
\int_{-1}^{1}\dee \mu 
\frac{1-\mu}{D_{\mu\mu}}\Bigg)
\nonumber\\
&&
\times
\frac{\partial{}}{\partial{z}}
V_z\frac{\partial{}}{\partial{z}}
\int_{-1}^{\mu}\dee\mu g,
\\
&&
Y_5(7)=\left(\nabla\cdot \bm{V}\right)
\frac{v}{2}
\int_{-1}^{1}\dee \mu\mu \Bigg(\int_{-1}^{\mu}\dee \mu \frac{1}{D_{\mu\mu}}
-\frac{1}{2}
\int_{-1}^{1}\dee \mu 
\frac{1-\mu}{D_{\mu\mu}}\Bigg)
\int_{-1}^{\mu}\dee\mu
\Bigg(\int_{-1}^{\mu}\dee \mu \frac{1}{D_{\mu\mu}}
-\frac{1}{2}
\int_{-1}^{1}\dee \mu 
\frac{1-\mu}{D_{\mu\mu}}\Bigg)
\nonumber\\
&&
\times
\frac{\partial{}}{\partial{z}}
\frac{1}{p^2}\frac{\partial{}}{\partial{p}}
\Bigg\{p^3\Bigg[\frac{3\mu-\mu^3+2}{6}
\nabla\cdot \bm{V}
+\frac{\mu^3-\mu}{2}\frac{\partial{V_z}}
{\partial{z}}
\Bigg]F\Bigg\},
\\
&&
Y_5(8)=\left(\nabla\cdot \bm{V}\right)
\frac{v}{2}
\int_{-1}^{1}\dee \mu\mu \Bigg(\int_{-1}^{\mu}\dee \mu \frac{1}{D_{\mu\mu}}
-\frac{1}{2}
\int_{-1}^{1}\dee \mu 
\frac{1-\mu}{D_{\mu\mu}}\Bigg)
\int_{-1}^{\mu}\dee\mu
\Bigg(\int_{-1}^{\mu}\dee \mu \frac{1}{D_{\mu\mu}}
-\frac{1}{2}
\int_{-1}^{1}\dee \mu 
\frac{1-\mu}{D_{\mu\mu}}\Bigg)
\nonumber\\
&&
\times
\frac{\partial{}}{\partial{z}}
\frac{1}{p^2}\frac{\partial{}}{\partial{p}}
\Bigg\{p^3\int_{-1}^{\mu}\dee\mu
\Bigg[\frac{1-\mu^2}{2}
\nabla\cdot \bm{V}
+\frac{3\mu^2-1}{2}\frac{\partial{V_z}}
{\partial{z}}
\Bigg]g\Bigg\},
\\
&&
Y_5(9)=-\left(\nabla\cdot \bm{V}\right)
\frac{v}{2}
\int_{-1}^{1}\dee \mu\mu \Bigg(\int_{-1}^{\mu}\dee \mu \frac{1}{D_{\mu\mu}}
-\frac{1}{2}
\int_{-1}^{1}\dee \mu 
\frac{1-\mu}{D_{\mu\mu}}\Bigg)
\int_{-1}^{\mu}\dee\mu
\Bigg(\int_{-1}^{\mu}\dee \mu \frac{1}{D_{\mu\mu}}
-\frac{1}{2}
\int_{-1}^{1}\dee \mu 
\frac{1-\mu}{D_{\mu\mu}}\Bigg)
\nonumber\\
&&
\times
\frac{\partial{}}{\partial{z}}
\frac{1-\mu^2}{2}
\mu\left(\nabla\cdot \bm{V}-
3\frac{\partial{V_z}}{\partial{z}}
\right)F,
\\
&&
Y_5(10)=-\left(\nabla\cdot \bm{V}\right)
\frac{v}{2}
\int_{-1}^{1}\dee \mu\mu \Bigg(\int_{-1}^{\mu}\dee \mu \frac{1}{D_{\mu\mu}}
-\frac{1}{2}
\int_{-1}^{1}\dee \mu 
\frac{1-\mu}{D_{\mu\mu}}\Bigg)
\int_{-1}^{\mu}\dee\mu
\Bigg(\int_{-1}^{\mu}\dee \mu \frac{1}{D_{\mu\mu}}
-\frac{1}{2}
\int_{-1}^{1}\dee \mu 
\frac{1-\mu}{D_{\mu\mu}}\Bigg)
\nonumber\\
&&
\times
\frac{\partial{}}{\partial{z}}
\frac{1-\mu^2}{2}
\mu\left(\nabla\cdot \bm{V}-
3\frac{\partial{V_z}}{\partial{z}}
\right)g,
\\
&&
Y_5(11)=-\left(\nabla\cdot \bm{V}\right)
\frac{v}{2}
\int_{-1}^{1}\dee \mu\mu \Bigg(\int_{-1}^{\mu}\dee \mu \frac{1}{D_{\mu\mu}}
-\frac{1}{2}
\int_{-1}^{1}\dee \mu 
\frac{1-\mu}{D_{\mu\mu}}\Bigg)
\int_{-1}^{\mu}\dee\mu
\Bigg(\int_{-1}^{\mu}\dee \mu \frac{1}{D_{\mu\mu}}
-\frac{1}{2}
\int_{-1}^{1}\dee \mu 
\frac{1-\mu}{D_{\mu\mu}}\Bigg)
\nonumber\\
&&
\times
\frac{\partial{}}{\partial{z}}
\frac{\partial{F}}{\partial{t}}(\mu+1),
\\
&&
Y_5(12)=-\left(\nabla\cdot \bm{V}\right)
\frac{v}{2}
\int_{-1}^{1}\dee \mu\mu \Bigg(\int_{-1}^{\mu}\dee \mu \frac{1}{D_{\mu\mu}}
-\frac{1}{2}
\int_{-1}^{1}\dee \mu 
\frac{1-\mu}{D_{\mu\mu}}\Bigg)
\int_{-1}^{\mu}\dee\mu
\Bigg(\int_{-1}^{\mu}\dee \mu \frac{1}{D_{\mu\mu}}
-\frac{1}{2}
\int_{-1}^{1}\dee \mu 
\frac{1-\mu}{D_{\mu\mu}}\Bigg)
\nonumber\\
&&
\times
\frac{\partial{}}{\partial{z}}
\int_{-1}^{\mu}\dee\mu
\frac{\partial{g}}{\partial{t}}.	
\end{eqnarray}
Considering the anisotropic distribution
function, we can obtain
\begin{eqnarray}
&&Y_5(1)
=-\left(\nabla\cdot \bm{V}\right)
\frac{v^2}{12D^2}
\frac{\partial^2{F}}{\partial{z^2}},
Y_5(2)=0, 
\nonumber\\
&&	
Y_5(3)=\left(\nabla\cdot \bm{V}\right)^2
\frac{v}{4D^2}
\frac{\partial{F}}{\partial{z}},
\nonumber\\
&&
Y_5(4)
=\left(\nabla\cdot \bm{V}\right)
\frac{\partial{V_z}}{\partial{z}}
\frac{v}{4D^2}
\frac{\partial{F}}{\partial{z}}
+V_z\left(\nabla\cdot \bm{V}\right)
\frac{v}{4D^2}
\frac{\partial^2{F}}{\partial{z^2}},
\nonumber\\
&&
Y_5(5)=0, Y_5(6)=0, 
\nonumber\\
&&
Y_5(7)=-\left(\nabla\cdot \bm{V}\right)^2
\frac{v}{12D^2}
\frac{1}{p^2}\frac{\partial{}}{\partial{p}}
\Bigg\{p^3\frac{\partial{F}}{\partial{z}}
\Bigg\},
\nonumber\\
&&
Y_5(8)=0, Y_5(9)=0, Y_5(10)=0, 
Y_5(11)=0, Y_5(12)=0.  
\end{eqnarray}
Combining the latter formulas gives
\begin{eqnarray}
Y_5&&
=Y_5(1)+\cdots+Y_5(12)
\nonumber\\
&&
=-\left(\nabla\cdot \bm{V}\right)
\frac{v^2}{12D^2}
\frac{\partial^2{F}}{\partial{z^2}}
+V_z\left(\nabla\cdot \bm{V}\right)
\frac{v}{4D^2}
\frac{\partial^2{F}}{\partial{z^2}}
\nonumber\\
&&
+
\Bigg[
\left(\nabla\cdot \bm{V}\right)^2
+\left(\nabla\cdot \bm{V}\right)
\frac{\partial{V_z}}{\partial{z}}
\Bigg]
\frac{v}{4D^2}
\frac{\partial{F}}{\partial{z}}
-\left(\nabla\cdot \bm{V}\right)^2
\frac{v}{12D^2}
\frac{1}{p^2}\frac{\partial{}}{\partial{p}}
p^3\frac{\partial{F}}{\partial{z}}.	
\end{eqnarray}
To continue, we find
\begin{eqnarray}
Y_6
=\sum_{n=1}^{12}Y_6(n)	
\end{eqnarray}
with
\begin{eqnarray}
	&&Y_6(1)=\frac{v}{8D^2}
	\left(
	\frac{\partial{V_z}}{\partial{z}}
	\frac{\partial{}}{\partial{z}}
	+
	V_z
	\frac{\partial^2{}}{\partial{z^2}}
	\right)
	\int_{-1}^{1}\dee \mu
	v\frac{\mu^2-1}{2}
	\frac{\partial{F}}{\partial{z}},
	\nonumber\\
	&&
	Y_6(2)=\frac{v}{8D^2}
	\left(
	\frac{\partial{V_z}}{\partial{z}}
	\frac{\partial{}}{\partial{z}}
	+
	V_z
	\frac{\partial^2{}}{\partial{z^2}}
	\right)
	\int_{-1}^{1}\dee \mu
	v\frac{\partial{}}{\partial{z}}
	\int_{-1}^{\mu}\dee\mu\mu g,
	\nonumber\\
	&&
	Y_6(3)=\frac{v}{8D^2}
	\left(
	\frac{\partial{V_z}}{\partial{z}}
	\frac{\partial{}}{\partial{z}}
	+
	V_z
	\frac{\partial^2{}}{\partial{z^2}}
	\right)
	\int_{-1}^{1}\dee \mu
	F\left(\nabla\cdot \bm{V}\right)(\mu+1),
	\nonumber\\
	&&
	Y_6(4)=\frac{v}{8D^2}
	\left(
	\frac{\partial{V_z}}{\partial{z}}
	\frac{\partial{}}{\partial{z}}
	+
	V_z
	\frac{\partial^2{}}{\partial{z^2}}
	\right)
	\int_{-1}^{1}\dee \mu
	V_z\frac{\partial{F}}{\partial{z}}
	(\mu+1),
	\nonumber\\
	&&
	Y_6(5)=\frac{v}{8D^2}
	\left(
	\frac{\partial{V_z}}{\partial{z}}
	\frac{\partial{}}{\partial{z}}
	+
	V_z
	\frac{\partial^2{}}{\partial{z^2}}
	\right)
	\int_{-1}^{1}\dee \mu
	\int_{-1}^{\mu}\dee\mu
	g\left(\nabla\cdot \bm{V}\right),
	\nonumber\\
	&&
	Y_6(6)=\frac{v}{8D^2}
	\left(
	\frac{\partial{V_z}}{\partial{z}}
	\frac{\partial{}}{\partial{z}}
	+
	V_z
	\frac{\partial^2{}}{\partial{z^2}}
	\right)
	\int_{-1}^{1}\dee \mu
	V_z\frac{\partial{}}{\partial{z}}
	\int_{-1}^{\mu}\dee\mu g,
	\nonumber\\
	&&
	Y_6(7)=-\frac{v}{8D^2}
	\left(
	\frac{\partial{V_z}}{\partial{z}}
	\frac{\partial{}}{\partial{z}}
	+
	V_z
	\frac{\partial^2{}}{\partial{z^2}}
	\right)
	\int_{-1}^{1}\dee \mu
	\frac{1}{p^2}\frac{\partial{}}{\partial{p}}
	\Bigg\{p^3\Bigg[\frac{3\mu-\mu^3+2}{6}
	\nabla\cdot \bm{V}
	+\frac{\mu^3-\mu}{2}\frac{\partial{V_z}}
	{\partial{z}}
	\Bigg]F\Bigg\},
	\nonumber\\
	&&
	Y_6(8)=-\frac{v}{8D^2}
	\left(
	\frac{\partial{V_z}}{\partial{z}}
	\frac{\partial{}}{\partial{z}}
	+
	V_z
	\frac{\partial^2{}}{\partial{z^2}}
	\right)
	\int_{-1}^{1}\dee \mu
	\frac{1}{p^2}\frac{\partial{}}{\partial{p}}
	\Bigg\{p^3\int_{-1}^{\mu}\dee\mu
	\Bigg[\frac{1-\mu^2}{2}
	\nabla\cdot \bm{V}
	+\frac{3\mu^2-1}{2}\frac{\partial{V_z}}
	{\partial{z}}
	\Bigg]g\Bigg\},
	\nonumber\\
	&&
	Y_6(9)=\frac{v}{8D^2}
	\left(
	\frac{\partial{V_z}}{\partial{z}}
	\frac{\partial{}}{\partial{z}}
	+
	V_z
	\frac{\partial^2{}}{\partial{z^2}}
	\right)
	\int_{-1}^{1}\dee \mu
	\frac{1-\mu^2}{2}
	\mu\left(\nabla\cdot \bm{V}-
	3\frac{\partial{V_z}}{\partial{z}}
	\right)F,
	\nonumber\\
	&&
	Y_6(10)=\frac{v}{8D^2}
	\left(
	\frac{\partial{V_z}}{\partial{z}}
	\frac{\partial{}}{\partial{z}}
	+
	V_z
	\frac{\partial^2{}}{\partial{z^2}}
	\right)
	\int_{-1}^{1}\dee \mu
	\frac{1-\mu^2}{2}
	\mu\left(\nabla\cdot \bm{V}-
	3\frac{\partial{V_z}}{\partial{z}}
	\right)g,
	\nonumber\\
	&&
	Y_6(11)=\frac{v}{8D^2}
	\left(
	\frac{\partial{V_z}}{\partial{z}}
	\frac{\partial{}}{\partial{z}}
	+
	V_z
	\frac{\partial^2{}}{\partial{z^2}}
	\right)
	\int_{-1}^{1}\dee \mu
	\frac{\partial{F}}{\partial{t}}
	(\mu+1),
	\nonumber\\
	&&
	Y_6(12)=
	\frac{v}{8D^2}
	\left(
	\frac{\partial{V_z}}{\partial{z}}
	\frac{\partial{}}{\partial{z}}
	+
	V_z
	\frac{\partial^2{}}{\partial{z^2}}
	\right)
	\int_{-1}^{1}\dee \mu
	\int_{-1}^{\mu}\dee\mu
	\frac{\partial{g}}{\partial{t}}.
\end{eqnarray}
Using the anisotropic distribution function
yields
\begin{eqnarray}
&&Y_6(1)
=
-
\frac{\partial{V_z}}{\partial{z}}
\frac{v^2}{12D^2}
\frac{\partial^2{F}}{\partial{z^2}},
Y_6(2)=0, 	
\nonumber\\
&&
Y_6(3)=\frac{v}{4D^2}
\frac{\partial{V_z}}{\partial{z}}
\left(\nabla\cdot \bm{V}\right)
\frac{\partial{F}}{\partial{z}}
+
\frac{v}{4D^2}
V_z\left(\nabla\cdot \bm{V}\right)
\frac{\partial^2{F}}{\partial{z^2}},
\nonumber\\
&&
Y_6(4)=
\frac{v}{4D^2}
\left(
\frac{\partial{V_z}}{\partial{z}}
\right)^2
\frac{\partial{F}}{\partial{z}}
+
\frac{3v}{4D^2}
V_z
\frac{\partial{V_z}}{\partial{z}}
\frac{\partial^2{F}}{\partial{z^2}},
\nonumber\\
&&
Y_6(5)=0, Y_6(6)=0, 
\nonumber\\
&&
Y_6(7)=
-\frac{v}{12D^2}
\left(\nabla\cdot \bm{V}\right)
\frac{\partial{V_z}}{\partial{z}}
\frac{1}{p^2}
\frac{\partial{}}{\partial{p}}
\Bigg(
p^3
\frac{\partial{F}}{\partial{z}}
\Bigg)
-\frac{v}{12D^2}V_z
\left(\nabla\cdot \bm{V}\right)
\frac{1}{p^2}
\frac{\partial{}}{\partial{p}}
\Bigg(
p^3
\frac{\partial^2{F}}{\partial{z^2}}
\Bigg),
\nonumber\\
&&
Y_6(8)=0, Y_6(9)=0, Y_6(10)=0,
Y_6(11)=0, Y_6(12)=0. 
\end{eqnarray}
With the latter formulas, we find
\begin{eqnarray}
Y_6&&=Y_6(1)+\cdots+Y_6(12)
\nonumber\\
&&
=
-
\frac{\partial{V_z}}{\partial{z}}
\frac{v^2}{12D^2}
\frac{\partial^2{F}}{\partial{z^2}}
\nonumber\\
&&
+V_z\Bigg[
\frac{v}{4D^2}
\left(\nabla\cdot \bm{V}\right)
+
\frac{3v}{4D^2}
\frac{\partial{V_z}}{\partial{z}}
\Bigg]
\frac{\partial^2{F}}{\partial{z^2}}
\nonumber\\
&&
+\frac{v}{4D^2}
\Bigg[
\frac{\partial{V_z}}{\partial{z}}
\left(\nabla\cdot \bm{V}\right)
+
\left(
\frac{\partial{V_z}}{\partial{z}}
\right)^2
\Bigg]
\frac{\partial{F}}{\partial{z}}
\nonumber\\
&&
-\frac{v}{12D^2}
\left(\nabla\cdot \bm{V}\right)
\frac{\partial{V_z}}{\partial{z}}
\frac{1}{p^2}
\frac{\partial{}}{\partial{p}}
\Bigg(
p^3
\frac{\partial{F}}{\partial{z}}
\Bigg)
\nonumber\\
&&
-\frac{v}{12D^2}V_z
\left(\nabla\cdot \bm{V}\right)
\frac{1}{p^2}
\frac{\partial{}}{\partial{p}}
\Bigg(
p^3
\frac{\partial^2{F}}{\partial{z^2}}
\Bigg)	
\end{eqnarray}
To proceed, we have
\begin{eqnarray}
Y_7
=\frac{v}{6D}
(\nabla\cdot \bm{V})
\frac{1}{p^2}\frac{\partial{}}{\partial{p}}p^3
\frac{\partial{F}}
{\partial{z}}, 
\end{eqnarray}
and 
\begin{eqnarray}
Y_8
=\sum_{n=1}^{12}Y_8(n)	
\end{eqnarray}
with
\begin{eqnarray}
	&&Y_8(1)=\frac{v}{4D}
	\frac{1}{p^2}
	\frac{\partial{}}{\partial{p}}
	\Bigg\{p^3
	\Bigg[
	\int_{-1}^{1}\dee \mu(1-\mu)
	\frac{1-\mu^2}{2}
	\nabla\cdot \bm{V}
	+
	\int_{-1}^{1}\dee \mu(1-\mu)
	\frac{3\mu^2-1}{2}\frac{\partial{V_z}}
	{\partial{z}}
	\Bigg]
	\nonumber\\
	&&
	\times
	\Bigg(\int_{-1}^{\mu}\dee \mu \frac{1}{D_{\mu\mu}}
	-\frac{1}{2}
	\int_{-1}^{1}\dee \mu 
	\frac{1-\mu}{D_{\mu\mu}}\Bigg)
	\frac{\partial{}}{\partial{z}}
	v\frac{\mu^2-1}{2}
	\frac{\partial{F}}{\partial{z}},
	\\
	&&
	Y_8(2)=
	\frac{v}{4D}
	\frac{1}{p^2}
	\frac{\partial{}}{\partial{p}}
	\Bigg\{p^3
	\Bigg[
	\int_{-1}^{1}\dee \mu(1-\mu)
	\frac{1-\mu^2}{2}
	\nabla\cdot \bm{V}
	+
	\int_{-1}^{1}\dee \mu(1-\mu)
	\frac{3\mu^2-1}{2}\frac{\partial{V_z}}
	{\partial{z}}
	\Bigg]
	\nonumber\\
	&&
	\times
	\Bigg(\int_{-1}^{\mu}\dee \mu \frac{1}{D_{\mu\mu}}
	-\frac{1}{2}
	\int_{-1}^{1}\dee \mu 
	\frac{1-\mu}{D_{\mu\mu}}\Bigg)
	\frac{\partial{}}{\partial{z}}
	v\frac{\partial{}}{\partial{z}}
	\int_{-1}^{\mu}\dee\mu\mu g,
	\\
	&&
	Y_8(3)=
	\frac{v}{4D}
	\frac{1}{p^2}
	\frac{\partial{}}{\partial{p}}
	\Bigg\{p^3
	\Bigg[
	\int_{-1}^{1}\dee \mu(1-\mu)
	\frac{1-\mu^2}{2}
	\nabla\cdot \bm{V}
	+
	\int_{-1}^{1}\dee \mu(1-\mu)
	\frac{3\mu^2-1}{2}\frac{\partial{V_z}}
	{\partial{z}}
	\Bigg]
	\nonumber\\
	&&
	\times
	\Bigg(\int_{-1}^{\mu}\dee \mu \frac{1}{D_{\mu\mu}}
	-\frac{1}{2}
	\int_{-1}^{1}\dee \mu 
	\frac{1-\mu}{D_{\mu\mu}}\Bigg)
	\frac{\partial{}}{\partial{z}}
	F\left(\nabla\cdot \bm{V}\right)(\mu+1),
	\\
	&&
	Y_8(4)=
	\frac{v}{4D}
	\frac{1}{p^2}
	\frac{\partial{}}{\partial{p}}
	\Bigg\{p^3
	\Bigg[
	\int_{-1}^{1}\dee \mu(1-\mu)
	\frac{1-\mu^2}{2}
	\nabla\cdot \bm{V}
	+
	\int_{-1}^{1}\dee \mu(1-\mu)
	\frac{3\mu^2-1}{2}\frac{\partial{V_z}}
	{\partial{z}}
	\Bigg]
	\nonumber\\
	&&
	\times
	\Bigg(\int_{-1}^{\mu}\dee \mu \frac{1}{D_{\mu\mu}}
	-\frac{1}{2}
	\int_{-1}^{1}\dee \mu 
	\frac{1-\mu}{D_{\mu\mu}}\Bigg)
	\frac{\partial{}}{\partial{z}}
	V_z\frac{\partial{F}}{\partial{z}}
	(\mu+1),
	\\
	&&
	Y_8(5)=
	\frac{v}{4D}
	\frac{1}{p^2}
	\frac{\partial{}}{\partial{p}}
	\Bigg\{p^3
	\Bigg[
	\int_{-1}^{1}\dee \mu(1-\mu)
	\frac{1-\mu^2}{2}
	\nabla\cdot \bm{V}
	+
	\int_{-1}^{1}\dee \mu(1-\mu)
	\frac{3\mu^2-1}{2}\frac{\partial{V_z}}
	{\partial{z}}
	\Bigg]
	\nonumber\\
	&&
	\times
	\Bigg(\int_{-1}^{\mu}\dee \mu \frac{1}{D_{\mu\mu}}
	-\frac{1}{2}
	\int_{-1}^{1}\dee \mu 
	\frac{1-\mu}{D_{\mu\mu}}\Bigg)
	\frac{\partial{}}{\partial{z}}
	\int_{-1}^{\mu}\dee\mu
	g\left(\nabla\cdot \bm{V}\right),
	\\
	&&
	Y_8(6)=
	\frac{v}{4D}
	\frac{1}{p^2}
	\frac{\partial{}}{\partial{p}}
	\Bigg\{p^3
	\Bigg[
	\int_{-1}^{1}\dee \mu(1-\mu)
	\frac{1-\mu^2}{2}
	\nabla\cdot \bm{V}
	+
	\int_{-1}^{1}\dee \mu(1-\mu)
	\frac{3\mu^2-1}{2}\frac{\partial{V_z}}
	{\partial{z}}
	\Bigg]
	\nonumber\\
	&&
	\times
	\Bigg(\int_{-1}^{\mu}\dee \mu \frac{1}{D_{\mu\mu}}
	-\frac{1}{2}
	\int_{-1}^{1}\dee \mu 
	\frac{1-\mu}{D_{\mu\mu}}\Bigg)
	\frac{\partial{}}{\partial{z}}
	V_z\frac{\partial{}}{\partial{z}}
	\int_{-1}^{\mu}\dee\mu g,
\\
	&&
	Y_8(7)=-
	\frac{v}{4D}
	\frac{1}{p^2}
	\frac{\partial{}}{\partial{p}}
	\Bigg\{p^3
	\Bigg[
	\int_{-1}^{1}\dee \mu(1-\mu)
	\frac{1-\mu^2}{2}
	\nabla\cdot \bm{V}
	+
	\int_{-1}^{1}\dee \mu(1-\mu)
	\frac{3\mu^2-1}{2}\frac{\partial{V_z}}
	{\partial{z}}
	\Bigg]
	\nonumber\\
	&&
	\times
	\Bigg(\int_{-1}^{\mu}\dee \mu \frac{1}{D_{\mu\mu}}
	-\frac{1}{2}
	\int_{-1}^{1}\dee \mu 
	\frac{1-\mu}{D_{\mu\mu}}\Bigg)
	\frac{\partial{}}{\partial{z}}
	\frac{1}{p^2}\frac{\partial{}}{\partial{p}}
	\Bigg\{p^3\Bigg[\frac{3\mu-\mu^3+2}{6}
	\nabla\cdot \bm{V}
	+\frac{\mu^3-\mu}{2}\frac{\partial{V_z}}
	{\partial{z}}
	\Bigg]F\Bigg\},
	\\
	&&
	Y_8(8)=-
	\frac{v}{4D}
	\frac{1}{p^2}
	\frac{\partial{}}{\partial{p}}
	\Bigg\{p^3
	\Bigg[
	\int_{-1}^{1}\dee \mu(1-\mu)
	\frac{1-\mu^2}{2}
	\nabla\cdot \bm{V}
	+
	\int_{-1}^{1}\dee \mu(1-\mu)
	\frac{3\mu^2-1}{2}\frac{\partial{V_z}}
	{\partial{z}}
	\Bigg]
	\nonumber\\
	&&
	\times
	\Bigg(\int_{-1}^{\mu}\dee \mu \frac{1}{D_{\mu\mu}}
	-\frac{1}{2}
	\int_{-1}^{1}\dee \mu 
	\frac{1-\mu}{D_{\mu\mu}}\Bigg)
	\frac{\partial{}}{\partial{z}}
	\frac{1}{p^2}\frac{\partial{}}{\partial{p}}
	\Bigg\{p^3\int_{-1}^{\mu}\dee\mu
	\Bigg[\frac{1-\mu^2}{2}
	\nabla\cdot \bm{V}
	+\frac{3\mu^2-1}{2}\frac{\partial{V_z}}
	{\partial{z}}
	\Bigg]g\Bigg\},
	\\
	&&
	Y_8(9)=
	\frac{v}{4D}
	\frac{1}{p^2}
	\frac{\partial{}}{\partial{p}}
	\Bigg\{p^3
	\Bigg[
	\int_{-1}^{1}\dee \mu(1-\mu)
	\frac{1-\mu^2}{2}
	\nabla\cdot \bm{V}
	+
	\int_{-1}^{1}\dee \mu(1-\mu)
	\frac{3\mu^2-1}{2}\frac{\partial{V_z}}
	{\partial{z}}
	\Bigg]
	\nonumber\\
	&&
	\times
	\Bigg(\int_{-1}^{\mu}\dee \mu \frac{1}{D_{\mu\mu}}
	-\frac{1}{2}
	\int_{-1}^{1}\dee \mu 
	\frac{1-\mu}{D_{\mu\mu}}\Bigg)
	\frac{\partial{}}{\partial{z}}
	\frac{1-\mu^2}{2}
	\mu\left(\nabla\cdot \bm{V}-
	3\frac{\partial{V_z}}{\partial{z}}
	\right)F,
	\\
	&&
	Y_8(10)=
	\frac{v}{4D}
	\frac{1}{p^2}
	\frac{\partial{}}{\partial{p}}
	\Bigg\{p^3
	\Bigg[
	\int_{-1}^{1}\dee \mu(1-\mu)
	\frac{1-\mu^2}{2}
	\nabla\cdot \bm{V}
	+
	\int_{-1}^{1}\dee \mu(1-\mu)
	\frac{3\mu^2-1}{2}\frac{\partial{V_z}}
	{\partial{z}}
	\Bigg]
	\nonumber\\
	&&
	\times
	\Bigg(\int_{-1}^{\mu}\dee \mu \frac{1}{D_{\mu\mu}}
	-\frac{1}{2}
	\int_{-1}^{1}\dee \mu 
	\frac{1-\mu}{D_{\mu\mu}}\Bigg)
	\frac{\partial{}}{\partial{z}}
	\frac{1-\mu^2}{2}
	\mu\left(\nabla\cdot \bm{V}-
	3\frac{\partial{V_z}}{\partial{z}}
	\right)g,
	\\
	&&
	Y_8(11)=
	\frac{v}{4D}
	\frac{1}{p^2}
	\frac{\partial{}}{\partial{p}}
	\Bigg\{p^3
	\Bigg[
	\int_{-1}^{1}\dee \mu(1-\mu)
	\frac{1-\mu^2}{2}
	\nabla\cdot \bm{V}
	+
	\int_{-1}^{1}\dee \mu(1-\mu)
	\frac{3\mu^2-1}{2}\frac{\partial{V_z}}
	{\partial{z}}
	\Bigg]
	\nonumber\\
	&&
	\times
	\Bigg(\int_{-1}^{\mu}\dee \mu \frac{1}{D_{\mu\mu}}
	-\frac{1}{2}
	\int_{-1}^{1}\dee \mu 
	\frac{1-\mu}{D_{\mu\mu}}\Bigg)
	\frac{\partial{}}{\partial{z}}
	\frac{\partial{F}}{\partial{t}}
	(\mu+1),
	\\
	&&
	Y_8(12)=
	\frac{v}{4D}
	\frac{1}{p^2}
	\frac{\partial{}}{\partial{p}}
	\Bigg\{p^3
	\Bigg[
	\int_{-1}^{1}\dee \mu(1-\mu)
	\frac{1-\mu^2}{2}
	\nabla\cdot \bm{V}
	+
	\int_{-1}^{1}\dee \mu(1-\mu)
	\frac{3\mu^2-1}{2}\frac{\partial{V_z}}
	{\partial{z}}
	\Bigg]
	\nonumber\\
	&&
	\times
	\Bigg(\int_{-1}^{\mu}\dee \mu \frac{1}{D_{\mu\mu}}
	-\frac{1}{2}
	\int_{-1}^{1}\dee \mu 
	\frac{1-\mu}{D_{\mu\mu}}\Bigg)
	\frac{\partial{}}{\partial{z}}
	\int_{-1}^{\mu}\dee\mu
	\frac{\partial{g}}{\partial{t}}
	\Bigg\}.
\end{eqnarray}
Using anisotropic distribution function,
we have
\begin{eqnarray}
&&Y_8(1)
=\frac{v^2}{60D^2}
\Bigg[
\nabla\cdot \bm{V}
+2
\frac{\partial{V_z}}
{\partial{z}}
\Bigg]
\frac{1}{p^2}
\frac{\partial{}}{\partial{p}}
\Bigg\{p^3
\frac{\partial^2{F}}{\partial{z^2}}
\Bigg\}, Y_8(2)=0, 
\nonumber\\
&&
Y_8(3)
=
-
\Bigg[
5
\left(\nabla\cdot \bm{V}\right)^2
+
3
\frac{\partial{V_z}}
{\partial{z}}
\left(\nabla\cdot \bm{V}\right)
\Bigg]
\frac{v}{72D^2}
\frac{1}{p^2}
\frac{\partial{}}{\partial{p}}
\Bigg\{p^3
\frac{\partial{F}}{\partial{z}}
\Bigg\},
\nonumber\\
&&
Y_8(4)
=
-
\Bigg[
5
\left(\nabla\cdot \bm{V}\right)
\frac{\partial{V_z}}{\partial{z}}
+
3
\left(\frac{\partial{V_z}}
{\partial{z}}\right)^2
\Bigg]
\frac{v}{72D^2}
\frac{1}{p^2}
\frac{\partial{}}{\partial{p}}
\Bigg\{p^3
\frac{\partial{F}}{\partial{z}}
\Bigg\}
\nonumber\\
&&
-
\Bigg[
5V_z
\left(\nabla\cdot \bm{V}\right)
+
3V_z
\frac{\partial{V_z}}
{\partial{z}}
\Bigg]
\frac{v}{72D^2}
\frac{1}{p^2}
\frac{\partial{}}{\partial{p}}
\Bigg\{p^3
\frac{\partial^2{F}}{\partial{z^2}}
\Bigg\},
\nonumber\\
&&	
Y_8(5)=0, Y_8(6)=0, 
\nonumber\\
&&	
Y_8(7)
=
\Bigg[
9
\left(\nabla\cdot \bm{V}\right)^2
+
\nabla\cdot \bm{V}
\left(\frac{\partial{V_z}}
{\partial{z}}\right)
+6
\left(\frac{\partial{V_z}}
{\partial{z}}\right)^2
\Bigg]
\frac{v}{120D^2}
\frac{1}{p^2}
\frac{\partial{}}{\partial{p}}
p^3
\frac{\partial{F}}{\partial{z}}
\nonumber\\
&&
+
\Bigg[
9
\left(\nabla\cdot \bm{V}\right)^2
+
\nabla\cdot \bm{V}
\left(\frac{\partial{V_z}}
{\partial{z}}\right)
+6
\left(\frac{\partial{V_z}}
{\partial{z}}\right)^2
\Bigg]
\frac{v}{360D^2}
\frac{1}{p^2}
\frac{\partial{}}{\partial{p}}
p^4
\frac{\partial{}}{\partial{p}}
\frac{\partial{F}}{\partial{z}},
\nonumber\\
&&
Y_8(8)=0, 
\nonumber\\
&&
Y_8(9)
=
\Bigg[
-\frac{4}{15}
\left(\nabla\cdot \bm{V}\right)^2
+\frac{24}{15}
\nabla\cdot \bm{V}
\frac{\partial{V_z}}{\partial{z}}
-
\frac{12}{5}
\left(
\frac{\partial{V_z}}{\partial{z}}
\right)^2
\Bigg]
\frac{v}{48D^2}
\frac{1}{p^2}
\frac{\partial{}}{\partial{p}}
\Bigg\{p^3
\frac{\partial{F}}{\partial{z}}
\Bigg\},
\nonumber\\
&&
Y_8(10)=0, Y_8(11)=0, Y_8(12)=0.
\end{eqnarray}
With the latter formulas, we obtain
\begin{eqnarray}
Y_8&&=Y_8(1)+\cdots+Y_8(12)
	\nonumber\\
	&&=+\frac{v^2}{60D^2}
	\Bigg[
	\nabla\cdot \bm{V}
	+2
	\frac{\partial{V_z}}
	{\partial{z}}
	\Bigg]
	\frac{1}{p^2}
	\frac{\partial{}}{\partial{p}}
	\Bigg\{p^3
	\frac{\partial^2{F}}{\partial{z^2}}
	\Bigg\}
	\nonumber\\
	&&
	-
	\Bigg[
	5V_z
	\left(\nabla\cdot \bm{V}\right)
	+
	3V_z
	\frac{\partial{V_z}}
	{\partial{z}}
	\Bigg]
	\frac{v}{72D^2}
	\frac{1}{p^2}
	\frac{\partial{}}{\partial{p}}
	\Bigg\{p^3
	\frac{\partial^2{F}}{\partial{z^2}}
	\Bigg\}
	\nonumber\\
	&&
	+
	\Bigg[
	-
	5
	\frac{\partial{V_z}}
	{\partial{z}}
	\left(\nabla\cdot \bm{V}\right)
	-
	3
	\left(
	\frac{\partial{V_z}}{\partial{z}}
	\right)^2
	\Bigg]
	\frac{v}{72D^2}
	\frac{1}{p^2}
	\frac{\partial{}}{\partial{p}}
	p^3
	\frac{\partial{F}}{\partial{z}}
	\nonumber\\
	&&
	+
	\Bigg[
	9
	\left(\nabla\cdot \bm{V}\right)^2
	+
	\nabla\cdot \bm{V}
	\left(\frac{\partial{V_z}}
	{\partial{z}}\right)
	+6
	\left(\frac{\partial{V_z}}
	{\partial{z}}\right)^2
	\Bigg]
	\frac{v}{360D^2}
	\frac{1}{p^2}
	\frac{\partial{}}{\partial{p}}
	p^4
	\frac{\partial{}}{\partial{p}}
	\frac{\partial{F}}{\partial{z}}.
\end{eqnarray}
In addition, we can find
\begin{eqnarray}
Y_9=0. 
\end{eqnarray}
To proceed, we have
\begin{eqnarray}
	Y_{10}
	=\sum_{n=1}^{12}Y_{10}(n)
\end{eqnarray}
with
\begin{eqnarray}
	&&Y_{10}(1)=-
	\left(\nabla\cdot \bm{V}-
	3\frac{\partial{V_z}}{\partial{z}}
	\right)
	\frac{v}{8D}
	\int_{-1}^{1}\dee \mu
	\left(\mu-\mu^3\right)
	\int_{-1}^{\mu}\dee \mu \frac{1}{D_{\mu\mu}}
	\frac{\partial{}}
	{\partial{z}}
	v\frac{\mu^2-1}{2}
	\frac{\partial{F}}{\partial{z}},
	\\
	&&Y_{10}(2)=
	-
	\left(\nabla\cdot \bm{V}-
	3\frac{\partial{V_z}}{\partial{z}}
	\right)
	\frac{v}{8D}
	\int_{-1}^{1}\dee \mu
	\left(\mu-\mu^3\right)
	\int_{-1}^{\mu}\dee \mu \frac{1}{D_{\mu\mu}}
	\frac{\partial{}}
	{\partial{z}}
	v\frac{\partial{}}{\partial{z}}
	\int_{-1}^{\mu}\dee\mu\mu g,
	\\
	&&
	Y_{10}(3)=-
	\left(\nabla\cdot \bm{V}-
	3\frac{\partial{V_z}}{\partial{z}}
	\right)
	\frac{v}{8D}
	\int_{-1}^{1}\dee \mu
	\left(\mu-\mu^3\right)
	\int_{-1}^{\mu}\dee \mu \frac{1}{D_{\mu\mu}}
	\frac{\partial{}}
	{\partial{z}}
	F\left(\nabla\cdot \bm{V}\right)(\mu+1),
	\\
	&&
	Y_{10}(4)=-
	\left(\nabla\cdot \bm{V}-
	3\frac{\partial{V_z}}{\partial{z}}
	\right)
	\frac{v}{8D}
	\int_{-1}^{1}\dee \mu
	\left(\mu-\mu^3\right)
	\int_{-1}^{\mu}\dee \mu \frac{1}{D_{\mu\mu}}
	\frac{\partial{}}
	{\partial{z}}
	V_z\frac{\partial{F}}{\partial{z}}
	(\mu+1),
	\\
	&&
	Y_{10}(5)=-
	\left(\nabla\cdot \bm{V}-
	3\frac{\partial{V_z}}{\partial{z}}
	\right)
	\frac{v}{8D}
	\int_{-1}^{1}\dee \mu
	\left(\mu-\mu^3\right)
	\int_{-1}^{\mu}\dee \mu \frac{1}{D_{\mu\mu}}
	\frac{\partial{}}
	{\partial{z}}
	\int_{-1}^{\mu}\dee\mu
	g\left(\nabla\cdot \bm{V}\right),
	\\
	&&
	Y_{10}(6)=-
	\left(\nabla\cdot \bm{V}-
	3\frac{\partial{V_z}}{\partial{z}}
	\right)
	\frac{v}{8D}
	\int_{-1}^{1}\dee \mu
	\left(\mu-\mu^3\right)
	\int_{-1}^{\mu}\dee \mu \frac{1}{D_{\mu\mu}}
	\frac{\partial{}}
	{\partial{z}}
	V_z\frac{\partial{}}{\partial{z}}
	\int_{-1}^{\mu}\dee\mu g,
	\\
	&&
	Y_{10}(7)=
	\left(\nabla\cdot \bm{V}-
	3\frac{\partial{V_z}}{\partial{z}}
	\right)
	\frac{v}{8D}
	\int_{-1}^{1}\dee \mu
	\left(\mu-\mu^3\right)
	\int_{-1}^{\mu}\dee \mu \frac{1}{D_{\mu\mu}}
	\frac{\partial{}}
	{\partial{z}}
	\frac{1}{p^2}\frac{\partial{}}{\partial{p}}
	\Bigg\{p^3\Bigg[\frac{3\mu-\mu^3+2}{6}
	\nabla\cdot \bm{V}
	+\frac{\mu^3-\mu}{2}\frac{\partial{V_z}}
	{\partial{z}}
	\Bigg]F\Bigg\},
	\\
	&&
	Y_{10}(8)=
	\left(\nabla\cdot \bm{V}-
	3\frac{\partial{V_z}}{\partial{z}}
	\right)
	\frac{v}{8D}
	\int_{-1}^{1}\dee \mu
	\left(\mu-\mu^3\right)
	\int_{-1}^{\mu}\dee \mu \frac{1}{D_{\mu\mu}}
	\frac{\partial{}}
	{\partial{z}}
	\frac{1}{p^2}\frac{\partial{}}{\partial{p}}
	\Bigg\{p^3\int_{-1}^{\mu}\dee\mu
	\Bigg[\frac{1-\mu^2}{2}
	\nabla\cdot \bm{V}
	+\frac{3\mu^2-1}{2}\frac{\partial{V_z}}
	{\partial{z}}
	\Bigg]g\Bigg\},
	\\
	&&
	Y_{10}(9)=-
	\left(\nabla\cdot \bm{V}-
	3\frac{\partial{V_z}}{\partial{z}}
	\right)
	\frac{v}{8D}
	\int_{-1}^{1}\dee \mu
	\left(\mu-\mu^3\right)
	\int_{-1}^{\mu}\dee \mu \frac{1}{D_{\mu\mu}}
	\frac{\partial{}}
	{\partial{z}}
	\frac{1-\mu^2}{2}
	\mu\left(\nabla\cdot \bm{V}-
	3\frac{\partial{V_z}}{\partial{z}}
	\right)F,
	\\
	&&
	Y_{10}(10)=-
	\left(\nabla\cdot \bm{V}-
	3\frac{\partial{V_z}}{\partial{z}}
	\right)
	\frac{v}{8D}
	\int_{-1}^{1}\dee \mu
	\left(\mu-\mu^3\right)
	\int_{-1}^{\mu}\dee \mu \frac{1}{D_{\mu\mu}}
	\frac{\partial{}}
	{\partial{z}}
	\frac{1-\mu^2}{2}
	\mu\left(\nabla\cdot \bm{V}-
	3\frac{\partial{V_z}}{\partial{z}}
	\right)g,
	\\
	&&
	Y_{10}(11)=-
	\left(\nabla\cdot \bm{V}-
	3\frac{\partial{V_z}}{\partial{z}}
	\right)
	\frac{v}{8D}
	\int_{-1}^{1}\dee \mu
	\left(\mu-\mu^3\right)
	\int_{-1}^{\mu}\dee \mu \frac{1}{D_{\mu\mu}}
	\frac{\partial{}}
	{\partial{z}}
	\frac{\partial{F}}{\partial{t}}
	(\mu+1),
	\\
	&&
	Y_{10}(12)=-
	\left(\nabla\cdot \bm{V}-
	3\frac{\partial{V_z}}{\partial{z}}
	\right)
	\frac{v}{8D}
	\int_{-1}^{1}\dee \mu
	\left(\mu-\mu^3\right)
	\int_{-1}^{\mu}\dee \mu \frac{1}{D_{\mu\mu}}
	\frac{\partial{}}
	{\partial{z}}
	\int_{-1}^{\mu}\dee\mu
	\frac{\partial{g}}{\partial{t}}.
\end{eqnarray}
Evaluating the latter formulas yields 
\begin{eqnarray}
&&Y_{10}(1)
=
\left(\nabla\cdot \bm{V}-
3\frac{\partial{V_z}}{\partial{z}}
\right)
\frac{v^2}{60D^2}
\frac{\partial^2{F}}{\partial{z^2}},
Y_{10}(2)=0, 
\nonumber\\
&&
Y_{10}(3)
=
\left(
-\left(\nabla\cdot \bm{V}\right)^2
+
3\frac{\partial{V_z}}{\partial{z}}
\left(\nabla\cdot \bm{V}\right)
\right)
\frac{v}{24D^2}
\frac{\partial{F}}{\partial{z}},
\nonumber\\
&&
Y_{10}(4)
=
\left[-\nabla\cdot \bm{V}
\frac{\partial{V_z}}
{\partial{z}}
+3
\left(
\frac{\partial{V_z}}{\partial{z}}
\right)^2
\right]
\frac{v}{24D^2}
\frac{\partial{F}}{\partial{z}}
+
V_z\left(-\nabla\cdot \bm{V}+
3
\frac{\partial{V_z}}{\partial{z}}
\right)
\frac{v}{24D^2}
\frac{\partial^2{F}}
{\partial{z^2}},
\nonumber\\
&&	
Y_{10}(5)=0, Y_{10}(6)=0, 
\nonumber\\
&&
Y_{10}(7)
=
\left(
\left(\nabla\cdot \bm{V}\right)^2
-
3\frac{\partial{V_z}}{\partial{z}}
\left(\nabla\cdot \bm{V}\right)
\right)
\frac{v}{72D^2}
\frac{1}{p^2}
\frac{\partial{}}{\partial{p}}
\Bigg\{p^3
\frac{\partial{F}}{\partial{z}}
\Bigg\}, 
\nonumber\\
&&
Y_{10}(8)=0, Y_{10}(9)=0, Y_{10}(10)=0,
Y_{10}(11)=0, Y_{10}(12)=0. 
\end{eqnarray}
With the latter equations, we can obtain
\begin{eqnarray}
	Y_{10}&&=Y_{10}(1)+\cdots+Y_{10}(12)
	\nonumber\\
	&&=
		\left(\nabla\cdot \bm{V}-
		3\frac{\partial{V_z}}{\partial{z}}
		\right)
		\frac{v^2}{60D}
		\frac{\partial^2{F}}{\partial{z^2}}
	\nonumber\\
	&&
	+
	V_z\left(-\nabla\cdot \bm{V}+
	3
	\frac{\partial{V_z}}{\partial{z}}
	\right)
	\frac{v}{24D^2}
	\frac{\partial^2{F}}
	{\partial{z^2}}
	\nonumber\\
	&&
	+\left(
	-\left(\nabla\cdot \bm{V}\right)^2
	+
	2\frac{\partial{V_z}}{\partial{z}}
	\left(\nabla\cdot \bm{V}\right)
	+3
	\left(
	\frac{\partial{V_z}}{\partial{z}}
	\right)^2
	\right)
	\frac{v}{24D^2}
	\frac{\partial{F}}{\partial{z}}
	\nonumber\\
	&&
	+\left(
	\left(\nabla\cdot \bm{V}\right)^2
	-
	3\frac{\partial{V_z}}{\partial{z}}
	\left(\nabla\cdot \bm{V}\right)
	\right)
	\frac{v}{72D^2}
	\frac{1}{p^2}
	\frac{\partial{}}{\partial{p}}
	\Bigg\{p^3
	\frac{\partial{F}}{\partial{z}}
	\Bigg\}. 
\end{eqnarray}
To continue, we have
\begin{eqnarray}
Y_{11}=0, Y_{12}=0.  
\end{eqnarray}

By setting $V\ll v$, $vT_{\mu}\ll L$,
the convection and diffusion effects
produced by background flow can be ignored.
Thus, the formulas of $Y_1\sim Y_{12}$
are simplified as
\begin{eqnarray}
	&&Y_1=\frac{v^2}{6D}
	\frac{\partial^2{F}}
	{\partial{z^2}}
	=\kappa_{\parallel}
	\frac{\partial^2{F}}
	{\partial{z^2}}
	\nonumber\\
	&&
	Y_2=\frac{v^2}{270D^2}
	\Bigg(
	73
	\nabla\cdot \bm{V}
	-39
	\frac{\partial{V_z}}
	{\partial{z}}
	\Bigg)
	\frac{1}{p^2}\frac{\partial{}}{\partial{p}}
	\left(p^3
	\frac{\partial^2{F}}
	{\partial{z^2}}\right)
	-\frac{\nabla\cdot\bm{V}}{3}
	v^2
	\frac{7}{36D^2}
	\frac{1}{p^2}
	\frac{\partial{}}{\partial{p}}
	\left(
	p^3
	\frac{\partial^2{F}}
	{\partial{z^2}}
	\right)
	\nonumber\\
	&&
	Y_4=
	-\left(\nabla\cdot \bm{V}\right)^2
	\frac{v}{12D^2}
	\frac{1}{p^2}\frac{\partial{}}{\partial{p}}
	\Bigg\{p^3\frac{\partial{F}}{\partial{z}}
	\Bigg\}
	\nonumber\\
	&&
	Y_5=
	-\frac{v}{12D^2}
	\left(\nabla\cdot \bm{V}\right)
	\frac{\partial{V_z}}{\partial{z}}
	\frac{1}{p^2}
	\frac{\partial{}}{\partial{p}}
	\Bigg(
	p^3
	\frac{\partial{F}}{\partial{z}}
	\Bigg)
	-\frac{v}{12D^2}V_z
	\left(\nabla\cdot \bm{V}\right)
	\frac{1}{p^2}
	\frac{\partial{}}{\partial{p}}
	\Bigg(
	p^3
	\frac{\partial^2{F}}{\partial{z^2}}
	\Bigg)
	\nonumber\\
	&&
	Y_6=\frac{v^2}{60D^2}
	\Bigg[
	\nabla\cdot \bm{V}
	+2
	\frac{\partial{V_z}}
	{\partial{z}}
	\Bigg]
	\frac{1}{p^2}
	\frac{\partial{}}{\partial{p}}
	\Bigg\{p^3
	\frac{\partial^2{F}}{\partial{z^2}}
	\Bigg\}
	\nonumber\\
	&&
	-
	\Bigg[
	5
	\left(\nabla\cdot \bm{V}\right)^2
	+
	3
	\frac{\partial{V_z}}
	{\partial{z}}
	\left(\nabla\cdot \bm{V}\right)
	\Bigg]
	\frac{v}{72D^2}
	\frac{1}{p^2}
	\frac{\partial{}}{\partial{p}}
	\Bigg\{p^3
	\frac{\partial{F}}{\partial{z}}
	\Bigg\}
	\nonumber\\
	&&
	-
	\Bigg[
	5
	\left(\nabla\cdot \bm{V}\right)
	\frac{\partial{V_z}}{\partial{z}}
	+
	3
	\left(\frac{\partial{V_z}}
	{\partial{z}}\right)^2
	\Bigg]
	\frac{v}{72D^2}
	\frac{1}{p^2}
	\frac{\partial{}}{\partial{p}}
	\Bigg\{p^3
	\frac{\partial{F}}{\partial{z}}
	\Bigg\}
	\nonumber\\
	&&
	-
	\Bigg[
	5V_z
	\left(\nabla\cdot \bm{V}\right)
	+
	3V_z
	\frac{\partial{V_z}}
	{\partial{z}}
	\Bigg]
	\frac{v}{72D^2}
	\frac{1}{p^2}
	\frac{\partial{}}{\partial{p}}
	\Bigg\{p^3
	\frac{\partial^2{F}}{\partial{z^2}}
	\Bigg\}
	\nonumber\\
	&&
	+\Bigg[
	9
	\left(\nabla\cdot \bm{V}\right)^2
	+
	\nabla\cdot \bm{V}
	\left(\frac{\partial{V_z}}
	{\partial{z}}\right)
	+6
	\left(\frac{\partial{V_z}}
	{\partial{z}}\right)^2
	\Bigg]
	\frac{v}{360D^2}
	\frac{1}{p^2}
	\frac{\partial{}}{\partial{p}}
	\Bigg\{p\frac{\partial{}}{\partial{p}}
	\Bigg(p^3
	\frac{\partial{F}}{\partial{z}}
	\Bigg)
	\Bigg\}
	\nonumber\\
	&&
	+\Bigg[
	-\frac{4}{15}
	\left(\nabla\cdot \bm{V}\right)^2
	+\frac{24}{15}
	\nabla\cdot \bm{V}
	\frac{\partial{V_z}}{\partial{z}}
	-
	\frac{12}{5}
	\left(
	\frac{\partial{V_z}}{\partial{z}}
	\right)^2
	\Bigg]
	\frac{v}{48D^2}
	\frac{1}{p^2}
	\frac{\partial{}}{\partial{p}}
	\Bigg\{p^3
	\frac{\partial{F}}{\partial{z}}
	\Bigg\}
	\nonumber\\
	&&
	Y_7=\left(
	\left(\nabla\cdot \bm{V}\right)^2
	-
	3\frac{\partial{V_z}}{\partial{z}}
	\left(\nabla\cdot \bm{V}\right)
	\right)
	\frac{v}{72D^2}
	\frac{1}{p^2}
	\frac{\partial{}}{\partial{p}}
	\Bigg\{p^3
	\frac{\partial{F}}{\partial{z}}
	\Bigg\}
	\nonumber\\
&&
Y_{10}=
\left(\nabla\cdot \bm{V}-
3\frac{\partial{V_z}}{\partial{z}}
\right)
\frac{v^2}{60D}
\frac{\partial^2{F}}{\partial{z^2}}
\nonumber\\
&&
+
V_z\left(-\nabla\cdot \bm{V}+
3
\frac{\partial{V_z}}{\partial{z}}
\right)
\frac{v}{24D^2}
\frac{\partial^2{F}}
{\partial{z^2}}
\nonumber\\
&&
+\left(
-\left(\nabla\cdot \bm{V}\right)^2
+
2\frac{\partial{V_z}}{\partial{z}}
\left(\nabla\cdot \bm{V}\right)
+3
\left(
\frac{\partial{V_z}}{\partial{z}}
\right)^2
\right)
\frac{v}{24D^2}
\frac{\partial{F}}{\partial{z}}
\nonumber\\
&&
+\left(
\left(\nabla\cdot \bm{V}\right)^2
-
3\frac{\partial{V_z}}{\partial{z}}
\left(\nabla\cdot \bm{V}\right)
\right)
\frac{v}{72D^2}
\frac{1}{p^2}
\frac{\partial{}}{\partial{p}}
\Bigg\{p^3
\frac{\partial{F}}{\partial{z}}
\Bigg\},
\nonumber\\
	&&
	Y_8=0,
	Y_9=0, 
	Y_{11}=0, Y_{12}=0.  
\end{eqnarray}

\setcounter{equation}{0}  

\section{Conversion formulas between the kinetic energy differential flux index and the momentum
spectrum index
of energetic particles}
\label{Conversion formulas}

The formula for
kinetic energy 
differential flux 
of energetic particles is 
\begin{eqnarray}
j(\varepsilon)=p^2F(p),
\label{j and F(p)}
\end{eqnarray}
which $\varepsilon$ is kinetic energy
of energetic particles,
$F(p)$ is momentum distribution 
function, 
$j(\varepsilon)$ is kinetic energy
differential flux,
and $p$ is momentum.
In the context of 
relativism, the 
formula for kinetic energy 
differential flux is shown as
follows
\begin{eqnarray}
j(\varepsilon)=A \varepsilon^{-\sigma}.
\label{energy spectrum}
\end{eqnarray}
Here, $A$ is a constant,
and $\sigma$ is the spectral 
index of kinetic energy differential flux
spectrum.
The kinetic energy equation
of energetic particles
is
\begin{eqnarray}
\varepsilon=\left(\gamma-1\right)m_0c^2
\label{kinetic energy}
\end{eqnarray} 
with the relativistic factor
\begin{eqnarray}
\gamma=\frac{1}{\sqrt{1-v^2/c^2}}.
\label{relativistic factor}
\end{eqnarray}
Here, $m_0$ is the 
rest mass, and $c$ is light speed.
Combining Equations 
(\ref{j and F(p)}),
(\ref{energy spectrum}), and (\ref{kinetic energy})
yields
\begin{eqnarray}
F(p)=A \left(\frac{m_0 v}{\sqrt{1-v^2/c^2}} 
\right)^{-2} 
\left[\left(
\frac{1}{\sqrt{1-v^2/c^2}}-1
\right)m_0c^2
\right]^{-\sigma}=g(v),
\label{g(v)}
\end{eqnarray}
which $g(v)$ speed is distribution 
function of energetic particles.
Here, the formula of
particle momentum is used 
\begin{eqnarray}
p=\gamma m_0 v.
\label{p with gamma}
\end{eqnarray}
Taking the logarithm of Equation
(\ref{g(v)}) gives
\begin{eqnarray}
\ln g(v)=\ln A 
-2\left[\ln m_0+\ln v-\frac{1}{2}\ln \left(1-\frac{v^2}{c^2}\right)\right]
-\sigma
\left[\ln m_0c^2
+\ln\left(
\frac{1}{\sqrt{1-v^2/c^2}}-1
\right)
\right].
\label{ln g(v)}
\end{eqnarray}
By setting $\ln v=x$, or
\begin{eqnarray}
v=e^x, 
\label{v and x}
\end{eqnarray}
Equation (\ref{ln g(v)}) becomes
\begin{eqnarray}
\ln g(x)=\ln A 
-2\left[\ln m_0+x-\frac{1}{2}\ln \left(1-\frac{e^{2x}}{c^2}\right)\right]
-\sigma 
\left[\ln m_0c^2
+\ln\left(
\frac{1}{\sqrt{1-e^{2x}/c^2}}-1
\right)
\right].
\label{ln g(v) with x}
\end{eqnarray}
The spectral index of speed 
spectrum
can be obatained from the latter 
equation 
\begin{eqnarray}
\frac{\dee \ln g(x)}{\dee x}
=-\omega
=-2\frac{1}{1-\frac{e^{2x}}{c^2}}
-\sigma 
\frac{1}{\frac{1}{\sqrt{1-e^{2x}/c^2}}
-1}
\frac{1}
{\left(1-e^{2x}/c^2\right)^{3/2}}
\frac{e^{2x}}{c^2}.
\end{eqnarray}
The latter equation can be rwritten as
\begin{eqnarray}
\omega
=\frac{2}{1-R}
+\sigma 
\frac{1}{\frac{1}{\sqrt{1-R}}-1}
\frac{1}
{\left(1-R\right)^{3/2}}R
\label{omega and delta with R}
\end{eqnarray}
with the parameter
\begin{eqnarray}
R=\frac{v^2}{c^2}.
\label{R}
\end{eqnarray}
Equation 
(\ref{omega and delta with R})
is the conversion formula between 
kinetic energy flux index $\sigma$
and the speed spectrum index $\omega$. 

In the following, we derive the
conversion formula between 
momentum spectrum index 
$\delta$
and speed spectrum index $\omega$.
We assume that 
energetic particles follow the momentum
spectrum
\begin{eqnarray}
F(p)=B p^{-\delta},
\end{eqnarray}
where $B$ is a constant, and $\delta$
is the spectral index of momentum
power law.   
Inserting Equation 
(\ref{p with gamma}) gives
\begin{eqnarray}
F(p)=B \left(\frac{m_0 v}
{\sqrt{1-v^2/c^2}} 
\right)^{-\delta}=g(v).
\end{eqnarray}
After taking logrithm of 
the latter equation, we have
\begin{eqnarray}
\ln g(v)&&=\ln B -\delta
\left[
\ln m_0+\ln v
-\frac{1}{2}
\ln \left(1-\frac{v^2}{c^2}
\right)
\right]
\nonumber\\
&&=\ln B -\delta 
\left[
\ln m_0+x
-\frac{1}{2}
\ln \left(1-\frac{e^{2x}}{c^2}
\right)
\right]
\end{eqnarray}
with the formula $v=e^x$. 
The spectral index of 
speed spectrum is
\begin{eqnarray}
\frac{\dee\ln g(v)}{\dee x}=-\omega
=-\delta\frac{1}{1-v^2/c^2}
=-\delta\frac{1}{1-R}.
\end{eqnarray}
The latter equation can be rewritten as
\begin{eqnarray}
\omega=\delta\frac{1}{1-R}.	
\label{omega and sigma}
\end{eqnarray}
Combining Equations
(\ref{omega and delta with R})
and 
(\ref{omega and sigma})
yields
\begin{eqnarray}
\sigma=2+\delta
\frac{R}{1-\sqrt{1-R}}.
\label{sigma and delta}
\end{eqnarray}

Next, we explore the properties
of parameter $R=v^2/c^2$. 
From the formula of kinetic energy 
$\varepsilon=(\gamma-1)m_0c^2$, we can obtain
\begin{eqnarray}
R=1-\frac{1}{\left(1+\frac{\varepsilon}{m_0c^2}
\right)^2}. 
\end{eqnarray}
Here, $m_0c^2$ is the rest energy 
of particles, and $\varepsilon$ is kinetic 
energy. 
For electron kinetic energy exceeding 1MeV,
we have $m_0c^2\approx0.5$ MeV, and 
\begin{eqnarray}
R\approx 1-\frac{1}
{\left(1+\frac{1}{0.5}\right)^2}
\approx \frac{8}{9}. 
\end{eqnarray}
When the kinetic energy 
of electron increases
starting from 1 MeV to $\infty$, 
the parameter $R$ increases from  
$8/9$ to 1. Thus, 
if the kinetic energy of electron 
is larger than 1 MeV, the parameter $R$ can be approximate to 1, and 
Equation 
(\ref{sigma and delta}) is simplified
as
\begin{eqnarray}
\delta=2+\sigma. 
\label{approximate formula}
\end{eqnarray}
The rest mass of proton is 
approximately equal to $938$ MeV.   
When the kinetic energy of proton
is twice its static energy,
the parameter $R=8/9$, and 
Equation (\ref{approximate formula})
approximately holds. 
That is, for a proton, when its kinetic 
energy is approximately 
larger than 2 GeV, 
the conversion formula for the spectral
indices of the kinetic enegy flux 
and momentum 
follow equation
(\ref{approximate formula}). 

In the context of non-relativism,
the relativistic factor $\gamma$
is approximately equal to 1, and 
the mass of a moving particle
is considered constant. 
For this case,  
the momentum of particle is
\begin{eqnarray}
p=mv,
\label{momentum and v for nonrelativism}
\end{eqnarray}
and the kinetic energy is
\begin{eqnarray}
\varepsilon=\frac{1}{2}mv^2=\frac{p^2}{2m}. 
\label{kinetic energy and momentum for nonrelativism}
\end{eqnarray}

By setting the differential flux 
of kinetic energy 
of energetic particle obeying 
the following the power law
\begin{eqnarray}
j(\varepsilon)=a \varepsilon^{-\sigma}
\end{eqnarray}
with the constant a, 
and using Equation (\ref{j and F(p)}),
we have
\begin{eqnarray}
F(p)=a p^{-2} (2m)^{\sigma}p^{-2\sigma}.
\end{eqnarray}
After taking logrithm of the latter 
equation, the latter equation 
becomes
\begin{eqnarray}
\ln F(p)=\ln a(2m)^{\sigma}
-2(1+\sigma)\ln p.
\end{eqnarray}  
To proceed, the index of momentum
spectrum can be derived
\begin{eqnarray}
\frac{\dee\ln F(p)}{\dee \ln p}=-\delta
=-2(1+\sigma),
\end{eqnarray}
which can be rewritten as
\begin{eqnarray}
\delta=2(1+\sigma).
\label{1}
\end{eqnarray}
Similarly, we can obtain the conversion
formula of spectral indices of 
momentum and speed spectrum
\begin{eqnarray}
\delta=\omega. 
\label{2}
\end{eqnarray} 
Combining equations (\ref{1})
and (\ref{2}) yields
\begin{eqnarray}
\omega=2(1+\sigma).
\end{eqnarray}

\setcounter{equation}{0}  

\section{The power laws for momentum transport equation up to the third-order
momentum derivatives}
\label{The power laws for momentum transport equation up to the third-order
	momentum derivatives}

Inserting Equation (\ref{g}) into 
Equations (\ref{Z4(2)}) yields 
\begin{eqnarray}
	Z_4(2)=\sum_{n=1}^{10}Z_4(2-n)
\end{eqnarray}
with $n=1, 2, 3, \cdots, 10$. 
Among them, the first and second 
terms are as follows
\begin{eqnarray}
	&&
	Z_4(2-1)=
	\frac{1}{2m^2}
	\frac{1}{p^2}\frac{\partial{}}{\partial{p}}p^5
	\Bigg[
	\int_{-1}^1\dee\mu
	\frac{1-\mu^2}{2}
	\nabla\cdot \bm{V}
	+\int_{-1}^1\dee\mu
	\frac{3\mu^2-1}{2}
	\frac{\partial{V_z}}
	{\partial{z}}
	\Bigg]
	\nonumber\\
	&&
	\times
	\Bigg(\int_{-1}^{\mu}\dee \mu \frac{1}{D_{\mu\mu}}
	-\frac{1}{2}
	\int_{-1}^{1}\dee \mu 
	\frac{1-\mu}{D_{\mu\mu}}\Bigg)
	\int_{-1}^{\mu}\dee\mu\mu 
	\Bigg(\int_{-1}^{\mu}\dee \mu \frac{1}{D_{\mu\mu}}
	-\frac{1}{2}
	\int_{-1}^{1}\dee \mu 
	\frac{1-\mu}{D_{\mu\mu}}\Bigg)
	\nonumber\\
	&&
	\times
	\int_{-1}^{\mu}\dee\mu\mu 
	\Bigg(\int_{-1}^{\mu}\dee \mu \frac{1}{D_{\mu\mu}}
	-\frac{1}{2}
	\int_{-1}^{1}\dee \mu 
	\frac{1-\mu}{D_{\mu\mu}}\Bigg)
	\frac{\partial^2{}}
	{\partial{z^2}}
	v\frac{\mu^2-1}{2}
	\frac{\partial{F}}{\partial{z}}
	\label{Z4(2-1)}
	\\
	&&
	Z_4(2-2)=
	\frac{1}{2m^2}
	\frac{1}{p^2}\frac{\partial{}}{\partial{p}}p^5
	\Bigg[
	\int_{-1}^1\dee\mu
	\frac{1-\mu^2}{2}
	\nabla\cdot \bm{V}
	+\int_{-1}^1\dee\mu
	\frac{3\mu^2-1}{2}
	\frac{\partial{V_z}}
	{\partial{z}}
	\Bigg]
	\nonumber\\
	&&
	\times
	\Bigg(\int_{-1}^{\mu}\dee \mu \frac{1}{D_{\mu\mu}}
	-\frac{1}{2}
	\int_{-1}^{1}\dee \mu 
	\frac{1-\mu}{D_{\mu\mu}}\Bigg)
	\int_{-1}^{\mu}\dee\mu\mu 
	\Bigg(\int_{-1}^{\mu}\dee \mu \frac{1}{D_{\mu\mu}}
	-\frac{1}{2}
	\int_{-1}^{1}\dee \mu 
	\frac{1-\mu}{D_{\mu\mu}}\Bigg)
	\nonumber\\
	&&
	\times
	\int_{-1}^{\mu}\dee\mu\mu 
	\Bigg(\int_{-1}^{\mu}\dee \mu \frac{1}{D_{\mu\mu}}
	-\frac{1}{2}
	\int_{-1}^{1}\dee \mu 
	\frac{1-\mu}{D_{\mu\mu}}\Bigg)
	\frac{\partial^2{}}
	{\partial{z^2}}
	v\frac{\partial{}}{\partial{z}}
	\int_{-1}^{\mu}\dee\mu\mu g.
	\label{Z4(2-2)}
\end{eqnarray}
Using Equations (\ref{pF/pz-s}), 
(\ref{p2F/pz2-s}), and 
(\ref{Z4(2-1)}), we obtain
\begin{eqnarray}
	Z_4(2-1)=\sum_{n=1}^{6}Z_4(2-1-n)
\end{eqnarray}
with $n=1, 2, \cdots, 6$. The 
$Z_4(2-1-1)$ is shown as follows
\begin{eqnarray}
	&&
	Z_4(2-1-1)=
	\frac{1}{2m^3}
	\frac{1}{\kappa_\parallel}
	\frac{1}{p^2}\frac{\partial{}}{\partial{p}}p^6
	\Bigg[
	\int_{-1}^1\dee\mu
	\frac{1-\mu^2}{2}
	\nabla\cdot \bm{V}
	+\int_{-1}^1\dee\mu
	\frac{3\mu^2-1}{2}
	\frac{\partial{V_z}}
	{\partial{z}}
	\Bigg]
	\nonumber\\
	&&
	\times
	\Bigg(\int_{-1}^{\mu}\dee \mu \frac{1}{D_{\mu\mu}}
	-\frac{1}{2}
	\int_{-1}^{1}\dee \mu 
	\frac{1-\mu}{D_{\mu\mu}}\Bigg)
	\int_{-1}^{\mu}\dee\mu\mu 
	\Bigg(\int_{-1}^{\mu}\dee \mu \frac{1}{D_{\mu\mu}}
	-\frac{1}{2}
	\int_{-1}^{1}\dee \mu 
	\frac{1-\mu}{D_{\mu\mu}}\Bigg)
	\nonumber\\
	&&
	\times
	\int_{-1}^{\mu}\dee\mu\mu 
	\Bigg(\int_{-1}^{\mu}\dee \mu \frac{1}{D_{\mu\mu}}
	-\frac{1}{2}
	\int_{-1}^{1}\dee \mu 
	\frac{1-\mu}{D_{\mu\mu}}\Bigg)
	\frac{\mu^2-1}{2}
	\frac{\partial{V_z}}{\partial{z}}
	\frac{\partial{F}}{\partial{z}}.
\end{eqnarray}
With Equations  
(\ref{p2F/pz2-s}), 
the latter equation becomes
\begin{eqnarray}
	Z_4(2-1-1)
	=\kappa_{ppp}
	\frac{1}{p^5}
	\frac{\partial{}}
	{\partial{p}}p^8
	\frac{\partial^2{F}}
	{\partial{p^2}}		
\end{eqnarray}
with
\begin{eqnarray}
	\kappa_{ppp}&&=
	\frac{1}{2m^3}
	\frac{1}{V_z}
	\frac{1}{\kappa_\parallel}
	\left(\kappa_{pp}^I
	+\kappa_{pp}^{II}
	\right)
	\Bigg[
	\int_{-1}^1\dee\mu
	\frac{1-\mu^2}{2}
	\nabla\cdot \bm{V}
	+\int_{-1}^1\dee\mu
	\frac{3\mu^2-1}{2}
	\frac{\partial{V_z}}
	{\partial{z}}
	\Bigg]
	\frac{\partial{V_z}}
	{\partial{z}}
	\nonumber\\
	&&
	\times
	\Bigg(\int_{-1}^{\mu}\dee \mu \frac{1}{D_{\mu\mu}}
	-\frac{1}{2}
	\int_{-1}^{1}\dee \mu 
	\frac{1-\mu}{D_{\mu\mu}}\Bigg)
	\int_{-1}^{\mu}\dee\mu\mu 
	\Bigg(\int_{-1}^{\mu}\dee \mu \frac{1}{D_{\mu\mu}}
	-\frac{1}{2}
	\int_{-1}^{1}\dee \mu 
	\frac{1-\mu}{D_{\mu\mu}}\Bigg)
	\nonumber\\
	&&
	\times
	\int_{-1}^{\mu}\dee\mu\mu 
	\Bigg(\int_{-1}^{\mu}\dee \mu \frac{1}{D_{\mu\mu}}
	-\frac{1}{2}
	\int_{-1}^{1}\dee \mu 
	\frac{1-\mu}{D_{\mu\mu}}\Bigg)
	\frac{\mu^2-1}{2}
	\nonumber\\
	&&=
	\frac{3D}{m^3}
	\frac{v}{V_z}
	\left(\kappa_{pp}^I
	+\kappa_{pp}^{II}
	\right)
	\Bigg[
	\int_{-1}^1\dee\mu
	\frac{1-\mu^2}{2}
	\nabla\cdot \bm{V}
	+\int_{-1}^1\dee\mu
	\frac{3\mu^2-1}{2}
	\frac{\partial{V_z}}
	{\partial{z}}
	\Bigg]
	\frac{\partial{V_z}}
	{\partial{z}}
	\nonumber\\
	&&
	\times
	\Bigg(\int_{-1}^{\mu}\dee \mu \frac{1}{D_{\mu\mu}}
	-\frac{1}{2}
	\int_{-1}^{1}\dee \mu 
	\frac{1-\mu}{D_{\mu\mu}}\Bigg)
	\int_{-1}^{\mu}\dee\mu\mu 
	\Bigg(\int_{-1}^{\mu}\dee \mu \frac{1}{D_{\mu\mu}}
	-\frac{1}{2}
	\int_{-1}^{1}\dee \mu 
	\frac{1-\mu}{D_{\mu\mu}}\Bigg)
	\nonumber\\
	&&
	\times
	\int_{-1}^{\mu}\dee\mu\mu 
	\Bigg(\int_{-1}^{\mu}\dee \mu \frac{1}{D_{\mu\mu}}
	-\frac{1}{2}
	\int_{-1}^{1}\dee \mu 
	\frac{1-\mu}{D_{\mu\mu}}\Bigg)
	\frac{\mu^2-1}{2}.
\end{eqnarray}
For the momentum transport 
equation 
\begin{eqnarray}
	\frac{\partial{F}}
	{\partial{t}}
	=\kappa_{ppp}
	\frac{1}{p^2}
	\frac{\partial{}}
	{\partial{p}}p^8
	\frac{\partial^2{F}}
	{\partial{p^2}}	
	\label{third-order momentum transport
		equation}
\end{eqnarray}
Comparing the formulas
\begin{eqnarray}
	\frac{1}{p^5}
	\frac{\partial{}}
	{\partial{p}}p^8
	\frac{\partial^2{F}}
	{\partial{p^2}}	
	=8p^2\frac{\partial^2{F}}
	{\partial{p^2}}	
	+p^3
	\frac{\partial^3{F}}
	{\partial{p^3}}	
\end{eqnarray}
and 
\begin{eqnarray}
	&&p^m
	\frac{\partial{}}
	{\partial{p}}
	p^n	
	\frac{\partial{}}
	{\partial{p}}
	p^{\rho}
	\frac{\partial{}}
	{\partial{p}}
	\left(
	p^{\delta}F
	\right)
	\nonumber\\
	&&
	=
	p^m
	\frac{\partial{}}
	{\partial{p}}
	p^n	
	\frac{\partial{}}
	{\partial{p}}
	p^{\rho}
	\Bigg(
	\delta p^{\delta-1}F
	+
	p^{\delta}
	\frac{\partial{F}}
	{\partial{p}}
	\Bigg)
	\nonumber\\
	&&
	=
	p^m
	\frac{\partial{}}
	{\partial{p}}
	p^n	
	\frac{\partial{}}
	{\partial{p}}
	\Bigg(
	\delta p^{\delta+\rho-1}F
	+
	p^{\delta+\rho}
	\frac{\partial{F}}
	{\partial{p}}
	\Bigg)
	\nonumber\\
	&&
	=
	p^m
	\frac{\partial{}}
	{\partial{p}}
	p^n	
	\Bigg[
	\delta 
	(\delta+\rho-1)
	p^{\delta+\rho-2}F
	+
	(2\delta+\rho)
	p^{\delta+\rho-1}
	\frac{\partial{F}}
	{\partial{p}}
	+
	p^{\delta+\rho}
	\frac{\partial^2{F}}
	{\partial{p^2}}
	\Bigg]
	\nonumber\\
	&&
	=
	p^m
	\frac{\partial{}}
	{\partial{p}}	
	\Bigg[
	\delta 
	(\delta+\rho-1)
	p^{\delta+\rho+n-2}F
	+
	(2\delta+\rho)
	p^{\delta+\rho+n-1}
	\frac{\partial{F}}
	{\partial{p}}
	+
	p^{\delta+\rho+n}
	\frac{\partial^2{F}}
	{\partial{p^2}}
	\Bigg]
	\nonumber\\
	&&
	=\delta(\delta+\rho-1)
	(\delta+\rho+n-2)
	p^{\delta+\rho+n+m-3}F
	\nonumber\\
	&&
	+\Bigg[
	\delta(\delta+\rho-1)
	+(2\delta+\rho)
	(\delta+\rho+n-1)
	\Bigg]
	p^{\delta+\rho+n+m-2}
	\frac{\partial{F}}
	{\partial{p}}
	\nonumber\\
	&&
	+(3\delta+2\rho+n)
	p^{\delta+\rho+n+m-1}
	\frac{\partial^2{F}}
	{\partial{p^2}}
	\nonumber\\
	&&
	+p^{\delta+\rho+n+m}
	\frac{\partial^3{F}}
	{\partial{p^3}}
\end{eqnarray}
yields
\begin{eqnarray}
	&&\delta(\delta+\rho-1)
	(\delta+\rho+n-2)=0,\\
	&&\delta(\delta+\rho-1)
	+(2\delta+\rho)
	(\delta+\rho+n-1)=0,\\
	&&3\delta+2\rho+n=8,\\
	&&\delta+\rho+n+m=3.
\end{eqnarray}
One set of solutions is
\begin{eqnarray}
	&&m=2,\\
	&&n=0,\\
	&&\rho=-5,\\
	&&\delta=6.
\end{eqnarray}
Thus, Equation (\ref{third-order momentum transport equation}) can rewritten 
\begin{eqnarray}
	\frac{\partial{F}}
	{\partial{t}}
	=\kappa_{ppp}
	p^2
	\left\{
	\frac{\partial^2{}}
	{\partial{p^2}}
	p^{-5}
	\left[
	\frac{\partial{}}
	{\partial{p}}
	\left(p^6F\right)
	\right]
	\right\}
\end{eqnarray}
For quasi-steady state,
we obtain
\begin{eqnarray}
	F\propto p^{-6}.
\end{eqnarray}
In addition, another solution set is
\begin{eqnarray}
	&&m=1,\\
	&&n=2,\\
	&&\rho=-6,\\
	&&\delta=6.
\end{eqnarray}
Correspondingly, the momentum transport 
equation is
\begin{eqnarray}
	\frac{\partial{F}}
	{\partial{t}}
	=\kappa_{ppp}
	p
	\left\{
	\frac{\partial{}}
	{\partial{p}}
	p^2
	\frac{\partial{}}
	{\partial{p}}
	p^{-6}
	\left[
	\frac{\partial{}}
	{\partial{p}}
	\left(p^6F\right)
	\right]
	\right\},
\end{eqnarray}
from which, for quasi-steady state,
the momentum power law can be found 
\begin{eqnarray}
	F\propto p^{-6}.
\end{eqnarray}

Similarly, from $Z_4(2-2-1)$ we can 
derive the fourth-order momentum derivative
term is
\begin{eqnarray}
	Z_4(2-2-1)
	=\kappa_{4p}
	\frac{1}{p^6}\frac{\partial{}}{\partial{p}}p^8
	\frac{\partial{}}{\partial{p}}
	p^2\frac{\partial^2{F}}
	{\partial{p^2}}	
\end{eqnarray}
with
\begin{eqnarray}
	\kappa_{4p}&&=
	\frac{6D^2}{m^4}
	\left(\kappa_{pp}^I
	+\kappa_{pp}^{II}
	\right)
	\Bigg[
	\int_{-1}^1\dee\mu
	\frac{1-\mu^2}{2}
	\nabla\cdot \bm{V}
	+\int_{-1}^1\dee\mu
	\frac{3\mu^2-1}{2}
	\frac{\partial{V_z}}
	{\partial{z}}
	\Bigg]
	\left(\nabla \cdot\bm{V}\right)
	\nonumber\\
	&&
	\times
	\Bigg(\int_{-1}^{\mu}\dee \mu \frac{1}{D_{\mu\mu}}
	-\frac{1}{2}
	\int_{-1}^{1}\dee \mu 
	\frac{1-\mu}{D_{\mu\mu}}\Bigg)
	\int_{-1}^{\mu}\dee\mu\mu 
	\Bigg(\int_{-1}^{\mu}\dee \mu \frac{1}{D_{\mu\mu}}
	-\frac{1}{2}
	\int_{-1}^{1}\dee \mu 
	\frac{1-\mu}{D_{\mu\mu}}\Bigg)
	\nonumber\\
	&&
	\times
	\int_{-1}^{\mu}\dee\mu\mu 
	\Bigg(\int_{-1}^{\mu}\dee \mu \frac{1}{D_{\mu\mu}}
	-\frac{1}{2}
	\int_{-1}^{1}\dee \mu 
	\frac{1-\mu}{D_{\mu\mu}}\Bigg)
	\int_{-1}^{\mu}\dee\mu\mu 
	\Bigg(\int_{-1}^{\mu}\dee \mu \frac{1}{D_{\mu\mu}}
	-\frac{1}{2}
	\int_{-1}^{1}\dee \mu 
	\frac{1-\mu}{D_{\mu\mu}}\Bigg)
	\frac{\mu^2-1}{2}
\end{eqnarray}
Comparing the latter formula 
\begin{eqnarray}
	\frac{1}{p^6}\frac{\partial{}}{\partial{p}}p^8
	\frac{\partial{}}{\partial{p}}
	p^2\frac{\partial^2{F}}
	{\partial{p^2}}
	=18p^2
	\frac{\partial^2{F}}
	{\partial{p^2}}
	+12p^3
	\frac{\partial^3{F}}
	{\partial{p^3}}
	+p^4
	\frac{\partial^4{F}}
	{\partial{p^4}}	
\end{eqnarray}
and the following equation 
\begin{eqnarray}
	&&p^m
	\frac{\partial{}}
	{\partial{p}}
	p^n	
	\frac{\partial{}}
	{\partial{p}}
	p^{\rho}
	\frac{\partial{}}
	{\partial{p}}
	p^{q}
	\frac{\partial{}}
	{\partial{p}}
	\left(
	p^{\delta}F
	\right)
	\nonumber\\
	&&
	=
	\delta (\delta+q-1)(\delta+q+\rho
	-2)
	(\delta+q+\rho+n-3)
	p^{\delta+q+\rho+n+m-4}F
	\nonumber\\	
	&&
	+
	\Bigg[
	\delta (\delta+q-1)(\delta+q+\rho
	-2)
	+\left[
	\delta (\delta+q-1)
	+(2\delta+q)(\delta+q+\rho-1)
	\right]
	(\delta+q+\rho+n-2)
	\Bigg]
	p^{\delta+q+\rho+n+m-3}
	\frac{\partial{F}}
	{\partial{p}}
	\nonumber\\	
	&&
	+
	\Bigg\{
	\left[
	\delta (\delta+q-1)
	+(2\delta+q)(\delta+q+\rho-1)
	\right]
	+\left[
	(2\delta+q)
	+(\delta+q+\rho)
	\right]
	(\delta+q+\rho+n-1)
	\Bigg\}
	p^{\delta+q+\rho+n+m-2}
	\frac{\partial^2{F}}
	{\partial{p^2}}
	\nonumber\\	
	&&
	+
	\Bigg\{
	\left[
	(2\delta+q)
	+(\delta+q+\rho)
	\right]
	+(\delta+q+\rho+n)
	\Bigg\}
	p^{\delta+q+\rho+n+m-1}
	\frac{\partial^3{F}}
	{\partial{p^3}}
	\nonumber\\	
	&&
	+
	p^{\delta+q+\rho+n+m}
	\frac{\partial^4{F}}
	{\partial{p^4}}
\end{eqnarray}
gives equations
\begin{eqnarray}
	&&\delta (\delta+q-1)(\delta+q+\rho
	-2)
	(\delta+q+\rho+n-3)=0,\\
	\label{1}
	&&\delta (\delta+q-1)(\delta+q+\rho
	-2)
	+\left[
	\delta (\delta+q-1)
	+(2\delta+q)(\delta+q+\rho-1)
	\right]
	(\delta+q+\rho+n-2)=0,\\
	&&\left[
	\delta (\delta+q-1)
	+(2\delta+q)(\delta+q+\rho-1)
	\right]
	+\left[
	(2\delta+q)
	+(\delta+q+\rho)
	\right]
	(\delta+q+\rho+n-1)=18,\\
	&&\delta+q+\rho+n+m-2=2,\\
	&&\left[
	(2\delta+q)
	+(\delta+q+\rho)
	\right]
	+(\delta+q+\rho+n)=12.
\end{eqnarray}
One solution set of the latter equations 
is $m=1, n=2, q=-6, \rho=0, \delta=7$. 
Thus, the fourth-order momentum transport
equation can be obtained
\begin{eqnarray}
	\frac{\partial{F}}
	{\partial{t}}=\kappa_{p4}
	p
	\frac{\partial{}}
	{\partial{p}}
	p^2
	\frac{\partial^2{}}
	{\partial{p^2}}
	p^{-6}
	\frac{\partial{}}
	{\partial{p}}
	\left(
	p^{7}F
	\right),	
\end{eqnarray}
which contain the following momentum
power law for quais-steady state
\begin{eqnarray}
	F\propto p^{-7}.
\end{eqnarray}

\end{appendices}

\end{CJK*}

\end{document}